\documentclass[11pt]{article}
\pdfoutput=1
\usepackage[dvipsnames]{xcolor}
\usepackage{amsmath}
\usepackage{amssymb}
\usepackage{braket}
\usepackage{caption}
\usepackage{comment} 
\usepackage[all,cmtip]{xy}
\usepackage{dsfont}
\usepackage{stackengine,scalerel}
\usepackage{float}
\restylefloat{table}
\usepackage{fancyhdr}
\usepackage{hyperref}

\hypersetup{
    unicode=true,          
    pdftoolbar=true,        
    pdfmenubar=true,        
    pdffitwindow=false,     
    pdfstartview={FitH},    
    pdftitle={My title},    
    pdfauthor={Author},     
    pdfsubject={Subject},   
    pdfcreator={Creator},   
    pdfproducer={Producer}, 
    pdfkeywords={keywords}, 
    pdfnewwindow=true,      
    colorlinks=false,       
    linkcolor=red,          
    citecolor=cayn,        
    filecolor=magenta,      
    urlcolor=cyan           
}
 
\newcommand\ocirc[1]{\ThisStyle{\ensurestackMath{%
  \stackon[1pt]{\SavedStyle#1}{\SavedStyle\kern.6\LMpt\circ}}}}

\newcommand{\xA}{{\mbox{\footnotesize(1)}}}
\newcommand{\xB}{{\mbox{\footnotesize(2)}}}
\newcommand{\delt}{\delta\,(\,{\mbox{\footnotesize{2}}}\,-\,{\mbox{\footnotesize 1}}\,)}
\newcommand{\deltP}{\delta^{\prime}\,(\,{\mbox{\footnotesize{2}}}\,-\,{\mbox{\footnotesize 1}}\,)}
\newcommand{\defeq}{\mathrel{\mathop:}=}

\newcommand{\sN}{\mathcal{N}}

\newcommand{\Es}{{\scriptsize\mathcal{E}}}

\newcommand{\Ms}{{\scriptsize\mathcal{M}}}
\newcommand{\Ns}{{\scriptsize\mathcal{N}}}

\def\be{\begin{equation}}
\def\ee{\end{equation}}

\def\ba{\begin{eqnarray}}
\def\ea{\end{eqnarray}}

\def\frac#1#2{{\textstyle{#1\over#2}}}
\def\Fontamici#1{{$\mathcal{#1}$}}
\def\go#1{{\mbox{{\scriptsize\Fontamici{#1}}}}}

\def\rgboo#1{\pdfliteral{#1 rg #1 RG}}
\def\pdfklink#1#2{%
	\noindent\pdfstartlink user
		{/Subtype /Link
		/Border [ 0 0 0 ]
		/A << /S /URI /URI (#2) >>}{\rgb{1 0 0}{#1}}%
	\pdfendlink}
\def\rgbo#1#2{\rgboo{#1}#2\rgboo{0 0 0}}
\def\rgb#1#2{\mark{#1}\rgbo{#1}{#2}\mark{0 0 0}}

\def\xxxlink#1{\pdfklink{[arXiv:#1]}{http://arXiv.org/abs/#1}}

\def\Gamma{\mathchar"0100}
\def\Delta{\mathchar"0101}
\def\Theta{\mathchar"0102}
\def\Lambda{\mathchar"0103}
\def\Xi{\mathchar"0104}
\def\Pi{\mathchar"0105}
\def\Sigma{\mathchar"0106}
\def\Upsilon{\mathchar"0107}
\def\Phi{\mathchar"0108}
\def\Psi{\mathchar"0109}
\def\Omega{\mathchar"010A}

\begin{document}

\title{{\bf\color{blue} Pre-potential in the $AdS_{5}\,\times\,S^{5}$ Type IIB superspace}\\[-2in]
{\normalsize August 2, 2016\hfill YITP-SB-16-33}\\[1.8in]}
\date{}
\author{Martin Pol\'a\v cek\footnote{\pdfklink{martin.polacek@stonybrook.edu}{mailto:martin.polacek@stonybrook.edu}},\ \ 
Warren Siegel\footnote{\pdfklink{siegel@insti.physics.sunysb.edu}{mailto:siegel@insti.physics.sunysb.edu},
	\pdfklink{http://insti.physics.sunysb.edu/\~{}siegel/plan.html}{http://insti.physics.sunysb.edu/\%7Esiegel/plan.html}}\\
{\textit{C. N. Yang Institute for Theoretical Physics}}\\ {\textit{State University of New York, Stony Brook, NY 11794-3840}}}

\maketitle

\begin{abstract}
\normalsize
We found the pre-potential in the superspace with $AdS_{5}\,\times\,S^{5}$ background. The pre-potential appears as part of the vielbeins, without derivatives. The space-cone gauge destroys the bulk Lorentz covariance, but still preserves boundary Lorentz covariance (projective superspace) $SO\,(\,3,\,1\,)\,\otimes\,SO\,(\,4\,)$, i.e. symmetries of boundary CFT are manifest.

\end{abstract}

\newpage

\tableofcontents

\section{\texorpdfstring{Introduction }{Introduction}}
In the paper \cite{natural} we obtained the curvature tensor (previously discovered in \cite{warren}) in a manifestly T-dual way. In the paper \cite{3d} we extended the techniques for the case of three dimensional $\sN\,=\,2$  T-dual extended superspace. There we correctly obtained the pre-potential (as a part of vielbein), structure of linearised dilaton and field equations.  The aim of this paper was to look at the full ten dimensional $\sN\,=\,2$ T-dually extended superspace in the flat and also in $AdS_{5}\,\times\,S^{5}$ background, i.e. IIB string theory expanded around $AdS_{5}\,\times\,S^{5}$ background. The $AdS$ was earlier analysed in superspace in papers \cite{one1}, \cite{one2}, \cite{implicat1} and \cite{implicat2}. In this paper we discovered the projective (and also the chiral) pre-potential to sit in a certain combination of $H_{S\,\widetilde{S}}$ and $H_{D\,\widetilde{D}}$. This was first obtained in the flat case and later generalised for the $AdS_{5}\,\times\,S^{5}$. We also performed the near horizon limit and derived the equation of motion for the pre-potential in that limit. This limit also picks out the projective pre-potential instead of the chiral pre-potential, even though both pre-potentials are valid bulk solutions. The projective and harmonic superspaces were earlier analysed in \cite{HHH1} and \cite{HHH2}.
 
\section{\texorpdfstring{Type II superspace, notation and motivation}{Type II superspace, notation and motivation}}
\subsection{\texorpdfstring{$10$ dimensional type II superspace}{10 dimensional type II superspace}}
We are closely following paper \cite{warren1}. In that paper the superspace was defined by two sets of non-degenerate super-Poincar\'e covariant derivatives (and their stringy generalisations). For convenience we are repeating the graded commutation relations for the flat space (string) covariant derivatives $\ocirc{\nabla}_{\go{M}}$ and symmetry generators $\widetilde{\nabla}_{\go{M}}$. The multi-index $\go{M}\,\in\,(\,S,\,D,\,P,\,\Omega,\,\Sigma)$. The indices $(\,S,\,D,\,P,\,\Omega,\,\Sigma\,)$ are multi-indices for (left and right): local Lorentz, supersymmetry, translation, and their dual generators. The whole approach is explicitly explained in \cite{warren1} and also in \cite{natural} and in foundational work \cite{warren} and later \cite{warrenNew}. The generators $\ocirc{\nabla}_{\go{M}}$ and $\widetilde{\nabla}_{\go{M}}$ satisfy:
\begin{eqnarray}
\label{covariant}
{[}\,\ocirc{\nabla}_{\go{M}}\,{{(1)}},\,\ocirc{\nabla}_{\go{N}}\,{{(2)}}\,{\}}&=
&i\,\ocirc{\eta}_{\go{M}\,\go{N}}\,\,\deltP\,+\,i\,f_{\go{M}\,\go{N}}^{\,\,\,\,\,\,\,\,\,\,\,\,\,\,\go{K}}\,\ocirc{\nabla}_{\go{K}}\,\delt\\
{[}\,\widetilde{\nabla}_{\go{M}}\,\,{{(1)}},\,\widetilde{\nabla}_{\go{N}}\,\,{{(2)}}\,{\}}&=&-i\,\widetilde{\eta}_{\go{M}\,\go{N}}\,\,\deltP\,-\,i\,f_{\go{M}\,\go{N}}^{\,\,\,\,\,\,\,\,\,\,\,\,\,\,\go{K}}\,\widetilde{\nabla}_{\go{K}}\,\delt\nonumber\\
{[}\,\widetilde{\nabla}_{\go{M}}\,\,{{(1)}},\,\ocirc{\nabla}_{\go{N}}\,\,{{(2)}}\,{\}}&=&0.\nonumber
\end{eqnarray}
The non-zero structure constants and the central charges for the left $10$ dimensional non-degenerate super-Poincar\'e affine algebra:
\begin{eqnarray}
\label{algebra}
{[}\,S_{\bf{m}\bf{n}}\,{{(1)}},\,S_{\bf{k}\bf{l}}\,{{(2)}}\,{]}&=
&-i\,\eta_{[\,\bf{m}\,[\,\bf{k}}\,S_{\bf{l}\,]\,\bf{n}\,]}\,\,\delta\,(\,{{{2}}}\,-\,{{ 1}}\,)\\
{[}\,S_{\bf{m}\bf{n}}\,{{(1)}},\,D_{\bf{\rho}}\,\,{{(2)}}\,{]}&=&\,-\,i\,\frac{1}{2}\,(\,\gamma_{\bf{m}\bf{n}}\,)^{\bf{\sigma}}{}_{\bf{\rho}}\,D_{\bf{\sigma}}\,\delt\nonumber\\
{[}\,S_{\bf{m}\bf{n}}\,\xA,\,P_{\bf{k}}\,\xB\,{]}&=&i\,\eta_{\bf{k}\,[\,\bf{m}}\,P_{\bf{n}\,]}\,\delt \nonumber\\
{[}\,S_{\bf{m}\bf{n}}\,\xA,\,\Omega^{\bf{\rho}}\,\xB\,{]}&=&\,i\,\frac{1}{2}\,(\,\gamma_{\bf{m}\bf{n}}\,)^{\bf{\rho}}{}_{\bf{\sigma}}\,\Omega^{\bf{\sigma}}\,\delt\nonumber\\
{[}\,S_{\bf{m}\bf{n}}\,\xA,\,\Sigma^{\bf{k}\bf{l}}\,\xB\,{]}&=
&i\,\delta_{\bf mn}{}^{\bf kl}\,\deltP\,-\,i\delta_{[\,\bf{m}}{}^{[\,\bf{k}}\,\eta_{\bf{n}\,]\,\bf{s}}\Sigma^{\bf{l}]\bf{s}}\,\delt\nonumber\\
{\{}\,D_{\bf{\rho}}\,\xA,\,D_{\bf{\sigma}}\xB\,{\}}&=&\,2\,(\,\gamma^{\bf{m}}\,)_{\bf{\rho\sigma}}\,P_{\bf{m}}\,\delt\nonumber\\
{[}\,D_{\bf{\rho}}\,\xA,\,P_{\bf{m}}\,\xB\,{]}&=&\,2\,(\,\gamma_{\bf{m}})_{\bf{\rho\sigma}}\,\Omega^{\bf{\sigma}}\,\delt\nonumber\\
{\{}\,D_{\bf{\rho}}\,\xA,\,\Omega^{\bf{\sigma}}\,\xB\,{\}}&=&\,i\,\delta_{\bf{\rho}}^{\bf{\sigma}}\,\deltP\,-\,i\,\frac{1}{4}\,(\,\gamma_{\,\bf{m}\bf{n}\,}\,)_{\,\,\,\bf{\rho}}^{\bf{\sigma}}\,\Sigma^{\bf{mn}}\,\delt\nonumber\\
{[}\,P_{\bf{m}}\,\xA,\,P_{\bf{n}}\,\xB\,{]}&=
&\,i\,\eta_{\bf mn}\,\deltP\,+\,i\,\eta_{\bf{m}\,\bf{h}}\,\eta_{\bf{n}\,\bf{s}}\Sigma^{\bf hs}\,\delt\nonumber\\
&{\xymatrix{\ar@/_/@{>}[r]&}}&\mbox{{ left algebra $\,\rightarrow\,$ $-$ right algebra}}\nonumber\\
{[}\,\mbox{left},\,\mbox{right}\,\}\,&=&\,0.\nonumber
\end{eqnarray}
As indicated above, the algebra for the right generators is the same up to the overall sign. We can assign the canonical dimensions to the generators: $\mbox{dim}\,(S,\,D,\,P,\,\Omega,\Sigma)\,=\,(0,\,\frac{1}{2},\,1,\,\frac{3}{2},\,2)$. The $S$ generators generate the algebra for $SO\,(\,9,1\,)\,\otimes\,SO(\,9,\,1\,)$, i.e. left and right local Lorentz transformations. The $D$ generate left and right SUSY transformation and $P$ left and right translations. The $\Omega$ and $\Sigma$ are the left and right dual currents (corresponding to $D$ and $S$), see also \cite{warren}.

The only nonvanishing terms in the metric and structure constants are (as can be guessed by dimensional analysis):
\be
\eta_{PP} , \ \eta_{S\Sigma}, \ \eta_{D\Omega} , ; \quad f_{SPP} \ , \  f_{SS\Sigma} \ , \ f_{DDP}\ , \ f_{SD\Omega}
\ee
where we have lowered the upper index on $f$ with $\eta$ to take advantage of its total (graded) antisymmetry, and used ``schematic" notation, replacing explicit indices with their type:
\be
\label{schem}
{{\go{M}}}
\,\defeq\,(\,_{MN},\,_{\mu},\,_M,\,^{\mu},\,^{\,MN\,}\,)
\,\defeq\,(\,S,\,D,\,P,\,\Omega,\,\Sigma\,)
\ee
Explicitly these are, for the left-handed algebra:
\be
(\eta)_{\bf mn}\,=\,\eta_{\bf mn}\,,\, \  (\eta)_{\bf mn}{}^{\bf pq} \,=\, \delta_{\bf mn}{}^{\bf pq} \,,\,(\eta)_{\bf \sigma}{}^{\bf \rho}\,=\,\delta_{\bf \sigma}^{\bf \rho} \\
\ee
\be
\label{ind}
f_{\bf mn}{}^{\bf p\,q}\,=\,-\,\delta_{\bf mn}{}^{\bf pq} \,,\, \ f_{\bf mn\,pq}{}^{\bf rs} \,=\, \eta_{\bf [m[p} \delta_{\bf q]n]}{}^{\bf rs} \,,\, f_{\bf \sigma\rho}{}^{\bf m}\,=\,2\,(\,\gamma^{\bf m}\,)_{\bf \sigma\rho}\,,\,f_{\bf mn\,\sigma}{}^{\bf \rho}\,=\,-\,\frac{1}{2}\,(\,\gamma_{\bf mn}\,)_{\bf \sigma}^{\bf \rho}
\ee

To proceed further we introduce the explicit notation for the left and right indices. We will call the left index to be the one indicated in algebra \ref{algebra} (i.e. without tilde). The right index will be with tilde: $(\mbox{left})\,\equiv\,\go M\,\equiv\,(S_{\bf{mn}},\,D_{\bf{\mu}},\,P_{\bf{m}},\,\Omega^{\bf{\mu}},\,\Sigma^{\bf{mn}})\,\equiv\,(S,\,D,\,P,\,\Omega,\,\Sigma)$ and $(\mbox{right})\,\equiv\,\widetilde{\go M}\,\equiv\,(S_{\bf{\widetilde{mn}}},\,D_{\bf{\widetilde{\mu}}},\,P_{\bf{\widetilde{m}}},\,\Omega^{\bf{\widetilde{\mu}}},\,\Sigma^{\bf{\widetilde{mn}}})\,\equiv\,(\widetilde{S},\,\widetilde{D},\,\widetilde{P},\,\widetilde{\Omega},\,\widetilde{\Sigma})$. (Note, we abuse the notation for indices a bit, compare with general indices definition in (\ref{schem}). From the context it should be clear which definition we are using). 

\subsection{\texorpdfstring{Gamma matrices}{Gamma matrices}}
The gamma matrices $(\gamma_{\bf{m}})_{\mu\,\nu}$ used in the algebra (\ref{algebra}) are the $16\otimes16$ block gamma matrices from $10$ dimensional $32\otimes32$ chiral representation:
\begin{eqnarray}
\label{gam}
\Gamma_{\bf{m}} = 
 \begin{pmatrix}
  0  &  (\gamma_{\bf{m}})^{\mu\,\nu}\\
  (\gamma_{\bf{m}})_{\mu\,\nu}  & 0 
 \end{pmatrix}&\mbox{where}&\{\,\Gamma_{\bf{m}},\,\Gamma_{\bf{n}}\,\}\,=\,2\,\eta_{\bf{m}\,\bf{n}}\,\delta.
\end{eqnarray} 
Moreover the block gamma matrices satisfy: 
\begin{eqnarray}
\label{vlast}
(\gamma_{\bf{m}})_{\mu\,\nu}\,=\,(\gamma_{\bf{m}})_{\nu\,\mu}\,\,\,\,\,||\,\,&(\gamma_{\bf{(m}})^{\mu\,\nu}\,(\gamma_{\bf{\,n)}})_{\nu\,\sigma}\,=\,2\,\eta_{\bf{m\,n}}\,\delta_{\sigma}^{\mu}\,\,\,\,\,||\,\,&(\gamma_{\bf{m}})_{(\mu\,\nu}(\gamma^{\bf{m}})_{\sigma)\,\lambda}\,=\,0
\end{eqnarray}

The IIB fermion generators (in algebra \ref{algebra}) are described by $16\,\oplus\,16$ chiral fermion generators (for left and right generators) with same $10$ dimensional chirality.  For the future use we need to look closer at the structure of the matrices $(\gamma_{\bf{m}})_{\mu\nu}$ from equation (\ref{gam}). The equation (\ref{gam}) gamma matrices could be constructed from $SO\,(\,9\,)$ gamma matrices or equivalently from $SO\,(\,8\,)$ gamma matrices and the chirality matrix.
We can go one step down and construct the $SO\,(\,8\,)$ gamma matrices from $SO\,(\,6\,)$ gamma matrices. For the $SO\,(\,6\,)$ gamma matrices we use the Majorana representation of those matrices (they are purely imaginary). Thus we can get the Majorana - Weyl representation of the $SO\,(\,8\,)$ gamma matrices.  

For the future reference we will define the following $16\,\otimes\,16$ matrix ${\widetilde{\Gamma}}_{5}$:
\be
\label{gam5}
\widetilde{\Gamma}_{5}\,\defeq\,\gamma_{{\bf{10}}}\,\prod \limits_{{\bf{m}}\,=\,1}^{4}\,\gamma_{{\bf{m}}}
\ee
where $\gamma_{{\bf{m}}}$ are block gamma matrices from (\ref{gam}). 

\subsection{\texorpdfstring{Space - cone basis and indices}{Space - cone basis and indices}}

Because in the future we will use the space - cone (and light - cone) basis we introduce it for the gamma matrices we constructed in previous subsection. We first notice that the block gamma matrices $\gamma_{{\bf{m}}}$ in equation (\ref{gam}) could have either upper indices  $(\gamma_{{\bf{m}}})^{\mu\,\nu}$ or lower indices $(\gamma_{{\bf{m}}})_{\mu\,\nu}$. From the construction it follows that those matrices are equal up to the sign. 

In the equation (\ref{gam}) let us further divide the (either upper or lower) $16$ dimensional index $\mu$ to $8\,\oplus\,8$ pieces (they are the $SO\,(\,8\,)$ chiral indices), thus we introduce $\mu\,\defeq\,(\,\mu,\,\mu^{\prime}\,)$. In another words we want to look how the block $\gamma_{{\bf{m}}}$ matrices look in the $SO\,(\,8\,)$ (Majorana - Weyl) basis. Furthermore we introduce the following space - cone combinations of the equation (\ref{gam}) block gamma matrices:
\begin{eqnarray}
\label{con}
(\gamma_{{\bf{+}}})_{\mu\,\nu}\,=\,\frac{1}{2}\,(\,\gamma_{{\bf{10}}}\,+\,\gamma_{{\bf{9}}}\,)_{\mu\,\nu}\,=
\begin{pmatrix}
  0  &  0\\
  0  & \delta_{\mu^{\prime}\,\nu^{\prime}}\end{pmatrix} \,\,\,||\,\,\,(\gamma_{{\bf{-}}})_{\mu\,\nu}\,=\,\frac{1}{2}\,(\,\gamma_{{\bf{10}}}\,-\,\gamma_{{\bf{9}}}\,)_{\mu\,\nu}\,=
\begin{pmatrix}
  \delta_{\mu\,\nu}  &  0\\
  0  & 0
 \end{pmatrix}\\
 (\gamma_{{\bf{+}}})^{\mu\,\nu}\,=\,\frac{1}{2}\,(\,\gamma_{{\bf{10}}}\,+\,\gamma_{{\bf{9}}}\,)^{\mu\,\nu}\,=
\begin{pmatrix}
  \delta^{\mu\,\nu} &  0\\
  0  & 0\end{pmatrix} \,\,\,||\,\,\,(\gamma_{{\bf{-}}})^{\mu\,\nu}\,=\,\frac{1}{2}\,(\,\gamma_{{\bf{10}}}\,-\,\gamma_{{\bf{9}}}\,)^{\mu\,\nu}\,=
\begin{pmatrix}
  0  &  0\\
  0  & \delta^{\mu^{\prime}\,\nu^{\prime}}
 \end{pmatrix}
 \nonumber
\end{eqnarray}
For the convenience we also write the remaining gamma matrices using the $SO\,(\,8\,)$ indices: 
\begin{eqnarray}
\label{gamei}
(\gamma_{\,{\bf{i}}})_{\mu\,\nu}\,=\,\begin{pmatrix}
  0  &  (\,\widetilde{\gamma}_{\,{\bf{i}}}\,)_{\mu\,\nu^{\prime}}\\
  (\,\widetilde{\gamma}_{\,{\bf{i}}}\,)_{\mu^{\prime}\,\nu}  & 0 \end{pmatrix} \,\,\,||\,\,\,(\gamma_{\,{\bf{i}}})^{\mu\,\nu}\,=\,\begin{pmatrix}
  0  &  (\,\widetilde{\gamma}_{\,{\bf{i}}}\,)^{\mu\,\nu^{\prime}}\\
  (\,\widetilde{\gamma}_{\,{\bf{i}}}\,)^{\mu^{\prime}\,\nu}  & 0 \end{pmatrix}
\end{eqnarray}
where the $\widetilde{\gamma}_{\,\bf{i}}$ are the $SO\,(\,8\,)$ gamma matrices.

In the above introduced $8\,\oplus\,8$ basis the (\ref{gam5}) looks like $\sigma_{3}\,\otimes\,{\mathds{1}}$ where ${\mathds{1}}$ is the $8\,\otimes\,8$ unit matrix and $\sigma_{3}$ is the Pauli matrix. 

\section{\texorpdfstring{The $AdS$ background in the T-dually extended superspace}{The AdS background in the T-dually extended superspace}}
\subsection{\texorpdfstring{Short review of the theory in curved background}{Short review of the theory in curved background}}
In the usual treatment of the theory (of T-dually extended superspaces) the curved background is introduced via vielbeins $E_{{{\go{A}}}}{}^{{{{\go{M}}}}}(X^{{{\go{N}}}})$, see papers \cite{warren}, \cite{natural}, \cite{3d}:
\be
\label{dur}
\Pi_{{{\go{A}}}}\xA\,=\,E_{{{\go{A}}}}{}^{{{{\go{M}}}}}(X^{{{\go{N}}}})\,\ocirc{\nabla}_{\go{M}}
\ee
where  $\ocirc{\nabla}_{\go{M}}$ are generators of the flat algebra (\ref{algebra}). The affine Lie algebra for the $\Pi_{{{\go{A}}}}\xA\,$ can be written as:
\begin{equation}
\begin{array}{cccccc}
\label{Lenuska1}
{[}\Pi_{{{\go{A}}}}\xA,\Pi_{{{\go{C}}}}\xB{]}\,\equiv\,-i\eta_{{{\go{A}}}{{\go{C}}}}\,\deltP-iT_{{{\go{A}}}{{\go{C}}}}{}^{{{\go{E}}}}\Pi_{{{\go{E}}}}\,\delt
\end{array}
\end{equation}
where $T_{{{\go{A}}}{{\go{C}}}}{}^{{{\go{E}}}}$ is a (super)stringy generalisation of torsion, see \cite{natural}:
\be
\label{torsion}
T_{{{\go{A}}}{{\go{C}}}}{}^{{{\go{E}}}}=E_{[{{\go{A}}}}{}^{{{\go{M}}}}(D_{{{\go{M}}}}E_{{{\go{C}}})}{}^{{{\go{N}}}})E^{-1}_{{{\go{N}}}}{}^{{{\go{E}}}}
+\frac{1}{2}\eta^{{{\go{E}}}{{\go{D}}}}E_{{{\go{D}}}}{}^{{{\go{M}}}}(D_{{{\go{M}}}}E_{[{{\go{A}}}|}{}^{{{\go{N}}}})E^{-1}_{{{\go{N}}}}{}^{{{\go{F}}}}\eta_{{{\go{F}}}|{{\go{C}}})}
+E_{{{\go{A}}}}{}^{{{\go{M}}}}E_{{{\go{C}}}}{}^{{{\go{N}}}}E^{-1}_{{{\go{P}}}}{}^{{{\go{E}}}}f_{{{\go{M}}}{{\go{N}}}}{}^{{{\go{P}}}}
\ee
where $[\,\go A\,|\,|\,\go C\,)$ indicates graded anti-symmetrization in only those indices. By $D_{\Ms}$ in (\ref{torsion}) and in the whole text we mean the group covariant derivatives of the (non-affine) part of algebra (\ref{algebra}): $[\,D_{\Ms},\,D_{\Ns}\,\}\,=\,i\,f_{\Ms\,\Ns}{}^{\Es}\,D_{\Es}$.
 
Note that the (super)Jacobi identities imply the total graded antisymmetry of the torsion, just as for the structure constants. Torsion (\ref{torsion}) can be identified with that of ``ordinary" curved-space covariant derivatives by use of the strong constraint, as explained in \cite{natural,warren}. 

We can set the coefficient of the Schwinger term to be the metric $\eta$; the vielbein is forced to obey the orthogonality constraints: 
\begin{equation}
\label{equation}
E_{{{\go{A}}}}{}^{{{\go{M}}}}\eta_{{{\go{M}}}{{\go{N}}}}\,E_{\,{{\go{C}}}}{}^{{{\go{N}}}}\,\equiv\,\eta_{{{\go{A}}}{{\go{C}}}}
\end{equation}
This choice  does not affect the physics, and simplifies many of the expressions. For example, it implies the total graded antisymmetry of the torsion, when the upper index is implicitly lowered with $\eta$:
\be
\label{t2}
T_{\go{A\,B\,C}} \,=\,\frac12 E_{[\,\go A\,|}{}^{\go M}(D_{\go M} E_{|\,\go B}{}^{\go N})E_{\go C\,)\,\go N}\,+\,E_{\go A}{}^{\go M} E_{\go B}{}^{\go N} E_{\go C}{}^{\go P} f_{\go{M\,N\,P}}
\ee
where we have used $E^{-1}_{\,\,\,{{\go{M}}}}{}^{\,{{\go{A}}}}\,=\,\eta^{{{\go{A\ B}}}}\eta_{{{\go{M\ N}}}}E_{\,{{\go{B}}}}{}^{{{\go{N}}}}$.  (Also note that in the first term the graded anti-symmetrization can be written as a cyclic sum without the $1/2$, since it is already graded antisymmetric in the last two indices.)  Thus, because of orthogonality, the vielbein is like (the exponential of) a super $2$-form, while the torsion is a super $3$-form; similarly, the Bianchi identities are a super $4$-form.

To solve the theory (in terms of pre-potential, to get physical fields and equation of motion) orthogonality condition (\ref{equation}) has to be solved explicitly (or at least at linearised level). Moreover there is a huge gauge group invariance that should be fixed:
\begin{eqnarray}
\label{gaugeg}
\delta_{\Lambda}\,\Pi_{\go A}\,=\,[\,-\,i\,\Lambda,\,\Pi_{\go A}\,\}&&\mbox{where}\,\,\Lambda\,\defeq\,\int\,d1\,\lambda^{\go M}\,(\,X\,)\,D_{\go M}
\end{eqnarray}
At the top of the orthogonality condition and the gauge invariance, we should include the torsion constraints. The torsion constraints are additional constraints on vielbein. They are imposed by putting some of the torsions in (\ref{t2}) to zero. Of course, not all torsions in (\ref{t2}) are zero. The relevant torsion constraints have been carefully analysed in \cite{warren1}. All possible constraints on torsions are coming from curved space version of the ${\bf{A\,B\,C\,D}}$ (first class) constraints: ${\bf{A}}$ Virasoro, (string world-sheet) diffeomorphism constraints, {\bf{B}} and {\bf{C}} and ${\bf{D}}$ are the first class fermionic $\kappa$ symmetry constraints, for further details see \cite{warrenTor}, \cite{wittenTwistor}, \cite{town}, \cite{berg}, \cite{shap} and \cite{warren1}. The rule of thumb is that at least the torsions of negative (10 dimensional) engineering dimension should be zero. 

We will not try to solve the full nonlinear version of the theory. We linearise the theory around some background. In previous papers \cite{natural}, \cite{3d} we linearised around the flat background. In this paper however we linearise the theory around the $AdS_{5}\,\times\,S^{5}$ background solution of the classical supergravity (reformulated in the language of the doubled algebra).    

After the linearisation we rewrite the orthogonality constraints (\ref{equation}) and torsions (\ref{t2}) using the vielbein expansion $E_{\go C}{}^{\go D}\,=\,\delta_{\go C}{}^{\go D}\,+\,E^{{\scriptsize (1)}}{}_{\go C}{}^{\go D}\,+\,{\mathcal{O}}\,(\,E^{{\scriptsize (2)}}\,)$. Let us for simplicity rename the first fluctuation $E^{{\scriptsize (1)}}{}_{\go C}{}^{\go D}\,\equiv\,H_{\go C}{}^{\go D}$. Then the equation (\ref{equation}) is just statement that: $H_{(\,{\go C}}{}^{\go D}\,\eta_{{\go \,D\,|\,E}\,]}\,\equiv\,H_{\go (\,C\,E\,]}\,=\,0$ and the structure of linearised torsion (\ref{t2}):
\begin{eqnarray}
\label{linear111}
T_{\go A\,\go B\,\go C}&=&f_{\go A\,\go B\,\go C}\,+\,T^{{\scriptsize (1)}}{}_{\go A\,\go B\,\go C}\,+\,{\mathcal{O}}\,(\,T^{{\scriptsize{(2)}}}\,)\\
\mbox{where}&&T^{{\scriptsize{(1)}}}{}_{\go A\,\go B\,\go C}\,\equiv\,\frac{1}{2}\,D_{ [\go A}\,H_{\go B\,\go C\,)}\,+\,\frac{1}{2}\,H_{\go [\,\go A\,}{}^{\go M}\,f_{\go B\,\go C\,)\,\go M}\nonumber
\end{eqnarray}

\subsection{\texorpdfstring{$AdS_{5}\,\times\,S^{5}$ background}{AdS\textfiveinferior{} x S\textfivesuperior{} background}}
In the expansion (\ref{linear111}) we need to have the concrete structure constants $f_{\go\,A\,B\,C}$ (i.e. vacuum values of torsions). We are interested in solving the theory (at least identifying the pre-potential) around this $AdS_{5}\,\times\,S^{5}$ background. The relevant structure constants for the T-dually extended superspace in the context of the $AdS_{5}\,\times\,S^{5}$ background were analysed in the last section of the paper \cite{warren1}. We are repeating them here for the convenience. In the next section we will embed this $AdS_{5}\,\times\,S^{5}$ version of T-dual algebra (see \cite{warren1}) to a certain larger algebra that will be actually used in computations. The relevant non-vanishing $AdS_{5}\,\times\,S^{5}$ torsions from \cite{warren1} are: 
\begin{eqnarray}
\label{naive}
\mbox{dim $0$}:\,\,&&T_{S\,S\,\Sigma}\,=\,f_{S\,S\,\Sigma}\,\,\,||\,\,\,T_{S\, D\, \Omega}\,=\,f_{S\, D\, \Omega}\,\,\,||\,\,\,T_{S\, P\, P}\,=\,f_{S\, P\, P}\,\,\,||\,\,\,T_{D\,D\,P}\,=\,f_{D\,D\,P}\nonumber\\
\mbox{dim $1$}:\,\,&&T_{D\,\widetilde{D}\,\Sigma}\,=\,R_{D\,\widetilde{D}\,\Sigma}\,\,\,||\,\,\,T_{P\,\widetilde{D}\,\Omega}\,=\,R_{P\,\widetilde{D}\,\Omega}\nonumber\\
\mbox{dim $2$}:\,\,&&T_{\Omega\,\Omega\,P}\,=\,R_{\Omega\,\Omega\,P}\,\,\,||\,\,\,T_{\Omega\,\Omega\,P}\,=\,R_{\Omega\,\Omega\,\widetilde{P}}\,\,\,||\,\,\,T_{P\,\widetilde{P}\,\Sigma}\,=\,R_{P\,\widetilde{P}\,\Sigma}\nonumber\\
\mbox{dim $3$}:\,\,&&T_{\Omega\,\widetilde{\Omega}\,\Sigma}\,=\,R_{\Omega\,\widetilde{\Omega}\,\Sigma}
\end{eqnarray}
note the left and right index notation introduced below (\ref{ind}).

The $f_{\go A\,\go B\,\go C}$ in (\ref{naive}) are usual flat superspace structure constants for the (flat) T-dually extended superspace with the (common) local Lorentz group $SO\,(\,5\,)\,\otimes\,SO\,(\,4,\,1\,)$. The nontrivial curvatures from table (\ref{naive}): $\,R_{D\,\widetilde{D}\,\Sigma}\,$,$\,R_{P\,\widetilde{D}\,\Omega}\,\dots$ are defined using the dimension $1$ torsion $T_{P\,\widetilde{D}\,\Omega}$. The $T_{P\,\widetilde{D}\,\Omega}\,\equiv\,T_{{\bf a}\,\widetilde{\alpha}}{}^{\beta}\,=\,\gamma_{{\bf a}\,\,\,{\alpha}\,{\sigma}}\,F^{\beta\,\widetilde{\sigma}}$, where the R-R field strength $F_{\Omega\,\widetilde{\Omega}}\,\equiv\,F^{\alpha\,\widetilde{\beta}}\,=\,\frac{1}{r_{AdS}}\,(\widetilde{\Gamma}_{5})^{\alpha\,\beta}$. Note that the $\widetilde{\Gamma}_{5}$ was defined in (\ref{gam5}) and the new parameter $r_{AdS}$ is the $AdS_{5}$ space radius (note, $r_{AdS}\,=\,r_{S}$ i.e. the radius of $S_{5}$ is the same of $AdS_{5}$, so that Ricci scalar $R\,=\,0$). More specifically some of the table  (\ref{naive}) curvatures:
\begin{eqnarray}
\label{kurniksopa}
\mbox{dim $1$}:\,\,&&R_{D\,\widetilde{D}\,\Sigma}\,\equiv\,R_{\alpha\,\widetilde{\beta}}{}^{\,\bf a\,b}\,=\,T_{\widetilde{\beta}}{}^{\,\sigma\,[{\bf a}}\,\gamma^{{\bf b}\,]}{}_{\alpha\,\sigma}\\
\mbox{dim $1$}:\,\,&&R_{P\,\widetilde{D}\,\Omega}\,\equiv\,R_{{\bf a}\,\widetilde{\alpha}}{}^{\beta}\,=\,T_{{\bf a}\,\widetilde{\alpha}}{}^{\beta}\,=\,\gamma_{{\bf a}\,\,\,{\alpha}\,{\sigma}}\,F^{\beta\,\widetilde{\sigma}}\nonumber\\
\mbox{dim $2$}:\,\,&&R_{P\,\widetilde{P}\,\Sigma}\,\equiv\,R_{{\bf a}\,\widetilde{{\bf b}}}{}^{\bf cd}\,\propto\,\,\,T_{\widetilde{{\bf b}}\,\alpha\,}{}^{\widetilde{\beta}}\,R_{\widetilde{\beta}\,\gamma\,}{}^{\bf cd}\,\gamma_{\bf a}{}^{\alpha\,\gamma}\nonumber\\
\mbox{dim $3$}:\,\,&&R_{\Omega\,\widetilde{\Omega}\,\Sigma}\,\equiv\,R^{\alpha\,\widetilde{\beta}}{}^{\,\,\,\bf{ab}} \,\propto\,\,\,\,(\,T^{\widetilde{\bf{d}}\,\,\widetilde{\beta}}{}_{\sigma}\,R_{\widetilde{\bf{d}}\,{\bf{e}}}{}^{\bf{ab}}\,+\,T_{\bf{e}}{}^{\widetilde{\beta}\,\nu}\,R_{\widetilde{\sigma}\,\nu}{}^{\bf{ab}})\,\gamma^{\bf{e}}{}^{\,\,\sigma\,\alpha} \nonumber
\end{eqnarray}
where the dim $2$ curvature is proportional with the constant $2^{-\,\frac{D}{2}\,+\,1}$ and the dim $3$ curvature is proportional with a constant $D$ (where $D$ is $10$ in our case). 

All the other curvatures (i.e. torsions) in (\ref{kurniksopa}) are obtained using the appropriate Bianchi identities (one can obtain basically all curvatures from $T_{P\,\widetilde{D}\,\Omega}$ using Bianchi identities). We note that the torsions (\ref{naive}) and curvatures (\ref{kurniksopa}) are consistent with torsions and curvatures given in the \cite{wess}. 

\subsection{\texorpdfstring{Extended $AdS_{5}\,\times\,S^{5}$ T-dual algebra}{Extended AdS\textfiveinferior{} x S\textfivesuperior{} T-dual algebra}}

To identify the pre-potential in the generalised vielbeins, i.e. solving the spectrum of the theory (on linearised level) we want to proceed as described in earlier papers \cite{warren1}, \cite{natural}, \cite{3d}. There the vielbeins were introduced as in (\ref{dur}) and linearised as above equation (\ref{linear111}). This procedure means the expansion of generally curved superspace around some (in those papers just a flat) background. Moreover the gauge invariance was completely fixed (in referenced papers the covariant gauge was considered) and after that the pre-potential was identified as a part of vielbein (acting on by derivatives, the physical spectrum is produced). 

Here we want to proceed in similar way. We want to introduce the vielbeins and linearise the theory around the $AdS_{5}\,\times\,S^{5}$ background. We have tried to use solely the algebra described in the previous sub-section, i.e. to take the algebra (\ref{naive}) and introduce the vielbeins, gauge fix and linearise. Even though we still believe that the pre-potential is sitting in that theory in some combination of vielbeins, it was not easy to identify it. The reason was that to identify the pre-potential we need to find a scalar contraction of some linear combination of vielbeins that is anihilated by the $D_{v}$ and $D_{\bar{v}}$ operators. The $D_{v}$ and $D_{\bar{v}}$ are certain combinations of  $D_{\alpha^{'}}$ and $D_{\widetilde{\alpha}^{'}}$ (see the index notation in (\ref{schem}) and above (\ref{con}), i.e. they are a particular chiral part of the $SO\,(\,8\,)$ chiral decomposition of the $16$ SUSY translations $D_{\alpha}$ defined at the beginning, see (\ref{algebra}) and section above (\ref{con})). 

Because the metric is in $H_{P\,\widetilde{P}}$ vielbein, we expected the pre-potential to be (at least a part of it) in $\mbox{Tr}\,H_{P\,\widetilde{P}}$. The problem with $\mbox{Tr}\,H_{P\,\widetilde{P}}$ is that it already has dimension $0$. To show that it is annihilated by  $D_{v}$ and $D_{\bar{v}}$ operators we would need to use torsion constraints of dimension $\frac{1}{2}$. Moreover we also know that the pre-potential has to be annihilated by the suitably defined ${\mathcal{P}}_{+}$ operator (${\mathcal{P}}_{+}\,\propto\,(\,{P_{+}}\,+\,{P_{\widetilde{+}}}\,)$ where ${P_{+}}\,\equiv\,D_{P_{+}}$ and ${P_{\widetilde{+}}}\,\equiv\,D_{P_{\widetilde{+}}}$), in a "light cone" basis introduced in (\ref{con}) and in the "near horizon limit" (defined later)). The high dimensionality of $\mbox{Tr}\,H_{P\,\widetilde{P}}$ then requires to use at least dimension $1$ torsion constraints  to prove that ${\mathcal{P}}_{+}$ vanishes (in the near horizon limit). That seemed to be problematic to analyse in the theory based just on the algebra (\ref{naive}) and (\ref{kurniksopa}) because of the degauging procedure. The degauging appears since the theory coming from algebras (\ref{naive}) and (\ref{kurniksopa}) is really coming from the full $SO\,(\,10\,)\,\otimes\,SO\,(\,10\,)$ so there are some missing Lorentz connections. By simple dimensional analysis its evident that the missing Lorentz connections are first appearing at dimension $1$ (for example the appearance of the full $H_{P\,S}$ in the dim $1$ torsion $T_{P\,P\,P}\,\propto\,\dots\,+\,H_{[\,P|\,S}\,f_{P\,P]\,\Sigma}\,+\,\dots$).

For that reason we extended the algebra (\ref{naive}) to include the original (Wick rotated) local Lorentz group $SO\,(\,10\,)\,\otimes\,SO\,(\,10\,)$. All the other structure constants and curvatures in (\ref{kurniksopa}) and (\ref{naive}) stay the same. Except now we have separate left local Lorentz $S_{\bf{ab}}$ generator together with the right local Lorentz generator $S_{\widetilde{\bf{ab}}}$, where ${\bf{a}}\,\in\,\{\,1,\,\dots\,10\}$. The common (Wick rotated) $SO\,(\,5\,)\,\otimes\,SO\,(\,5\,)$ Lorentz group of original $AdS_{5}\,\times\,S^{5}$ algebra (\ref{naive}) is then the subgroup in the diagonal $SO\,(\,10\,)$ subgroup of $SO\,(\,10\,)\,\otimes\,SO\,(\,10\,)$. The extension procedure can be viewed from the different picture. We could start with the full $10$ dimensional string superspace as introduced in \cite{warren} and \cite{mach}. Then introduce the curved version of that space via vielbeins and then linearise around some background as described earlier. The extension of (\ref{naive}) and (\ref{kurniksopa}) to full $SO\,(\,10\,)\,\otimes\,SO\,(\,10\,)$ is then just a choice of some particular background that is consistent with the original $AdS_{5}\,\times\,S^{5}$. This has an advantage that now we have a natural place where to put the troubling (part) of the pre-potential  $\mbox{Tr}\,H_{P\,\widetilde{P}}$. Because of the additional $S_{\bf{ab}}$ and $S_{\widetilde{\bf{ab}}}$ we can have $\mbox{Tr}\,H_{{\bf{+a}}\,\widetilde{\bf{+b}}}\,\equiv\,\mbox{Tr}\,H_{S\,\widetilde{S}}$ instead of the $\mbox{Tr}\,H_{P\,\widetilde{P}}$. The $\mbox{Tr}\,H_{P\,\widetilde{P}}\,\equiv\,H_{{\bf{a}}\,\widetilde{\bf{b}}}$ is related to  $\mbox{Tr}\,H_{S\,\widetilde{S}}\,\equiv\,\mbox{Tr}\,H_{{\bf{+a}}\,\widetilde{\bf{+b}}}$ by an action of $D_{P_{-}}\,\equiv\,P_{-}$ and $D_{P_{\widetilde{-}}}\,\equiv\,P_{\widetilde{-}}$ (they are both invertible operators). The vielbein $H_{S\,\widetilde{S}}$ is of the dimension $-\,2$ and so there is no need to use the higher dimensional torsion constraints. Moreover in the full $SO\,(\,10\,)\,\otimes\,SO\,(\,10\,)$ theory we do not have to do the degauging procedure. 

This extension comes with the cost. The mixed pieces of the $AdS$ algebra (\ref{kurniksopa}) are breaking the explicit $SO\,(\,10\,)\,\otimes\,SO\,(\,10\,)$ invariance (they are not the $SO\,(\,10\,)\,\otimes\,SO\,(\,10\,)$ invariant tensors). We still have present the full $D_{S}\,\equiv\,S$ generators. Those derivatives could hit the (non-invariant) curvatures. The solution of this is to keep the explicit mixed curvature dependence (as generic mixed curvatures) till the $S$ derivatives are not being explicitly evaluated. We will describe this procedure in detail later.   

\section{\texorpdfstring{Gauge fixing}{Gauge fixing}}
We want to fix the space-cone gauge (T-dual super space-cone gauge) for the first fluctuation $H_{\go A\,\go B}$, i.e. like in the usual light-cone we have $D_{P_{-}}\,\equiv\,P_{-}$ operator invertible, now we have $P_{-}$ and $P_{\widetilde{-}}$ invertible (where $D_{P_{\widetilde{-}}}\,\equiv\,P_{\widetilde{-}}$). 

First we look at the gauge variation (\ref{gaugeg}) more closely and at the linearised level:
\be
\label{variat}
\delta_{\Lambda}\,H_{\go A \go B}\,=\,D_{[\go A}\,\lambda_{\go B\,)}\,+\,f_{\go A\,\go B}{}^{\go C}\,\lambda_{\go C}
\ee
In the light-cone gauge we in general pick a vielbein with an $P_{-}$ or $P_{\widetilde{-}}$ index, put that vielbein to zero. In order to maintain that gauge we need to fix the particular gauge parameter. For simplicity we call $P_{-}\,\equiv\,-$ and $P_{\widetilde{-}}\,\equiv\,\widetilde{-}$ then:
\begin{eqnarray}
\label{fixingss}
&H_{-\,\go A}&\,=\,0\,\,\,\Rightarrow\,\,\,\delta_{\Lambda}\,H_{-\,\go A}\,=\,0\,\,\,\Rightarrow\,\,\,P_{-}\,\lambda_{\go A}\,-\,D_{\go A}\,\lambda_{-}\,+f_{-\,\go A}{}^{\go C}\,\lambda_{\go C}\,=\,0\\
&\mbox{then}&\,\,\lambda_{\go A}\,=\,\frac{1}{P_{-}}\,(\,D_{\go A}\,\lambda_{-}\,+\,f_{-\,\go A}{}^{\go C}\,\lambda_{\go C}\,)\nonumber
\end{eqnarray}
Note that there are more possibilities to fix the particular gauge parameters $\lambda_{\go A}$. To fix $\lambda_{\go A}$ we could also put $H_{\widetilde{-}\go A}\,=\,0$ and use the invertibility of $P_{\widetilde{-}}$. Of course we can not fix some gauge parameter twice. We have to decide which vielbeins we are going to fix in this ``double'' light-cone gauge. 

We picked the approach where we used the mixed vielbeins to vanish by the double light-cone gauge fixing, i.e. we put $H_{\widetilde{-}\, \go A}\,=\,0$ for $ \go A\,\in \{\,S,\,D,\,P,\,\Omega,\,\Sigma\,\}\,\equiv\,\mbox{left part of algebra}$. Together with  $H_{-\,\widetilde{ \go A}}\,=\,0$ for $\widetilde{ \go A}\,\in \{\,\widetilde{S},\,\widetilde{D},\,\widetilde{P},\,\widetilde{\Omega},\,\widetilde\Sigma{}\,\}\,\equiv\,\mbox{right part of algebra}$. By that choice we fully fix the gauge parameters $\lambda_{\go A}$ and $\lambda_{\widetilde{\go A}}$ in terms of $\lambda_{-}$. That parameter can be fixed by the gauge invariance of the gauge invariance.

The motivation for the previously described mixed left right light-cone gauge fixing came from the flat space (just the extended $AdS_{5}\,\times\,S^{5}$ space with $r_{AdS}\,\rightarrow\,\infty$, i.e. the flat $SO\,(\,10\,)\,\otimes\,SO\,(\,10\,)$ T-dual superspace). After picking this type of the light-cone gauge the mixed torsion constraints of the type $T_{\widetilde{-}\,\go A\,\go B}\,=\,0$ are as algebraic as possible: 
\begin{eqnarray}
\label{torzor}
T_{\widetilde{-}\,\go A\,\go B}&=&P_{\widetilde{-}}\,H_{\go A\,\go B}\,+D_{\go B}\,H_{\widetilde{-}\,\go A}\,+D_{\go A}\,H_{\go B\,\widetilde{-}}\,+H_{\widetilde{-}\,\go M}\,\eta^{\go M\,\go N}f_{\go A\,\go B\,\go N}\,+H_{[\go\,A\,|\,\go M}\eta^{\go M\,\go N}f_{\go B\,)\,\widetilde{-}\,\go N}\\
\label{torzor1}
T_{\widetilde{-}\,\go A\,\go B}&=&P_{\widetilde{-}}\,H_{\go A\,\go B}\,=\,0\,\,\,\Rightarrow\,\,\,H_{\go A\,\go B}\,=\,0\,\,\mbox{for}\,\go A,\,\go B\,\,:\,\,T_{\widetilde{-}\,\go A\,\go B}\,=\,0
\end{eqnarray}
where we used our mixed light-cone gauge and $r_{AdS}\,\rightarrow\,\infty$ of extended algebra in (\ref{torzor1}).

The same as in (\ref{torzor}) and (\ref{torzor1}) holds if one fully swaps left and right indices. For finite $r_{AdS}$ we can have the mixed structure constants nonzero (i.e. $f_{\go A\,\go B\,\widetilde{-}}\,\neq\,0$) and so we would have a right hand side in (\ref{torzor1}). Note also that there could be the contribution from $S$ derivatives hitting the mixed structure constants. Even though the right hand side in (\ref{torzor1}) is not generally vanishing for finite $r_{AdS}$ we found that the mixed left-right light-cone gauge is still useful in the $AdS_{5}\,\times\,S^{5}$ case. 

\section{\texorpdfstring{Torsion constraints}{Torsion constraints}}
\subsection{\texorpdfstring{$AdS_{5}\,\times\,S^{5}$ curvatures and $D_{S}$ derivatives}{AdS\textfiveinferior{} x S\textfivesuperior{} curvatures and D S derivatives}}
As we noted in the introduction section. Because we have enhanced our superspace, we have to take special care when the local Lorentz derivatives $D_{S}\,\equiv\,S$ are hitting the mixed curvatures (\ref{kurniksopa}). This problem arises because the curvatures in (\ref{kurniksopa}) are not full $SO\,(\,10\,)\,\otimes\,SO\,(\,10\,)$ invariant.  The solution is to keep the non-invariant torsions (\ref{kurniksopa}) generic and explicitly act by the $D_{S}\,\equiv\,S$ derivatives on those torsions. Only after this explicit $S$ action we can evaluate those torsions (or curvatures) and be fixed as in (\ref{kurniksopa}). 

Let's take an example, from the equation (\ref{torzor}) we can see that in the $AdS_{5}\,\times\,S^{5}$ case in the mixed light-cone gauge the $H_{{\go A}{\go B}}$ is determined as:
\begin{eqnarray}
\label{expl}
H_{{\go A}{\go B}}\,=\,\frac{1}{P_{\widetilde{-}}}\,H_{[\,{\go A}|\,{\go M}}\,\eta^{\go M \go N}\,f_{{\go B})\,{\widetilde{-}}\,{\go N}}
\end{eqnarray}
In many instances in this paper we use similar relation as in (\ref{expl}) to fix some particular vielbein in terms of another vielbeins. If all structure constants $f$ would be $SO\,(\,10\,)\,\otimes\,SO\,(\,10\,)$ invariant tensors then there is not a problem and we can treat the $f$ structure constants as genuine constants also with respect to the $S$ derivatives. In our case however the mixed $f$ structure constants (that we call also the curvatures) in (\ref{kurniksopa}) explicitly break the $SO\,(\,10\,)\,\otimes\,SO\,(\,10\,)$  local Lorentz invariance down to the $SO\,(\,5\,)\,\otimes\,SO\,(\,5\,)$ local Lorentz (as it should be in the $AdS_{5}\,\times\,S^{5}$ case). One possibility is to restrict our superspace local Lorentz invariance (the $S$ derivatives) to $SO\,(\,5\,)\,\otimes\,SO\,(\,5\,)$. Then we would return back to the $PSU\,(\,2,2\,|\,4\,)$ that we wanted to avoid in the first place (in order to have $H_{{\bf{+a}}\,\widetilde{\bf{+b}}}$ instead of $H_{{\bf{a}}\,{\widetilde{\bf{b}}}}$). The alternative, that we picked, is to work with the full $SO\,(\,10\,)\,\otimes\,SO\,(\,10\,)$ local Lorentz group. But then the structure constants that are breaking that invariance are not invariant tensors and so the action of those $S$ derivatives on the mixed structure constants has to be accounted for. So we should keep the mixed $f$ structure constants and when needed explicitly act by the $S$ derivatives on them. We will evaluate them as the very last step in our calculations. Let's look at the example in (\ref{expl}) and look at the action of ${S}_{\widetilde{\bf{+a}}}$, where ${\bf{a}}\,\in\,\{\,1\,\dots\,8\}$:
\begin{eqnarray}
\label{action}
S_{\widetilde{\bf{+a}}}\,H_{\go A\,\go B}&=&S_{\widetilde{\bf{+a}}}\,{\Big{(}}\,\frac{1}{P_{\widetilde{-}}}\,H_{[\,{\go A}|\,{\go M}}\,\eta^{\go M \go N}\,{\Big{)}}\,f_{{\go B})\,{\widetilde{-}}\,{\go N}}\,+\,\frac{1}{P_{\widetilde{-}}}\,H_{[\,{\go A}|\,{\go M}}\,\eta^{\go M \go N}\,{\Big{(}}\,S_{\widetilde{\bf{+a}}}\,f_{{\go B})\,{\widetilde{-}}\,{\go N}}\,{\Big{)}}\\
&=&S_{\widetilde{\bf{+a}}}\,{\Big{(}}\,\frac{1}{P_{\widetilde{-}}}\,H_{[\,{\go A}|\,{\go M}}\,\eta^{\go M \go N}\,{\Big{)}}\,f_{{\go B})\,{\widetilde{-}}\,{\go N}}\,+\,\eta_{\bf{+-}}\,\frac{1}{P_{\widetilde{-}}}\,H_{[\,{\go A}|\,{\go M}}\,\eta^{\go M \go N}\,f_{{\go B})\,{\widetilde{\bf{a}}}\,{\go N}}\nonumber
\end{eqnarray} 
We will evaluate the $f_{{\go B}\,{\widetilde{\bf{a}}}\,{\go N}}$ in the second equation just after all the (possibly future) $S$ derivatives have already acted. We also should bear in mind that whenever we are acting by the $S$ derivative on some vielbein (that is determined by another vielbeins) there might be the above described issue. The second term can (and it will) nontrivially contribute to our calculations.
\subsection{\texorpdfstring{Torsion constraints and $H_{S\,S}$ vielbein}{Torsion constraints and H S S vielbein}}
\label{torconI}
The torsion constraints are (mainly) given by the curved version of the ${\bf{A\,B\,C\,D}}$ (first class) constraints, see \cite{warrenTor} and \cite{warren1}. There are further constraints called $\widetilde{T}_{\go A}\,=\,0$ coming from requirement of partial integration in the presence of the dilaton measure, see \cite{warren}, and \cite{3d}. There is also a strong constraint: on every field in the double field theory one has to require $D^{\go A}\,D_{\go A}\,=\,0$. There are also a dimensional reduction constraints, as we see later. 

Our aim is to analyse the necessary constraints consistent with the above constraints by which we can identify the pre-potential. Following the analysis given in \cite{sconf}, we identify the pre-potential as a scalar super-field (given by some super-trace of possibly a combination of vielbeins), that is annihilated by certain combination of the $D_{\nu^{\prime}}$ and $D_{\widetilde{\nu}^{\prime}}$. The precise combination of $D_{\nu^{\prime}}$ and $D_{\widetilde{\nu}^{\prime}}$ is also going to be determined from the constraints.
 
As usual, we start to eliminate the lowest dimensional vielbeins. The vielbeins of the lowest dimension are $H_{S\,S},\,\,H_{S\,\widetilde{S}},\,H_{\widetilde{S}\,\widetilde{S}}$. They are of the dimension $-\,2$ (we mean the ten dimensional dimension). Using equations (\ref{torzor}) and (\ref{torzor1}) for indices $\go A\,=\,S$ and $\go B\,=\,S$ (also after change $\mbox{left}\,\leftrightarrow\,\mbox{right}$) we immediately get that $H_{S\,S}\,=\,0\,=\,H_{\widetilde{S}\,\widetilde{S}}$. Note, that even in the extended $AdS_{5}\,\times\,S^{5}$ superspace the structure constant $f_{S\,\widetilde{\go A}\,\go B}\,=\,0\,\rightarrow\,f_{S\,\widetilde{-}\,\go B}\,=\,0$. 

We mention an important observation that will help us simplify future calculations. As we saw in previous sub-chapter, we should keep the mixed structure constants generic and evaluate them at the end. Note however, that the mixed structure constants with the $S$ indices are always zero (like the one we considered here: $f_{\go S \widetilde{\go A} \go B}$). The action of $S$ derivatives on them results in the mixed structure constants again with the $S$ index and such are zero after the evaluation. So specifically, we can evaluate the mixed structure constants with the $S$ indices to zero even before acting by $S$ derivatives on them. 

The mixed vielbein $H_{S\,\widetilde{S}}$ is not all zero and the claim is that the part of the pre-potential is in this particular vielbein. To see which parts are possibly nonzero we rewrite the $H_{S\,\widetilde{S}}$ in the double light-cone components: 
\begin{eqnarray}
\label{song}
H_{S\,\widetilde{S}}&\equiv&\{\,H_{\bf{+a}\,\widetilde{\bf{+c}}},\,H_{\bf{+a}\,\widetilde{\bf{-b}}},\,H_{{\bf{-a}}{\widetilde{\bf{-b}}}},\,H_{\bf{+a}\,\widetilde{\bf{bc}}},\,H_{\bf{-a}\,\widetilde{\bf{bc}}},\,H_{\bf{ab}\,\widetilde{\bf{cd}}},\,H_{+-\,\widetilde{+-}},\,H_{\bf{+-}\,\widetilde{\bf{+a}}}\\
&&H_{\bf{+-}\,\widetilde{\bf{-a}}},\,H_{\bf{+-}\,\widetilde{\bf{ab}}}\}\,\oplus\,\,\mbox{swap}\nonumber
\end{eqnarray}
where we might swap left index with the right index in (\ref{song}). Also note that in all previous we have ${\bf{a}}\,\in\,\{\,{{1}},\dots,\,{{8}}\,\}$. We remind that $P_{+}\,\equiv\,{\bf{+}}\,\propto\,{{10}\,+\,{{9}}}$ and $P_{-}\,\equiv\,{\bf{-}}\,\propto\,{{10}\,-\,{{9}}}$.

We want to use analog of equations (\ref{torzor}) and (\ref{torzor1}) for the mixed $H_{S\,\widetilde{S}}$ and $r_{AdS}\,\neq\,\infty$: 
\begin{eqnarray}
T_{\widetilde{-}\,S\,\widetilde{S}}&=&0\,=\,P_{\widetilde{-}}\,H_{S\,\widetilde{S}}\,+\,D_{\widetilde{S}}\,H_{\widetilde{-}\,S}\,+\,D_{S}\,H_{\widetilde{S}\,\widetilde{-}}\,+\,H_{[\,\widetilde{-}|\,\go M}\,\eta^{\go M\,\go N}\,f_{S\,\widetilde{S}\,]\,\go N}\label{mixos}\label{pomixos}
\end{eqnarray}
In equation (\ref{pomixos}) we have term $H_{\widetilde{-}\,S}$ zero by gauge choice.  The vielbein $H_{\widetilde{S}\,\widetilde{-}}$ is proportional to $f_{S\,\widetilde{-}\,\go N}$. We see that this term is zero after evaluating $f_{S\,\widetilde{-}\,\go N}\,=\,0$ by (\ref{kurniksopa}). Using the (\ref{torzor}) and (\ref{torzor1}) for $\go A\,=\,S$ and $\go B\,=\,-$ and keeping $f_{-\,\widetilde{-}\,S}$ nonzero, we get $H_{\go S\, -}\,=\,\frac{1}{P_{\widetilde{-}}}\,f_{-\,\widetilde{-}\,{\go M}}\,\eta^{\go M \go N}\,H_{S\,{\go N}}$ and similarly for $H_{\widetilde{\go S}\,\widetilde{-}}$ and so the third term in (\ref{pomixos}) is also fixed. The (\ref{pomixos}) is then:
\be
\label{posero}
T_{\widetilde{-}\,S\,\widetilde{S}}\,=\,0\,=\,P_{\widetilde{-}}\,H_{S\,\widetilde{S}}\,+D_{S}{\Big{(}}\,\frac{1}{P_{-}}\,\,f_{-\,\widetilde{-}\,{\go M}}\,\eta^{\go M \go N}\,H_{\widetilde{S}\,{\go N}}\,{\Big{)}}\,+\,H_{\,S\,\go M}\,\eta^{\go M\,\go N}\,f_{\widetilde{S}\,\widetilde{-}\,\go N}
\ee
Using equations (\ref{torzor}), (\ref{torzor1}), (\ref{posero}) and the mixed light-cone gauge together with keeping the mixed structure constants and evaluating the explicit actions of the $S$ and $\widetilde{S}$ derivatives we get the first important result for the structure of the $H_{\go S {\widetilde{\go S}}}$ vielbein (in the $AdS$ case), see table (\ref{hrncek1}) in Appendices. 
From the table (\ref{hrncek1}) we can see that the possibly nonzero $H_{S\,\widetilde{S}}$ in (\ref{posero}) are those for which $f_{\widetilde{S}\,\widetilde{-}\,\go N}$ are nonzero (after evaluation of mixed structure constants). By simple $\mbox{left}\,\leftrightarrow\,\mbox{right}$ swap we get that for $H_{S\,\widetilde{S}}$ to be nonzero also $f_{S\,-\,\go N}$ has to be nonzero. From the $[\,S,\,P\,]$ part of $SO\,(\,10\,)\,\otimes\,SO\,(\,10\,)$ extended algebra of (\ref{algebra}) we can see the only possibility: $H_{\bf{+a}\,\widetilde{\bf{+c}}}\,\neq\,0$. All the other components of $H_{S\,\widetilde{S}}\,=\,0$ by (\ref{posero}) and table (\ref{hrncek1}) after evaluation. This is a first hint that we are on the right track. The $H_{\bf{+a}\,\widetilde{\bf{+c}}}\,\neq\,0$ is the only nonzero part (after evaluation of mixed structure constants) of $H_{S\,\widetilde{S}}$ piece, it has a scalar trace and we can easily relate it to $H_{{\bf{a}}\,\widetilde{{\bf{b}}}}\,\equiv\,H_{P\,\widetilde{P}}$, where we expect the part of the pre-potential to be (the symmetric part corresponds to the metric). 

To see how $H_{P\,\widetilde{P}}$ is related to $H_{S\,\widetilde{S}}$ consider the third relation from table (\ref{hrncek1}) and after evaluation of mixed structure constants:
\begin{eqnarray}
\label{vidlak}
(\,1\,+\,\frac{1}{2\,(r_{AdS})^{2}}\,\frac{1}{P_{{-}}\,P_{\widetilde{-}}}\,)\,H_{{\bf{+a}}\widetilde{\bf{+b}}}\,=\,\frac{1}{P_{-}}\,H_{{\bf{a}}\,\widetilde{\bf{+b}}}
\end{eqnarray}
We want to reduce $H_{\bf{a}\,\widetilde{\bf{+b}}}$ further to get $H_{\bf{a}\,\widetilde{b}}$. One can na{\"i}vely expect to just hit $H_{{\bf{a}}\,\widetilde{{\bf{+b}}}}$ with $P_{\widetilde{-}}$ to get rid of the $\widetilde{S}$ index (or alternatively hit by $P_{-}$ the $H_{{\bf{+a}}\,\widetilde{\bf{b}}}$). It works but one has to be more careful since in the $AdS_{5}\,\times\,S^{5}$ space one has the mixed structure constant $f_{{\bf{-}}\,\widetilde{\bf{b}}\,\go N}\,\neq\,0$. To see what is this structure constant (after the mixed structure constants evaluation) we remind that in (\ref{kurniksopa}) we saw that dimension 2 structure constant is given as $R_{P\,\widetilde{P}\,\Sigma}\,\equiv\,R_{{\bf a}\,\widetilde{{\bf b}}}{}^{\bf cd}\,\propto\,\,\,T_{\widetilde{{\bf b}}\,\alpha\,}{}^{\widetilde{\beta}}\,R_{\widetilde{\beta}\,\gamma\,}{}^{\bf cd}\,\gamma_{\bf a}{}^{\alpha\,\gamma}$. We have to be careful with the indices in the $R_{{\bf a}\,\widetilde{{\bf b}}}{}^{\bf cd}$. The $\Sigma$ indices ${}^{\bf{cd}}$ in (\ref{kurniksopa}) were indices for $SO\,(\,5\,)\,\otimes\,SO\,(\,4,\,1\,)$ local Lorentz group (or its Wick rotated version $SO\,(\,5\,)\,\otimes\,SO\,(\,5\,)$). But the index $\go N$ in $f_{{\bf{-}}\,\widetilde{\bf{b}}\,\go N}\,\neq\,0$ includes the indices for the full $SO\,(\,10\,)\,\otimes\,SO\,(\,10\,)$. We have already made the claim that the original local Lorentz group $SO\,(\,5\,)\,\otimes\,SO\,(\,5\,)$ is in the {\textit{diagonal}} subgroup of the $SO\,(\,10\,)\,\otimes\,SO\,(\,10\,)$, i.e. we have the following (at the level of algebras) $so\,(\,10\,)\,\oplus\,so\,(\,10\,)\,\equiv\frac{1}{2}\,(so\,(\,10\,)\,+\,so\,(\,10\,))\,\oplus\,\frac{1}{2}\,(so\,(\,10\,)\,-\,so\,(\,10\,))\,\defeq\,so(\,10\,)_{\bf{D}}\,\oplus\,so(\,10\,)_{\bf{Off}}$. (The meaning of previous is to do the operations on basis. The $so\,(\,10\,)\,-\,so\,(\,10\,)$ means for example to combine e.g. Lorentz generators like: $S\,-\,{\widetilde{S}}\,=\,S_{\bf{Off}}$ and similarly for another generators). Now the $so(\,5\,)\,\oplus\,so(\,5\,)\,\hookrightarrow\,so(\,10\,)_{\bf{D}}$. 
Let us write the last sequence of algebras more precisely. Using indices, the diagonal subgroup (subalgebra) is $so(\,10\,)_{\bf{D}}\,\equiv\,S^{\bf{D}}{}_{\bf{ab}}\,\defeq\,\frac{1}{2}\,(\,S_{\scriptsize{\bf{ab}}}\,+\,S_{\widetilde{\bf{ab}}}\,)\,=\,(\,S^{{\bf{D}}}{}_{\bf{ij}},\,S^{{\bf{D}}}{}_{\bf{kl}},\,S^{{\bf{D}}}{}_{\bf{ik}}\,)$, where ${\bf{i}}\,\in\,\{\,{{10}},\,{{1}},\,{{2}},\,{{3}},\,{{4}}\,\}$ and ${\bf{k}}\,\in\,\{\,{{5}},\,{{6}},\,{{7}},\,{{8}},\,{{9}}\,\}$ and the ${\bf{a}}\,=\,(\,{\bf{i}},\,{\bf{k}}\,)\,\equiv\,\{\,{{1}},\dots,\,{{10}}\,\}$. The $S^{{\bf{D}}}{}_{\bf{ij}}$ and $S^{{\bf{D}}}{}_{\bf{kl}}$ are the generators of the $SO\,(\,5\,)\,\otimes\,SO\,(\,5\,)$. Moreover, the previous definitions give precise embedding of those operators. 

Defining $S^{{\bf{D}}}{}_{\bf{ij}}$ and $S^{{\bf{D}}}{}_{\bf{kl}}$ we can see that the structure constant $f_{{\bf{a}}\,\widetilde{\bf{b}}\,\go N}$ is either 0 or given by the appropriate $R_{P\,\widetilde{P}\,\Sigma_{\bf{D}}}$.  We included a small subindex to the $\Sigma$ coordinate just to remind us that the $\Sigma$ coordinate is now for the $SO\,(\,5\,)\,\otimes\,SO\,(\,5\,)$ diagonal subgroup of $SO\,(\,10\,)_{\bf{D}}$ only. 

Finally, taking definitions of $R_{{\bf a}\,\widetilde{{\bf b}}}{}^{\bf cd}$ and table (\ref{kurniksopa}) and our definitions we can see that $f_{{\bf{a}}\,\widetilde{\bf{b}}\,\go N}$ is coming from the mixed commutator $[\,P_{\bf{a}},\,P_{\widetilde{\bf{b}}}\,]\propto\begin{cases} 
      \hfill (\frac{1}{r_{AdS}})^2\,S^{\bf{D}}{}_{\bf{ab}}    \hfill & \text{ if {\bf{a}} \& {\bf{b}}}\,\in\,\{\,{{10}},\,{{1}},\,{{2}},\,{{3}},\,{{4}}\,\}  \\
      \hfill -\,(\frac{1}{r_{AdS}})^2\,S^{\bf{D}}{}_{\bf{ab}}    \hfill & \text{ if {\bf{a}} \& {\bf{b}}}\,\in\,\{\,{{5}},\,{{6}},\,{{7}},\,{{8}},\,{{9}}\,\} \\
      \hfill 0&\text{otherwise}\\
  \end{cases}$
The proportionality constant is ${\bf{c}}_{1}\,=\,-\,2$. With the previous definition and with $P_{-}\,\equiv\,-\,=\,\frac{1}{2}\,(\,{{10}}\,-\,{{9}}\,)$ and ${\widetilde{\bf{b}}\,\in\,\{\,{{5}},\,{{6}},\,{{7}},\,{{8}},\,{{9}}\,\}}$ we will get the $[\,P_{-},P_{\widetilde{{\bf{b}}}}\,]\,\propto\,S^{\bf{D}}{}_{{\bf{9}}\,{\bf{b}}}\,=\,\frac{1}{2}\,(\,S_{{\bf{9}}\,{\bf{b}}}\,+\,S_{\widetilde{{\bf{9}}\,{\bf{b}}}}\,)$ we can also use $P_{\bf{9}}\,=\,(\,P_{+}\,-\,P_{-}\,)$ and so $[\,P_{-},P_{\widetilde{{\bf{b}}}}\,]\,\propto\,(\,S_{{\bf{+}}\,{\bf{b}}}\,+\,S_{\widetilde{\bf{+}\,{\bf{b}}}}\,-\,S_{{\bf{-}}\,{\bf{b}}}\,-\,S_{\widetilde{\bf{-}\,{\bf{b}}}}\,)$. Knowing the last relation we can proceed and hit the result of (\ref{vidlak}) by $P_{\widetilde{-}}$ and relate $H_{{\bf{+a}}\,\widetilde{{\bf{+b}}}}$ with $H_{{\bf{a}}\,\widetilde{{\bf{b}}}}$ for the evaluated version. For non-evaluated version we need to do the same for the non-evaluated version of (\ref{vidlak}), that is the third top equation in table (\ref{hrncek1}). For vielbein $H_{\widetilde{\bf{+b}}\,{\bf{a}}}$ needed in this procedure we get:
\begin{eqnarray}
\label{sedlak}
0\,=\,T_{{P}\,\widetilde{P}\,\widetilde{S}}\,\equiv\,T_{\bf{a}\,\widetilde{-}\,\widetilde{\bf{+b}}}&=&P_{[\,{\bf{a}}}\,H_{\widetilde{\bf{-}}\,\widetilde{\bf{+b}}\,]}\,+\,H_{[\,{\bf{a}}\,|\,\go M}\,\eta^{\go M\,\go N}\,f_{\widetilde{\bf{-}}\,\widetilde{\bf{+b}}\,]\,\go N}\\
&=&\,-\,P_{\widetilde{-}}\,H_{\widetilde{\bf{+b}}\,{\bf{a}}}\,-\,P_{{\bf{a}}}\,H_{\widetilde{-}\,\widetilde{\bf{+b}}}\,-\,H_{\widetilde{\bf{+b}}\,\go M}\,\eta^{\go M\,\go N}\,f_{{\bf{a}}\,\widetilde{-}\,\go N}\,-\,H_{{\bf{a}\,\go M}}\,\eta^{\go M\,\go N}\,f_{\widetilde{-}\,\widetilde{\bf{+b}}\,\go N}\nonumber
\end{eqnarray}
The term $H_{\widetilde{\bf{-}}\,{\bf{a}}}$ vanishes because of mixed light-cone gauge, the term $H_{\widetilde{-}\,\widetilde{\bf{+b}}}$ is fixed by the torsion $T_{-\,\widetilde{\bf{-}}\,\widetilde{\bf{+b}}}\,=0$ and use of mixed light-cone gauge  similarly as in equation (\ref{homedepo}). By that we get:
\begin{eqnarray}
\label{rolex}
H_{\widetilde{\bf{-}}\,\widetilde{\bf{+b}}}\,=\,-\,f_{-\,\widetilde{-}\,\go M}\,\eta^{\go M \go N}\,\frac{1}{P_{-}}\,H_{\widetilde{\bf{+b}}\,\go N}\,\rightsquigarrow\,0
\end{eqnarray}
The last term in (\ref{sedlak}) is just $\eta_{-\,+}\,\eta_{{\bf{a}}\,\go N}$, the analog term as in (\ref{vidlak}). The extra mixed term (after evaluation) in (\ref{sedlak}) is $H_{\bf{+a}\,\go M}\,\eta^{\go M\,\go N}\,f_{\widetilde{\bf{b}}\,-\,\go N}\,\propto\,(\,\frac{1}{r_{AdS}}\,)^{2}\,H_{{\bf{+a}}\,\widetilde{\bf{+b}}}$.  Plugging (\ref{rolex}) into (\ref{sedlak}) and then the result (that is the fixed vielbein $H_{\widetilde{\bf{+b}}\,\bf{a}}$) into the third top equation in table (\ref{hrncek1}) we obtain the non-evaluated relation between $H_{{\bf{+a}}\,\widetilde{\bf{+b}}}$:
\begin{eqnarray}
\label{speky}
P_{{-}}\,H_{{\bf{+a}}\,\widetilde{\bf{+b}}}&=&-\,f_{-\,\widetilde{-}\,\go M}\,\eta^{\go M \go N}\,S_{\widetilde{\bf{+b}}}\,(\,\frac{1}{P_{\widetilde{-}}}\,H_{{\bf{+a}}\,\go N}\,)\,-\,f_{-\,\widetilde{\bf{b}}\,\go M}\,\eta^{\go M\, \go N}\,\frac{1}{P_{\widetilde{-}}}\,H_{{\bf{+a}} \go N}\,\\
&&-\,f_{-\,\widetilde{-}\,\go M}\,\eta^{\go M \go N}\,\frac{1}{P_{\widetilde{-}}}\,P_{\bf{a}}\,\frac{1}{P_{-}}\,H_{\widetilde{\bf{+b}}\,\go N}\,+\,f_{\widetilde{-}\,\bf{a}\,\go M}\,\eta^{\go M \go N}\,\frac{1}{\,P_{\widetilde{-}}}\,H_{\widetilde{\bf{+b}}\,\go N}\,+\,\frac{1}{P_{\widetilde{-}}}\,H_{{\bf{a}}\,\widetilde{\bf{b}}}\nonumber
\end{eqnarray}
after evaluation of the mixed structure constants in (\ref{speky}) we get:
\begin{eqnarray}
\label{redneck}
H_{{\bf{a}}\,\widetilde{\bf{b}}}\,=\,{\Big{(}}\,P_{{-}}\,P_{\widetilde{-}}\,+\,\frac{1}{(r_{AdS})^{2}}\,{\Big{)}}\,H_{{\bf{+a}}\,\widetilde{\bf{+b}}}
\end{eqnarray}

\subsection{\texorpdfstring{Torsion constraints and $H_{D\,S}$ vielbein}{Torsion constraints and H D S vielbein}}
\label{vareska2}
To identify what combination of vielbeins gives the pre-potential, we first repeat the properties we are looking for. We are looking for combination of vielbeins (of the low dimension), that has a scalar contraction and is annihilated by certain combination of $D_{\alpha^{\prime}}$  and $D_{\widetilde{\alpha}^{\prime}}$ (see indices defined above (\ref{con})). Moreover the combination has to be annihilated by the properly defined ${\mathcal{P}}_{+}$ operator in the $R\,\rightarrow\,0$ limit (still to be defined). 

To start, we have one nontrivial hint. We showed that the vielbein $H_{{\bf{+a}}\,\widetilde{\bf{+b}}}$ is nonzero and is related to the $H_{{\bf{a}}\,\widetilde{\bf{b}}}$. So we can examine what is the action of the $D_{\alpha^{\prime}}$ on $H_{{\bf{+a}}\,\widetilde{\bf{+b}}}$, i.e. we look at the torsion constraint:
\begin{eqnarray}
\label{fnukadlo}
T_{D\,S\,\widetilde{S}}\,\equiv\,T_{\alpha^{\prime}\,{\bf{+a}}\,\widetilde{\bf{+b}}}\,=\,0&=&D_{[\alpha^{\prime}}\,H_{{\bf{+a}}\,\widetilde{\bf{+b}})}\,+\,H_{[\alpha^{\prime}\,|\,\go M}\,\eta^{\go M \go N}\,f_{\,{\bf{+a}}\,\widetilde{\bf{+b}})\,\go N}\\
&=&D_{\alpha^{\prime}}\,H_{{\bf{+a}}\,\widetilde{\bf{+b}}}\,+\,S_{\widetilde{\bf{+b}}}\,H_{\alpha^{\prime}\,{\bf{+a}}}\,+\,S_{\bf{+a}}\,H_{\widetilde{\bf{+b}}\,\alpha^{\prime}}\,+\,H_{\widetilde{\bf{+b}}\,\go M}\,\eta^{\go M\,\go N}\,f_{\alpha^{\prime}\,{\bf{+a}}\,\go N}\nonumber\\
\label{fulltime}
&=&D_{\alpha^{\prime}}\,H_{{\bf{+a}}\,\widetilde{\bf{+b}}}\,+\,S_{\widetilde{\bf{+b}}}\,H_{\alpha^{\prime}\,{\bf{+a}}}\,+\,S_{\bf{+a}}\,H_{\widetilde{\bf{+b}}\,\alpha^{\prime}}\,+\,\frac{1}{2}\,(\gamma_{\bf{+a}})_{\alpha^{\prime}}{}^{\beta}\,H_{\widetilde{\bf{+b}}\,\beta}
\end{eqnarray}
In the (\ref{fulltime}) we can see various terms with the $S$ derivatives. If we could evaluate mixed structure constants before an action of $S$ derivatives, those $S$ terms in (\ref{fulltime}) would vanish (because the relevant vielbeins are proportional to vanishing mixed constants as we will see). Now they will nontrivially contribute.  We note again that we still have the $f_{\widetilde{S}\,D\,\go N}\,\equiv\,f_{\widetilde{\bf{+b}}\,\alpha^{\prime}\,\go N}\,=\,0\,=\,f_{S\,\widetilde{S}\,\go N}\,\equiv\,f_{{\bf{+a}}\,\widetilde{\bf{+b}}\,\go N}$. The vielbeins $H_{\alpha^{\prime}\,{\bf{+a}}}$ and $H_{\alpha^{\prime}\,\widetilde{\bf{+b}}}$ are fixed by torsion constraints $T_{\widetilde{P}\,D\,S}\,=\,T_{\widetilde{-}\,\alpha^{\prime}\,{\bf{+a}}}\,=\,0\,=\,T_{P\,D\,\widetilde{S}}\,=\,T_{-\,\alpha^{\prime}\,\widetilde{\bf{+b}}}$ and some few other torsion constraints, as will be shown. We note that as in (\ref{torzor}) and (\ref{torzor1}) we almost always have the strategy to use invertibility of $P_{-}\,\mbox{and}\,P_{\widetilde{-}}$ together with our mixed left-right light cone gauge to eliminate/fix vielbeins. Sometimes its not enough and we need to explore some further torsion constraints. Let's look at the already mentioned set of torsion constraints: 
\begin{eqnarray}
\label{hudry}
T_{\widetilde{P}\,D\,S}\,=\,T_{\widetilde{-}\,\alpha^{\prime}\,{\bf{+a}}}\,=\,0&=&P_{[\widetilde{-}}\,H_{\alpha^{\prime}\,{\bf{+a}}]}\,+\,H_{[\widetilde{-}\,|\,\go M}\,\eta^{\go M \go N}\,f_{\alpha^{\prime}\,{\bf{+a}}]\,\go N}\\
\label{dudry}
&=&P_{\widetilde{-}}\,H_{\alpha^{\prime}\,{\bf{+a}}}\,+\,H_{\bf{+a}\,\go M}\,\eta^{\go M \go N}\,f_{\widetilde{-}\,\alpha^{\prime}\,\go N}\\
\Rightarrow&&H_{\alpha^{\prime}\,{\bf{+a}}}\,=\,f_{\widetilde{-}\,\alpha^{\prime}\,\go M}\,\eta^{\go M \go N}\,\frac{1}{P_{\widetilde{-}}}\,H_{{\bf{+a}}\,\go N}\,\rightsquigarrow\,0
\end{eqnarray}
in (\ref{hudry}) we used just the mixed light cone gauge and $f_{S\,\widetilde{P}\,\go N}\,=\,0$ (in flat case and also in $AdS$). To evaluate the last term in (\ref{dudry}) that is present only in $AdS$ case we have to take the $AdS$ curvature $T_{\widetilde{P}\,D\,\widetilde{\Omega}}\,\equiv\,T_{\widetilde{\bf{a}}\,\alpha}{}^{\widetilde{\beta}}\,\equiv\,f_{\widetilde{{\bf{a}}}\,\alpha}{}^{\widetilde{\beta}}\,=\,\frac{1}{r_{AdS}}\,(\gamma_{\bf{a}})_{\alpha\,\nu}\,\widetilde{\Gamma}_{5}{}^{\nu\,\beta}$ as discussed above (\ref{kurniksopa}). For our specific indices we have $\frac{1}{r_{AdS}}\,(\gamma_{\bf{-}})_{\alpha^{\prime}\,\nu^{\prime}}\,\widetilde{\Gamma}_{5}{}^{\nu^{\prime}\,\beta}$ but the $(\gamma_{\bf{-}})_{\alpha^{\prime}\,\nu^{\prime}}\,=\,0$ as we can see in the construction of the light cone basis for the gamma matrices in (\ref{con}). The vielbein $H_{\alpha^{\prime}\,{\bf{-a}}}$ can be fixed in almost the same set of equations as $H_{\alpha^{\prime}\,{\bf{+b}}}$. Fixing the vielbein $H_{\alpha^{\prime}\,\widetilde{\bf{+b}}}$ (and similarly $H_{\alpha^{\prime}\,\widetilde{\bf{-b}}}$) is also similar but a bit more profound. For that we first examine torsion analogous to (\ref{fnukadlo}) but for $H_{\alpha^{\prime}\,\widetilde{\bf{+b}}}$ vielbein: 
\begin{eqnarray}
\label{hena}
T_{P\,D\,\widetilde{S}}\,\equiv\,T_{{\bf{-}}\,\alpha^{\prime}\,\widetilde{\bf{+b}}}\,=\,0&=&P_{[-}\,H_{\alpha^{\prime}\,\widetilde{\bf{+b}}]}\,+\,H_{[-\,|\,\go M}\,\eta^{\go M\,\go N}\,f_{\alpha^{\prime}\,\widetilde{\bf{+b}}]\,\go N}\\
&=&P_{-}\,H_{\alpha^{\prime}\,\widetilde{\bf{+b}}}\,+\,S_{\widetilde{\bf{+b}}}\,H_{-\,\alpha^{\prime}}\,+\,H_{\widetilde{\bf{+b}}\,\go M}\,\eta^{\go M\,\go N}\,f_{-\,\alpha^{\prime}\,\go N}\nonumber\\
\label{puzdro}
\Rightarrow&&H_{\alpha^{\prime}\,\widetilde{\bf{+b}}}\,=\,-\,\frac{1}{P_{-}}\,S_{\widetilde{\bf{+b}}}\,H_{-\,\alpha^{\prime}}
\end{eqnarray}
The (\ref{hena}) structure constant $f_{-\,\alpha^{\prime}\,\go N}\,\propto\,(\gamma_{-})_{\alpha^{\prime}\,\nu^{\prime}}$ but as before that particular piece of gamma matrix is zero (remember the non-mixed structure constants are not breaking the $SO\,(\,10\,)\,\otimes\,SO\,(\,10\,)$ so we can evaluate them without any concern). The other term in (\ref{hena}) is $H_{-\,\alpha^{\prime}}$. That is fixed by the dim $\frac{1}{2}$ torsion constraint $T_{-\,\widetilde{-}\,\alpha^{\prime}}\,=\,0$: 
\begin{eqnarray}
\label{wintech}
T_{P\,\widetilde{P}\,D}\,=\,T_{-\,\widetilde{-}\,\alpha^{\prime}}\,=\,0&=&P_{[-}\,H_{\widetilde{-}\,\alpha^{\prime}\,)}\,+\,H_{[-\,\go M}\,\eta^{\go M \go N}\,f_{\widetilde{-}\,\alpha^{\prime}\,)\,\go N}\\
&=&P_{\widetilde{-}}\,H_{\alpha^{\prime}\,-}\,+\,H_{\alpha^{\prime}\,\go M}\,\eta^{\go M \go N}\,f_{-\,\widetilde{-}\,\go N}\\
\label{ondrej}
\Rightarrow&&H_{\alpha^{\prime}\,-}\,=\,-\,\frac{1}{P_{\widetilde{-}}}\,f_{-\,\widetilde{-}\,\go M}\,\eta^{\go M \go N}\,H_{\alpha^{\prime}\,\go N}\,\rightsquigarrow\,0
\end{eqnarray}
We used in (\ref{wintech}) the mixed light cone gauge, also the fact that $f_{-\,\alpha^{\prime}\,\go M}\propto\,(\gamma_{-})_{\alpha^{\prime}\,\beta^{\prime}}\,=\,0$. We note that the (\ref{wintech}) evaluates to zero because the mixed structure constant $f_{-\,\widetilde{-}\,\go N}\,=\,0$. Moreover by the light-cone gauge the (\ref{wintech}) term $H_{-\,\go M}\,\eta^{\go M \go N}\,f_{\widetilde{-}\,\alpha^{\prime}\,\go N}\,=\,0$ even in the non-evaluated regime. The reason is that the  structure constant $f_{\widetilde{-}\,\alpha^{\prime}\,\go N}$ is zero after evaluation and the action of whatever $S$ on this structure constant produces either zero or the right $D$ index $\equiv\,\widetilde{D} $  (after the summation with the vielbein) i.e. the vielbein $H_{-\,\widetilde{D}}$ that is again zero by the light-cone gauge. Combining (\ref{puzdro}) and (\ref{ondrej}) we can get a fixed version of the $H_{\alpha^{\prime}\,\widetilde{\bf{+b}}}$. By the similar equations as above we can fix $H_{\alpha^{\prime}\,\widetilde{\bf{-b}}}$. That result and more detailed analysis is shown in the Appendix, see table (\ref{hrncek2}).

In the Appendix, we also derived the equations (\ref{verizon}) and (\ref{vilda}). Those are the actions of the $S$ and $\widetilde{S}$ derivatives that we need in the equation (\ref{fulltime}). Putting the results from (\ref{verizon}) and (\ref{vilda}) into (\ref{fulltime}) we get fixing of the $D_{\alpha^{\prime}}\,H_{\bf{+a}\,\widetilde{\bf{+b}}}$, we note that this is an important result:
\begin{eqnarray}
\label{kukura1}
D_{\alpha^{\prime}}\,H_{{\bf{+a}}\,\widetilde{\bf{+b}}}&=&\,-\,\frac{1}{g}\,\frac{1}{(r_{AdS})\,P_{\widetilde{-}}}\,(\gamma_{{\bf{b}}})_{\alpha^{\prime}\,\sigma}\,(\widetilde{\Gamma}_{5})^{\sigma\,\beta}\,H_{\widetilde{\beta}\,{\bf{+a}}}\,+\,\frac{1}{2\,g}\,(\,1\,-\,\frac{1}{f}\,\frac{1}{(r_{AdS})^{2}\,P_{\widetilde{-}}\,P_{-}}\,)(\gamma_{\bf{+a}})_{\alpha^{\prime}}{}^{\beta}\,H_{\beta\,\widetilde{\bf{+b}}}\,\nonumber\\
&&+\,\frac{1}{f\,g}\,\frac{1}{2\,(r_{AdS})^{3}\,P_{-}\,(P_{\widetilde{-}})^{2}}\,\,(\gamma_{\bf{a}})_{\alpha^{\prime}\,\nu}\,(\widetilde{\Gamma}_{5})^{\nu\,\beta}\,H_{\widetilde{\beta}\,{\bf{+b}}}
\end{eqnarray}
where $f$ and $g$ are defined as follows:
\begin{eqnarray}
\label{kukura2}
f&\defeq&(\,1\,-\,\frac{1}{2\,(r_{AdS})^{2}\,P_{\widetilde{-}}\,P_{-}}\,)\\
g&\defeq&(\,1\,-\,\frac{1}{f}\,\frac{1}{2\,(r_{AdS})^{2}\,P_{\widetilde{-}}\,P_{-}\,}\,) \nonumber
\end{eqnarray}
Changing $\mbox{left}\,\leftrightarrow\,\mbox{right}$ in (\ref{kukura1}) we get the equation for $D_{\widetilde{\alpha}^{\prime}}\,H_{{\bf{+a}}\,\widetilde{\bf{+b}}}$. There is one simplification we can make in equations (\ref{kukura1}). Because only half of the block diagonal $\gamma_{+}$ matrix is nonzero and is proportional to the $\delta$ for the nonzero part. The $(\gamma_{\bf{+a}})_{\alpha^{\prime}}{}^{\beta}\,=\,\delta_{\alpha^{\prime}\,\nu^{\prime}}\,(\gamma_{\bf{a}})^{\nu^{\prime}\,\beta}\,\equiv\,(\gamma_{\bf{a}})_{\alpha^{\prime}}{}^{\beta}$.

The observation from (\ref{kukura1}) and its $\mbox{left}\,\leftrightarrow\,\mbox{right}$ swap is that the action of the $D_{\alpha^{\prime}}$ and $D_{\widetilde{\alpha}^{\prime}}$ on $H_{{\bf{+a}}\,\widetilde{\bf{+b}}}$ is producing two new vielbeins $H_{\beta\,\widetilde{\bf{+a}}}$ and $H_{\widetilde{\beta}\,{\bf{+a}}}$. This hints that we need some another vielbein, such that the action of $D_{\alpha^{\prime}}$ and $D_{\widetilde{\alpha}^{\prime}}$ on it will effectively subtract the fields $H_{\beta\,\widetilde{\bf{+a}}}$ and $H_{\widetilde{\beta}\,{\bf{+a}}}$. We found such a vielbein, but before giving it we will look at the flat case superspace first to give a motivation. After that we will generalise it to the $AdS_{5}\,\times\,S^{5}$ background.

\section{\texorpdfstring{Flat space solution}{Flat space solution}}
\subsection{\texorpdfstring{Flat space diagram}{Flat space diagram}}
To see what could be possibly a missing vielbein that will subtract vielbeins in (\ref{kukura1}) (and its $\mbox{left}\,\leftrightarrow\,\mbox{right}$ change) we first solve the same problem in flat space background. That is the extended superspace with $r_{AdS}\,\rightarrow \infty$. Note that in flat superspace ($r_{AdS}\,\rightarrow\,\infty$) the relation (\ref{kukura1}) simplifies significantly, because there are no $r_{AdS}$ dependent parts. The surviving part after $r_{AdS}\,\rightarrow\,\infty$ is just the second term on the right hand side of (\ref{kukura1}) with $g\,=\,1$, i.e. $\frac{1}{2}\,(\gamma_{\bf{+a}})_{\alpha^{\prime}}{}^{\beta}\,H_{\beta\,\widetilde{\bf{+b}}}$. 

Let us therefore further examine an action of $D_{\alpha^{\prime}}$ and $D_{\widetilde{\alpha}^{\prime}}$ on $H_{\beta\,\widetilde{\bf{+a}}}$ and  $H_{\widetilde{\beta}\,{\bf{+a}}}$ respectively:
\begin{eqnarray}
\label{priehradka}
T_{D\,D\,\widetilde{S}}\,\equiv\,T_{\alpha^{\prime}\,\beta\,\widetilde{\bf{+a}}}\,=\,0&=&D_{[\alpha^{\prime}}\,H_{\beta\,\widetilde{\bf{+a}})}\,+\,H_{[\alpha^{\prime}\,\go M}\,\eta^{\go M \go N}\,f_{\beta\,\widetilde{\bf{+a}}\,)\,\go N}\\
&=&D_{\alpha^{\prime}}\,H_{\beta\,\widetilde{\bf{+a}}}\,+\,S_{\widetilde{\bf{+a}}}\,H_{\alpha^{\prime}\,\beta}\,-\,D_{\beta}\,H_{\widetilde{\bf{+a}}\,\alpha^{\prime}}\,+\,H_{\alpha^{\prime}\,\go M}\,\eta^{\go M \go N}\,f_{\beta\,\widetilde{\bf{+a}}\,\go N}\,\nonumber\\
&&+\,H_{\widetilde{\bf{+a}}\,\go M}\,\eta^{\go M \go N}\,f_{\alpha^{\prime}\,\beta\,\go N}\,-\,H_{\beta\,\go M}\,\eta^{\go M \go N}\,f_{\widetilde{\bf{+a}}\,\alpha^{\prime}\,\go N}\nonumber
\end{eqnarray}
The mixed terms in the $f$ part of (\ref{priehradka}) are zero (note they are zero also in the $AdS$ background). The structure constant $f_{\alpha^{\prime}\,\beta\,\go N}\,=\,2\,(\gamma_{\bf{a}})_{\alpha^{\prime}\,\beta}\,\delta^{\bf{a}}{}_{\go N}$ (the same is in the $AdS$ background). The vielbein $H_{\widetilde{\bf{+a}}\,\alpha^{\prime}}\,\rightsquigarrow\,0$ by the table (\ref{hrncek2}). Then the equation (\ref{priehradka}) can be rewritten as:
\begin{eqnarray}
\label{gulasik}
0\,=\,D_{\alpha^{\prime}}\,H_{\beta\,\widetilde{\bf{+a}}}\,+\,S_{\widetilde{\bf{+a}}}\,H_{\alpha^{\prime}\,\beta}\,+\,2\,(\gamma^{\bf{c}})_{\alpha^{\prime}\,\beta}\,H_{\widetilde{\bf{+a}}\,{\bf{c}}}
\end{eqnarray}

To evaluate the only $S$ derivative term in (\ref{gulasik}) i.e. $S_{\widetilde{\bf{+a}}}\,H_{\alpha^{\prime}\,\beta}$ we would need to work a bit, in the $AdS$ superspace. The whole $AdS$ analysis of the actions of $D_{\alpha^{\prime}}$ and $D_{\widetilde{\alpha}^{\prime}}$ on $H_{\beta\,\widetilde{\bf{+a}}}$ and $H_{\widetilde{\beta}\,{\bf{+a}}}$ is done in the Appendix, see equations (\ref{hrozienko}) till (\ref{latenight}). In this section we would need only $r_{AdS}\,\rightarrow\,\infty$ limit of that analysis. 

In the Appendix we derived the equations (\ref{vonadlo1}) and (\ref{latenight}). Those equations are telling us that in the $AdS$ case (and so also in the flat case) the actions of $D_{\widetilde{\alpha}^{\prime}}$ and $D_{\alpha^{\prime}}$ result in a combination of $H_{{\bf{+a}}\,\widetilde{\bf{+b}}}$ and $H_{\alpha\,\widetilde{\beta}}$. This is actually a hint that we should add the trace of $H_{\alpha\,\widetilde{\beta}}$ to the trace of $H_{{\bf{+a}}\,\widetilde{\bf{+b}}}$ in order to subtract an action of a linear combination of $D_{\alpha^{\prime}}$ and $D_{\widetilde{\alpha}^{\prime}}$ (future $D^{v}$ derivative) on trace $H_{{\bf{+a}}\,\widetilde{\bf{+b}}}$. In the rest of this paragraph and next chapter we will look at how the pre-potential is built up in a flat space limit, i.e. we consider equations (\ref{vonadlo1}) and (\ref{vonadlo2}) and (\ref{latenight}) and (\ref{latenight1}) in the limit $r_{AdS}\,\rightarrow\,\infty$. We find equations that are fixing pre-potential and vanishing $D_{w}$ derivative.

Thus we repeat the flat space limits of the $D_{\widetilde{\alpha}^{\prime}}$ and $D_{\alpha^{\prime}}$ actions on $H_{\beta\,\widetilde{\bf{+a}}}$ and $H_{\widetilde{\beta}\,{\bf{+a}}}$ respectively, i.e. the equations (\ref{latenight}) and (\ref{latenight1}) in $r_{AdS}\,\rightarrow\,\infty$ limit: 
\begin{eqnarray}
\label{dur1}
0&=&D_{\widetilde{\alpha}^{\prime}}H_{\beta\,\widetilde{\bf{+a}}}\,+\,\frac{1}{2}\,(\gamma_{\bf{a}})_{\alpha^{\prime}}{}^{\nu}H_{\beta\,\widetilde{\nu}}\\
\label{dur2}
0&=&D_{\alpha^{\prime}}H_{\widetilde{\beta}\,{\bf{+a}}}\,+\,\frac{1}{2}\,(\gamma_{\bf{a}})_{\alpha^{\prime}}{}^{\nu}H_{\nu\,\widetilde{\beta}}
\end{eqnarray}
Similarly, we can also look at the equations (\ref{kukura1}) and (\ref{vonadlo1}) and (\ref{vonadlo2}) (in the flat space limit) and together with (\ref{dur1}) and (\ref{dur2}) we can observe the following interesting flat space diagram:
\begin{equation}
\label{scheme1}
\xymatrix{
H_{{\bf{+a}}\,{\widetilde{\bf{+b}}}} \ar[d]^{D_{\widetilde{\alpha}^{\prime}}} \ar[r]^{D_{\alpha^{\prime}}} & H_{\beta\,\widetilde{\bf{+b}}} \ar[d]^{D_{\widetilde{\alpha}^{\prime}}} \ar[r]^{D_{\alpha^{\prime}}} & H_{{\bf{c}}\,\widetilde{\bf{+b}}}\\
H_{{\bf{+a}}\,\widetilde{\beta}} \ar[r]^{D_{\alpha^{\prime}}} \ar[d]^{D_{\widetilde{\alpha}^{\prime}}}& H_{\beta\,\widetilde{\beta}}\\
H_{{\bf{+a}}\,\widetilde{\bf{c}}}
}
\end{equation}
The scheme (\ref{scheme1}) is nice and actually tells us what we should do next. Recall that the nodes $H_{{\bf{+a}}\,\widetilde{\bf{+b}}}$ and $H_{{\bf{c}}\,\widetilde{\bf{+b}}}$ and $H_{{\bf{+a}}\,\widetilde{\bf{c}}}$ could be identified by the use of invertible operators $P_{-}$ and $P_{\widetilde{-}}$ see (\ref{vidlak}). Our original aim was find a field that has a scalar trace and could possibly subtract actions of $D_{\alpha^{\prime}}$ and $D_{\widetilde{\alpha}^{\prime}}$ on $H_{{\bf{+a}}\,\widetilde{\bf{+b}}}$. We will see that the missing field is exactly $H_{\beta \widetilde{\beta}}$ (in the flat space, the only nonzero part of $H_{D\,\widetilde{D}}$). The diagram (\ref{scheme1}) suggests what to do. We calculate the remaining arrows and fill the square. 

To fill the remaining arrows we need to calculate the action of $D_{\alpha^{\prime}}$ and $D_{\widetilde{\alpha}^{\prime}}$ on $H_{\alpha\,\widetilde{\beta}}$ together with some another arrows that will be discussed later. We consider the dimension $\frac{1}{2}$ torsion constraint $T_{D\,D\,\widetilde{D}}\,\equiv\,T_{\alpha^{\prime}\,\beta\,\widetilde{\sigma}}\,=\,0$. We note again that for now on we are working in the flat space. Later we will generalise the procedure for the $AdS$ space:
\begin{eqnarray}
\label{hydinka}
T_{D\,D\,\widetilde{D}}\,\equiv\,T_{\alpha^{\prime}\,\beta\,\widetilde{\sigma}}\,=\,0&=&D_{(\alpha^{\prime}}H_{\beta\,\widetilde{\sigma})}\,+\,H_{(\alpha^{\prime}\,|\,\go M}\,\eta^{\go M \go N}\,f_{\beta\,\widetilde{\sigma}\,)\,\go N}\\
\label{hydinka2}
&=&D_{\alpha^{\prime}}H_{\beta\,\widetilde{\sigma}}\,+\,2\,(\gamma^{\bf{a}})_{\alpha^{\prime}\,\beta}\,H_{\widetilde{\sigma}\,{\bf{a}}}
\end{eqnarray}
where we used that $H_{\alpha^{\prime}\,\beta}\,=\,H_{\alpha^{\prime}\,\widetilde{\sigma}}\,=\,0$ in flat space (see (\ref{bukvica}) and (\ref{zmena}) and do flat space limit). We also have a $\mbox{left}\,\leftrightarrow\,\mbox{right}$ swap of (\ref{hydinka2}). The vielbein $H_{\widetilde{\sigma}\,{\bf{c}}}$ in (\ref{hydinka2}) is related to $H_{{\bf{+c}}\,\widetilde{\sigma}}$. For that consider torsion constraint $T_{P\,S\,\widetilde{D}}\,\equiv\,T_{-\,{\bf{+c}}\,\widetilde{\sigma}}\,=\,0$:  
\begin{eqnarray}
T_{P\,S\,\widetilde{D}}\,\equiv\,T_{-\,{\bf{+c}}\,\widetilde{\sigma}}\,=\,0&=&P_{[-}H_{{\bf{+c}}\,\widetilde{\sigma})}\,+\,H_{[-\,|\,\go M}\,\eta^{\go M \go N}\,f_{{\bf{+c}}\,\widetilde{\sigma}\,)\,\go N}\\
\label{carpenter1}
&=&P_{-}H_{{\bf{+c}}\,\widetilde{\sigma}}\,+\,\eta_{+-}\,H_{\widetilde{\sigma}\,{\bf{c}}}
\end{eqnarray}
where we used left-right light-cone gauge, together with $H_{-\,{\bf{+c}}}\,=\,0$ that is shown in the Appendix and holds even in $AdS$, see (\ref{fishtail}). 

To fill the diagram (\ref{scheme1}) we need to calculate two more torsion constraints that are providing the actions of $D_{\alpha^{\prime}}$ on $H_{{\bf{+a}}\,\widetilde{\bf{c}}}$ and on $H_{\sigma\,\widetilde{\bf{c}}}$. We first consider $T_{D\,S\,\widetilde{P}}\,\equiv\,T_{\alpha^{\prime}\,{\bf{+a}}\,\widetilde{\bf{c}}}\,=\,0$:
\begin{eqnarray}
\label{uber1}
T_{D\,S\,\widetilde{P}}\,\equiv\,T_{\alpha^{\prime}\,{\bf{+a}}\,\widetilde{\bf{c}}}\,=\,0&=&D_{[\alpha^{\prime}}H_{{\bf{+a}}\,\widetilde{\bf{c}})}\,+\,H_{[\alpha^{\prime}\,|\,\go M}\eta^{\go M \go N}\,f_{{\bf{+a}}\,\widetilde{\bf{c}})\,\go N}\\
\label{uber2}
&=&D_{\alpha^{\prime}}H_{{\bf{+a}}\,\widetilde{\bf{c}}}\,+\,\frac{1}{2}\,(\gamma_{\bf{+a}})_{\alpha^{\prime}}{}^{\sigma}\,H_{\widetilde{\bf{c}}\,\sigma}\\
\label{uber3}
&=&D_{\alpha^{\prime}}H_{{\bf{+a}}\,\widetilde{\bf{c}}}\,+\,\frac{1}{2}\,(\gamma_{\bf{a}})_{\alpha^{\prime}}{}^{\sigma}\,H_{\widetilde{\bf{c}}\,\sigma}
\end{eqnarray}
where we used that $H_{\alpha^{\prime}\,{\bf{+a}}}\,=\,0$ (holds even in the $AdS$, see table (\ref{hrncek2})). We also used that $H_{\alpha^{\prime}\,\widetilde{\bf{c}}}\,=\,0$ (that is enough in a flat space to have $S_{\bf{+a}}\,H_{\alpha^{\prime}\,\widetilde{\bf{c}}}\,=\,0$). To see that $H_{\alpha^{\prime}\,\widetilde{\bf{c}}}\,=\,0$ we use the torsion $T_{\widetilde{P}\,\widetilde{S}\,D}\,\equiv\,T_{\widetilde{-}\,\widetilde{\bf{+c}}\,\alpha^{\prime}}\,=\,0$:
\begin{eqnarray}
T_{\widetilde{P}\,\widetilde{S}\,D}\,\equiv\,T_{\widetilde{-}\,\widetilde{\bf{+c}}\,\alpha^{\prime}}\,=\,0&=&P_{[\widetilde{-}}H_{\widetilde{\bf{+c}}\,\alpha^{\prime})}\,+\,H_{[\widetilde{-}\,|\,\go M}\,\eta^{\go M \go N}\,f_{\widetilde{\bf{+c}}\,\alpha^{\prime}\,)\,\go N}\\
\label{smutno}
&=&P_{\widetilde{-}}H_{\widetilde{\bf{+c}}\,\alpha^{\prime}}\,+\,\eta_{+-}\,H_{\alpha^{\prime}\,\widetilde{\bf{c}}}
\end{eqnarray}
and previously we saw that $H_{\widetilde{\bf{+c}}\,\alpha^{\prime}}\,=\,0$ (even in the $AdS$ case, see table (\ref{hrncek2})). From (\ref{smutno}) in the flat case follows that $H_{\alpha^{\prime}\,\widetilde{\bf{c}}}\,=\,0$. Examining the (\ref{smutno}) in the $AdS$ case one also finds that $H_{\alpha^{\prime}\,\widetilde{\bf{c}}}\,=\,0$ (after evaluation). The (\ref{uber3}) however could have some additional term in the $AdS$ case. The structure constant $f_{\alpha^{\prime}\,\widetilde{\bf{c}}\,\go N}\,\neq\,0$ and so the term proportional to that structure constant in the $AdS$ case is $\frac{1}{r_{AdS}}\,(\gamma_{\bf{c}})_{\alpha^{\prime}\,\sigma}\,(\widetilde{\Gamma}_{5})^{\sigma\,\nu}\,H_{{\bf{+a}}\,\widetilde{\nu}}$. That term is nonzero in the $AdS$ case. Moreover, in the (\ref{uber1}) one finds one more $AdS$ term, coming from evaluated action $S_{\bf{+a}}\,H_{\alpha^{\prime}\,\widetilde{\bf{c}}}$. Those terms are not of a big concern right now (doing the flat space first), we will see them later in the section where we generalise to $AdS$ case.

Last torsion constraint to examine in order to fill the (\ref{scheme1}) is the one that determines the action of $D_{\alpha^{\prime}}$ on $H_{\beta\,\widetilde{\bf{c}}}$. Consider therefore the dimension $\frac{1}{2}$ torsion $T_{D\,D\,\widetilde{P}}\,\equiv\,T_{\alpha^{\prime}\,\beta\,\widetilde{\bf{c}}}\,=\,0$:
\begin{eqnarray}
T_{D\,D\,\widetilde{P}}\,\equiv\,T_{\alpha^{\prime}\,\beta\,\widetilde{\bf{c}}}\,=\,0&=&D_{[\alpha^{\prime}}H_{\beta\,\widetilde{\bf{c}})}\,+\,H_{[\alpha^{\prime}\,|\,\go M}\,\eta^{\go M \go N}\,f_{\beta\,\widetilde{\bf{c}}\,)\,\go N}\\
\label{last}
&=&D_{\alpha^{\prime}}\,H_{\beta\,\widetilde{\bf{c}}}\,+\,2\,(\gamma^{\bf{a}})_{\alpha^{\prime}\,\beta}\,H_{\widetilde{\bf{c}}\,{\bf{a}}}
\end{eqnarray}
where we used the $H_{\alpha^{\prime}\,\beta}\,=\,0$ (holds also in $AdS$ after the evaluation) and $H_{\widetilde{\bf{c}}\,\alpha^{\prime}}\,=\,0$ (also holds in $AdS$ after the evaluation). In the $AdS$ case in the equation (\ref{last}) we have two additional terms. They come from $f_{\beta\,\widetilde{\bf{c}}\,\go N}\,\neq\,0$ and also $f_{\alpha^{\prime}\,\widetilde{\bf{c}}\,\go N}\,\neq\,0$. Those terms will be further analysed in future sections, let just write their structure as $\frac{1}{r_{AdS}}\,(\gamma_{\bf{c}})_{\beta\,\nu^{\prime}}\,(\widetilde{\Gamma}_{5})^{\nu^{\prime}\,\sigma^{\prime}}\,H_{\widetilde{\sigma}^{\prime}\,\alpha^{\prime}}$ and $\frac{1}{r_{AdS}}\,(\gamma_{\bf{c}})_{\alpha^{\prime}\,\nu}\,(\widetilde{\Gamma}_{5})^{\nu\,\sigma}\,H_{\widetilde{\sigma}\,\beta}$. The vielbein $H_{\widetilde{\sigma}^{\prime}\,\alpha^{\prime}}\,=\,0$ (in flat case and also in $AdS$ after the evaluation) as can be calculated from torsion constraints $T_{-\,\widetilde{\sigma}^{\prime}\,\alpha^{\prime}}\,=\,0$ and $T_{\widetilde{-}\,-\,\widetilde{\sigma}^{\prime}}\,=\,0$ and the use of the double light-cone gauge. The term $H_{\widetilde{\sigma}\,\beta}$ is nonzero ($H_{\widetilde{\sigma}\,\beta}$ vielbein is a part of a pre-potential). 

We can add results of (flat space) equations (\ref{uber3}) and (\ref{last}) together with (\ref{hydinka2}) and their $\mbox{left}\,\leftrightarrow\,\mbox{right}$ swaps to the diagram (\ref{scheme1}) and find the following square diagram:
\begin{equation}
\label{scheme2}
\xymatrix{
H_{{\bf{+a}}\,{\widetilde{\bf{+b}}}} \ar[d]^{D_{\widetilde{\alpha}^{\prime}}} \ar[r]^{D_{\alpha^{\prime}}} & H_{\beta\,\widetilde{\bf{+b}}} \ar[d]^{D_{\widetilde{\alpha}^{\prime}}} \ar[r]^{D_{\alpha^{\prime}}} & H_{{\bf{c}}\,\widetilde{\bf{+b}}} \ar[d]^{D_{\widetilde{\alpha}^{\prime}}}\\
H_{{\bf{+a}}\,\widetilde{\beta}} \ar[r]^{D_{\alpha^{\prime}}} \ar[d]^{D_{\widetilde{\alpha}^{\prime}}}& H_{\beta\,\widetilde{\beta}} \ar[r]^{D_{\alpha^{\prime}}} \ar[d]^{D_{\widetilde{\alpha}^{\prime}}}  &H_{{\bf{c}}\,\widetilde{\beta}} \ar[d]^{D_{\widetilde{\alpha}^{\prime}}}\\
H_{{\bf{+a}}\,\widetilde{\bf{c}}} \ar[r]^{D_{\alpha^{\prime}}} & H_{\beta\,\widetilde{\bf{c}}} \ar[r]^{D_{\alpha^{\prime}}} & H_{\bf{c}\,\widetilde{\bf{c}}} 
}
\end{equation}

As we saw before the (\ref{scheme2}) nodes $\{\,H_{{\bf{+a}}\,\widetilde{\bf{+b}}},\,H_{{\bf{c}}\,\widetilde{\bf{+b}}},\,H_{{\bf{+a}}\,\widetilde{\bf{c}}},\,H_{{\bf{c}}\,\widetilde{\bf{c}}}\,\}$ should be identified (as one node). We proved that using various torsion constraints, mixed light-cone gauge and invertibility of $P_{-}$ and $P_{\widetilde{-}}$. The same way the nodes  $\{\,H_{{\bf{+a}}\,\widetilde{\beta}},\,H_{{\bf{c}}\,\widetilde{\beta}}\,\}$ and independently nodes $\{\,H_{\beta\,\widetilde{\bf{+b}}},\,H_{\beta\,\widetilde{\bf{c}}}\,\}$ should be identified (as two independent nodes). The vielbein $H_{\beta\,\widetilde{\beta}}$ is then just a single node.  After the described identifications the diagram (\ref{scheme2}) could be rewritten in the simpler and more informative form.
\begin{equation}
\label{scheme3}
\xymatrix{
&H_{{\bf{+a}}\,\widetilde{\bf{+b}}}\ar@/^1pc/[rd]^{D_{\alpha^{\prime}}} \ar@{-->}@/^0.5pc/[ld]^{D_{\widetilde{\alpha}^{\prime}}}&\\
H_{{\bf{+a}}\,\widetilde{\beta}}\ar@{-->}@/^1pc/[ru]^{D_{\widetilde{\alpha}^{\prime}}} \ar@/^0.5pc/[rd]^{D_{{\alpha}^{\prime}}}&&H_{\beta\,\widetilde{\bf{+b}}} \ar@/^0.5pc/[lu]^{D_{{\alpha}^{\prime}}} \ar@{-->}@/^1pc/[ld]^{D_{\widetilde{\alpha}^{\prime}}}\\
&H_{\beta\,\widetilde{\beta}}\ar@{-->}@/^0.5pc/[ru]^{D_{\widetilde{\alpha}^{\prime}}} \ar@/^1pc/[lu]^{D_{\alpha^{\prime}}}&
}
\end{equation}
Note, the dashed arrows stand for action of $D_{\widetilde{\alpha}^{\prime}}$ and solid arrows stand for action of $D_{\alpha^{\prime}}$. From the nice flat space diagram (\ref{scheme3}) it is obvious that in order to have a vanishing derivative we have to combine $D_{\alpha^{\prime}}$ with $D_{\widetilde{\alpha}^{\prime}}$ and that combination should act on the combination of traces of $H_{{\bf{+a}}\,\widetilde{\bf{+b}}}$ with $H_{\beta\,\widetilde{\beta}}$.

\subsection{\texorpdfstring{The ${\mathbb{H}}$ matrix}{The H matrix}}
The diagram (\ref{scheme3}) could be rewritten in the matrix form. The observation is that each action of the derivatives in the (\ref{scheme3}) is given by some matrix. The derivatives are mixing fields just as in (\ref{scheme3}). Let us introduce the $2\,\otimes\,2$ block matrix $\mathbb{H}$:
\begin{eqnarray}
\label{hmat}
{\mathbb{H}} \defeq 
 \begin{pmatrix}
  H_{{\bf{+a}}\,\widetilde{\bf{+b}}}  &  H_{{\bf{+a}}\,\widetilde{\beta}}\\
  H_{\beta\,\widetilde{\bf{+b}}}  & H_{\beta\,\widetilde{\beta}} 
 \end{pmatrix}
\end{eqnarray} 
The action of $D_{\alpha^{\prime}}$ is then given as the left action of some constant (up to $P_{-}$ operator) block off diagonal matrix $\mathcal{\Gamma_{\alpha^{\prime}}}$: 
\begin{eqnarray}
\label{vintage1}
D_{\alpha^{\prime}}
 \begin{pmatrix}
  H_{{\bf{+a}}\,\widetilde{\bf{+b}}}  &  H_{{\bf{+a}}\,\widetilde{\beta}}\\
  H_{\beta\,\widetilde{\bf{+b}}}  & H_{\beta\,\widetilde{\beta}} 
 \end{pmatrix}&\equiv&\begin{pmatrix}
  0  &  \frac{1}{2}(\gamma_{{\bf{a}}})_{\alpha^{\prime}}{}^{\sigma}\\
  2\,P_{-}\,(\gamma^{{\bf{c}}})_{\alpha^{\prime}\,\beta}  & 0
 \end{pmatrix}\,\begin{pmatrix}
  H_{{\bf{+c}}\,\widetilde{\bf{+b}}}  &  H_{{\bf{+c}}\,\widetilde{\beta}}\\
  H_{\sigma\,\widetilde{\bf{+b}}}  & H_{\sigma\,\widetilde{\beta}} 
 \end{pmatrix}\\
 &&\nonumber\\
D_{\alpha^{\prime}}\,{\mathbb{H}}&=&\Gamma_{\alpha^{\prime}}\,{\mathbb{H}}
\end{eqnarray}
The action of $D_{\widetilde{\alpha}^{\prime}}$ on $\mathbb{H}$ is given as a right action of similar matrix $\Gamma_{\widetilde{\alpha}^{\prime}}$:
\begin{eqnarray}
\label{vintage2}
D_{\widetilde{\alpha}^{\prime}}
 \begin{pmatrix}
  H_{{\bf{+a}}\,\widetilde{\bf{+b}}}  &  H_{{\bf{+a}}\,\widetilde{\beta}}\\
  H_{\beta\,\widetilde{\bf{+b}}}  & H_{\beta\,\widetilde{\beta}} 
 \end{pmatrix}&\equiv&\begin{pmatrix}
  H_{{\bf{+a}}\,\widetilde{\bf{+c}}}  &  H_{{\bf{+a}}\,\widetilde{\sigma}}\\
  H_{\beta\,\widetilde{\bf{+c}}}  & H_{\beta\,\widetilde{\sigma}} 
 \end{pmatrix}\,\begin{pmatrix}
  0  &  2\,P_{\widetilde{-}}\,(\gamma^{{\bf{c}}})_{\alpha^{\prime}\,\beta}\\
  \frac{1}{2}\,(\gamma_{{\bf{b}}})_{\alpha^{\prime}}{}^{\sigma}  & 0
 \end{pmatrix}\\
 &&\nonumber\\
D_{\widetilde{\alpha}^{\prime}}\,{\mathbb{H}}&=&{\mathbb{H}}\,\Gamma_{\widetilde{\alpha}^{\prime}}
\end{eqnarray}
Now we will proceed to the main step. We arbitrarily linearly combine $D_{\alpha^{\prime}}$ and $D_{\widetilde{\alpha}^{\prime}}$, i.e. we multiply the $D_{\widetilde{\alpha}^{\prime}}$ with some unknown nonsingular matrix $\mathcal{M}_{\alpha^{\prime}}{}^{\beta^{\prime}}$:
\be
\label{dv}
{\mathcal{D}}_{v}\,\equiv\,{\mathcal{D}}^{v}{}_{\alpha^{\prime}}\,\defeq\,(\,D_{\alpha^{\prime}}\,-\,{\mathcal{M}}_{\alpha^{\prime}}{}^{\beta^{\prime}}\,D_{\widetilde{\beta}^{\prime}}\,)
\ee
Moreover we impose that in the matrix version of $D_{\widetilde{\alpha}^{\prime}}$ action the matrix ${\mathcal{M}}$ acts as follows:
\be
\label{act}
{\mathcal{M}}_{\alpha^{\prime}}{}^{\beta^{\prime}}\,\Gamma_{\widetilde{\beta}^{\prime}}\,\defeq\,{\mathbb{A}}\,\Gamma_{\widetilde{\alpha}^{\prime}}\,{\mathbb{B}}
\ee
for some nonsingular matrices ${\mathbb{A}}$ and ${\mathbb{B}}$. Combining (\ref{vintage1}) and (\ref{vintage2}) together with (\ref{dv}) and (\ref{act}) we get:
\begin{eqnarray}
\label{slniecko}
{\mathcal{D}}^{v}{}_{\alpha^{\prime}}\,{\mathbb{H}}\,&=&\Gamma_{\alpha^{\prime}}\,{\mathbb{H}}\,-\,{\mathbb{H}}\,{\mathbb{A}}\,\Gamma_{\widetilde{\alpha}^{\prime}}\,{\mathbb{B}}\,\,\,\,\,\,\,\,\,\,/\,{\mathbb{B}}^{-1}\,\\
{\mathcal{D}}^{v}{}_{\alpha^{\prime}}\,{\mathbb{H}}\,{\mathbb{B}}^{-1}&=&\Gamma_{\alpha^{\prime}}\,{\mathbb{H}}\,{\mathbb{B}}^{-1}\,-\,{\mathbb{H}}\,{\mathbb{A}}\,\Gamma_{\widetilde{\alpha}^{\prime}}\,\,\,\,\,\,\,\,\,\,/\,{\mbox{Str}}\,\\
\label{toaletiak}
{\mathcal{D}}^{v}{}_{\alpha^{\prime}}\,\mbox{Str}\,(\,{\mathbb{H}}\,{\mathbb{B}}^{-1}\,)&=&\mbox{Str}\,{\Big{(}}(\,{\mathbb{B}}^{-1}\,\Gamma_{\alpha^{\prime}}\,-\,{\mathbb{A}}\,\Gamma_{\widetilde{\alpha}^{\prime}}\,)\,{\mathbb{H}}\,{\Big{)}}
\end{eqnarray}
by the $\mbox{Str}$ we mean the super-trace. We put to zero the $\mbox{Str}\,{\Big{(}}\,(\,{\mathbb{B}}^{-1}\,\Gamma_{\alpha^{\prime}}\,-\,{\mathbb{A}}\,\Gamma_{\widetilde{\alpha}^{\prime}}\,)\,{\mathbb{H}}\,{\Big{)}}\,=\,0$ by finding the suitable matrices ${\mathbb{B}}$ and $\mathbb{A}$ and the matrix ${\mathcal{M}}$. By that we get the equation: 
\be
\label{rovnica}
{\mathcal{D}}^{v}{}_{\alpha^{\prime}}\,\mbox{Str}\,(\,{\mathbb{H}}\,{\mathbb{B}}^{-1}\,)\,=\,0
\ee
thus the equation (\ref{rovnica}) defines the $\mbox{Str}\,(\,{\mathbb{H}}\,{\mathbb{B}}^{-1}\,)$ as the scalar field on which particular combination of $D_{\alpha^{\prime}}$ and $D_{\widetilde{\alpha}^{\prime}}$ now called ${\mathcal{D}}^{v}{}_{\alpha^{\prime}}$ vanishes. So, we found a pre-potential $V\,\defeq\,\mbox{Str}\,(\,{\mathbb{H}}\,{\mathbb{B}}^{-1}\,)$. We note that even though the equation (\ref{toaletiak}) might seem easy to solve just by putting ${\mathbb{B}^{-1}}\,=\,{\mathbb{A}}$. It is not that simple since $\Gamma_{\alpha^{\prime}}\,\neq\,\Gamma_{\widetilde{\alpha}^{\prime}}$. Therefore some more involved solution has to be found. 

\subsection{\texorpdfstring{Solution via the gamma matrix identity}{Solution via the gamma matrix identity}}
We solve the equation (\ref{toaletiak}) using the following identity:
\begin{eqnarray}
\label{iden}
{\mathcal{A}}_{\alpha^{\prime}}{}^{\sigma^{\prime}}\,{\mathcal{B}}_{\bf{a}}{}^{\bf{c}}\,{\mathcal{C}}_{\beta}{}^{\nu}\,(\gamma_{\bf{c}})_{\sigma^{\prime}\,\nu}\,=\,(\gamma_{\bf{a}})_{\alpha^{\prime}\,\beta}
\end{eqnarray}
has two $SO\,(\,4\,)$ invariant solutions: 
\begin{eqnarray}
\label{chir}
{\mathbb{I.}}&:&{\mathcal{A}_{\alpha^{\prime}}{}^{\beta^{\prime}}}\,=\,\delta_{\alpha^{\prime}}{}^{\beta^{\prime}}\,\,\,\,||\,\,\,\,{\mathcal{C}_{\alpha}{}^{\beta}}\,=\,\delta_{\alpha}{}^{\beta}\,\,\,\,||\,\,\,\,{\mathcal{B}}_{\bf{a}}{}^{\bf{b}}\,=\,\delta_{\bf{a}}{}^{\bf{b}}\\
\label{proj}
{\mathbb{II.}}&:&{\mathcal{A}_{\alpha^{\prime}}{}^{\beta^{\prime}}}\,=\,(\widetilde{\Gamma}_{5})_{\alpha^{\prime}}{}^{\beta^{\prime}}\,\,\,\,||\,\,\,\,{\mathcal{C}_{\alpha}{}^{\beta}}\,=\,(\widetilde{\Gamma}_{5})_{\alpha}{}^{\beta}\,\,\,\,||\,\,\,\,{\mathcal{B}}_{\bf{a}}{}^{\bf{b}}\,=\,(\widetilde{\Gamma}_{5})_{\bf{a}}{}^{\bf{b}}
\end{eqnarray}
The solution (\ref{chir}) is trivial, the solution (\ref{proj}) is based on property of the $\widetilde{\Gamma}_{5}$ matrix:  $[\,\widetilde{\Gamma}_{5},\,\gamma_{\bf{a}}\,]\,=\,0\,\,\,\mbox{for}\,\,\,{\bf{a}}\,\in\,\{\,{{10}},\,{{1}},\,{{2}},\,{{3}},\,{{4}}\,\}$ and $\{\,\widetilde{\Gamma}_{5},\,\gamma_{\bf{a}}\,\}\,=\,0\,\,\,\mbox{for}\,\,\,{\bf{a}}\,\in\,\{\,{{5}},\,{{6}},\,{{7}},\,{{8}},\,{{9}}\,\}$. The previous follows directly from the definition of $\widetilde{\Gamma}_{5}$, see (\ref{gam5}). The new matrix $(\widetilde{\Gamma}_{5})_{\bf{a}}{}^{\bf{b}}$ in (\ref{proj}) is defined by the (\ref{iden}) to fix the signs. Note that the indices ${\bf{a}}$ in (\ref{iden}) have a range: ${\bf{a}}\,\in\,\{\,{{1}},\,\dots\,,{{8}}\}$. 

Next, we look explicitly at the equation: 
\be
\label{aqua}
\mbox{Str}\,{\Big{(}}(\,{\mathbb{B}}^{-1}\,\Gamma_{\alpha^{\prime}}\,-\,{\mathbb{A}}\,\Gamma_{\widetilde{\alpha}^{\prime}}\,)\,{\mathbb{H}}\,{\Big{)}}\,\equiv\,{\mbox{Str}}\,{\mathbb{X}}_{\alpha^{\prime}}\,=\,0
\ee
let's rename the members of the matrix ${\mathbb{H}}$:
\begin{eqnarray}
\label{sovicka}
{\mathbb{H}}\,\equiv
 \begin{pmatrix}
  H_{{\bf{+a}}\,\widetilde{\bf{+b}}}  &  H_{{\bf{+a}}\,\widetilde{\beta}}\\
  H_{\beta\,\widetilde{\bf{+b}}}  & H_{\beta\,\widetilde{\beta}} 
 \end{pmatrix}&\equiv&\begin{pmatrix}
  H_{S\,\widetilde{S}}  &  H_{S\,\widetilde{D}}\\
  H_{D\,\widetilde{S}}  & H_{D\,\widetilde{D}} 
 \end{pmatrix}
 \end{eqnarray}
Let us define the matrices ${\mathbb{A}}$ and ${\mathbb{B}^{-1}}$ to be block diagonal matrices. This is a consistent choice with the fact that we want to have a pre-potential build out of $H_{S\,\widetilde{S}}$ and $H_{D\,\widetilde{D}}$. The pre-potential is in (\ref{rovnica}) given as $\mbox{Str}\,(\,{\mathbb{H}}\,{\mathbb{B}}^{-1}\,)$. We do not want to mix in some off diagonal ${\mathbb{H}}$ fields by the action of ${\mathbb{B}^{-1}}$. Thus ${\mathbb{A}}$ and ${\mathbb{B}^{-1}}$ are:
 \begin{eqnarray}
\label{lenUsko}
{\mathbb{A}}&\equiv
 &\begin{pmatrix}
  A_{S\,\widetilde{S}}  &  0\\
  0  & A_{D\,\widetilde{D}} 
 \end{pmatrix}
 \,\,\,\,\,\,\,||\,\,\,\,\,\,\,{\mathbb{B}}^{-1}\equiv
 \begin{pmatrix}
  B^{-1}{}_{S\,\widetilde{S}}  &  0\\
  0  & B^{-1}{}_{D\,\widetilde{D}} 
 \end{pmatrix}
 \end{eqnarray}
 With definitions (\ref{lenUsko}) we get the equation (\ref{aqua}) into the following matrix equation:
\begin{eqnarray}
\label{vlozticka}
{\mbox{Str}}\,{\Big{(}}\,
\begin{pmatrix}
  \frac{1}{2}\,B^{-1}{}_{S\,\widetilde{S}}\,\gamma\,H_{D\,\widetilde{S}}  &  \dots\\
  \dots  & 2\,P_{-}\,B^{-1}{}_{D\,\widetilde{D}}\,\gamma\,H_{S\,\widetilde{D}} 
 \end{pmatrix}\,-\,\begin{pmatrix}
  2\,P_{\widetilde{-}}\,A{}_{S\,\widetilde{S}}\,\gamma\,H_{D\,\widetilde{S}}  &  \dots\\
  \dots  &\,\frac{1}{2}\,A{}_{D\,\widetilde{D}}\,\gamma\,H_{S\,\widetilde{D}} 
 \end{pmatrix}\,{\Big{)}}=\,0
 \end{eqnarray}
 Then from (\ref{vlozticka}) we get two equations (since fields $H_{D\,\widetilde{S}}$ and $H_{S\,\widetilde{D}}$ are independent):
 \begin{eqnarray}
 \frac{1}{2}\,B^{-1}{}_{S\,\widetilde{S}}\,-\,2\,P_{\widetilde{-}}\,A_{S\,\widetilde{S}}&=&0\,\Rightarrow\,A{}_{S\,\widetilde{S}}\,=\,\frac{1}{4\,P_{\widetilde{-}}}\,B^{-1}{}_{S\,\widetilde{S}}\\
 P_{-}\,B^{-1}{}_{D\,\widetilde{D}}\,-\,A_{D\,\widetilde{D}}&=&0\,\Rightarrow\,A{}_{D\,\widetilde{D}}\,=\,4\,P_{-}\,B^{-1}{}_{D\,\widetilde{D}} 
 \end{eqnarray}
Now we are prepared to examine the equation (\ref{act}) using the ${\mathbb{A}}$ and ${\mathbb{B}}$ constructed above. Then the matrix equation (\ref{act}) can be (schematically) written: 
\begin{eqnarray}
\label{stream}
{\mathcal{M}}
\begin{pmatrix}
 0 &  2\,P_{\widetilde{-}}\,\gamma\\
  \frac{1}{2}\,\gamma & 0 
 \end{pmatrix}\,=\,\begin{pmatrix}
 \frac{1}{\,4\,P_{\widetilde{-}}}\,B^{-1}{}_{S\,\widetilde{S}} &  0\\
  0 & 4\,P_{-}\,B^{-1}{}_{D\,\widetilde{D}}
 \end{pmatrix}\,\begin{pmatrix}
 0 & 2\,P_{\widetilde{-}}\,\gamma\\
  \frac{1}{2}\,\gamma & 0 
 \end{pmatrix}\,\begin{pmatrix}
 B_{S\,\widetilde{S}} &  0\\
  0 & B_{D\,\widetilde{D}} 
 \end{pmatrix}
\end{eqnarray}
\begin{eqnarray}
\label{streamIN}
{\mathcal{M}}
\begin{pmatrix}
 0 &  2\,P_{\widetilde{-}}\,\gamma\\
  \frac{1}{2}\,\gamma & 0 
 \end{pmatrix}\,=\,\begin{pmatrix}
0 & \frac{1}{2}\,B^{-1}{}_{S\,\widetilde{S}}\,\gamma\,B_{D\,\widetilde{D}}\\
2\,P_{-}\,B^{-1}{}_{D\,\widetilde{D}}\,\gamma\,B_{S\,\widetilde{S}} & 0
 \end{pmatrix}
\end{eqnarray}
We now do the following re-scalings ${\mathcal{M}}\,\rightarrow\,\frac{1}{\lambda}\,{\mathcal{M}}$ and  $B_{S\,\widetilde{S}}\,\rightarrow\,\Delta\,B_{S\,\widetilde{S}}$ and $B_{D\,\widetilde{D}}\,\rightarrow\,\rho\,B_{D\,\widetilde{D}}$. Rescaled ${\mathcal{M}}$ and $B_{S\,\widetilde{S}}$ and $B_{D\,\widetilde{D}}$ belong to one of the two solutions of identity (\ref{iden}). Then we get the version of (\ref{streamIN}):
\begin{eqnarray}
\label{streamOUT}
{\mathcal{M}}
\begin{pmatrix}
 0 &  2\,P_{\widetilde{-}}\,\gamma\\
  \frac{1}{2}\,\gamma & 0 
 \end{pmatrix}\,=\,\begin{pmatrix}
0 & \frac{1}{2}\,\frac{\lambda\,\rho}{\Delta}\,B^{-1}{}_{S\,\widetilde{S}}\,\gamma\,B_{D\,\widetilde{D}}\\
2\,\frac{\lambda\,\Delta}{\rho}\,P_{-}\,B^{-1}{}_{D\,\widetilde{D}}\,\gamma\,B_{S\,\widetilde{S}} & 0
 \end{pmatrix}
\end{eqnarray}
Now we want the $\lambda$ and $\rho$ and $\Delta$ to satisfy:
\begin{eqnarray}
\label{solsol}
\frac{\lambda\,\rho}{\Delta}\,=\,4\,P_{\widetilde{-}}\,\,\,\,\,\,\,\,\mbox{and}\,\,\,\,\,\,\,\,\,\frac{\lambda\,\Delta}{\rho}\,=\,\frac{1}{4\,P_{-}}\,\,\,\,\,\,\,\,\Rightarrow\,\,\,\lambda\,=\,\pm\,\sqrt{\frac{P_{\widetilde{-}}}{P_{-}}}\,\,\,\,\,\mbox{and}\,\,\,\,\,\,\frac{\rho}{\Delta}\,=\,\pm\,4\,\sqrt{P_{\widetilde{-}}\,P_{-}}
\end{eqnarray}
The (\ref{streamOUT}) is just a matrix equation:
\begin{eqnarray}
\label{house}
{\mathcal{M}}\,
\begin{pmatrix}
 0 &  P_{\widetilde{-}}\,\gamma\\
  \gamma & 0 
 \end{pmatrix}\,=\,\begin{pmatrix}
0 &A^{-1}{}_{S\,\widetilde{S}}\,P_{\widetilde{-}}\,\gamma\,A_{D\,\widetilde{D}}\\
A^{-1}{}_{D\,\widetilde{D}}\,\gamma\,A_{S\,\widetilde{S}} & 0
\end{pmatrix}
\end{eqnarray}
that can be solved by (\ref{iden}).
Even though we saw the appearance of the nasty square roots in the (\ref{solsol}) and so in the definition of  ${\mathcal{D}}^{v}{}_{\alpha^{\prime}}$ and in the pre-potential via super-trace of  ${\mathbb{H\,B}}^{-1}$. We will see in the $AdS$ case solution that there is a way how to get rid of it.

\section{\texorpdfstring{$AdS_{5}\,\times\,S^{5}$ solution}{AdS\textfiveinferior{} x S\textfivesuperior{} solution}}

\subsection{\texorpdfstring{$AdS_{5}\,\times\,S^{5}$ diagram}{AdS\textfiveinferior{} x S\textfivesuperior{} diagram}}
In the previous sub-sections we saw how to find the pre-potential in the flat case. We are really interested in the $AdS$ case. Along the way we analysed the flat case in the previous sub-sections we mentioned also changes one has to make in the $AdS$ case. We repeat them here again since they are scattered over the previous flat case sub-sections and in the Appendix. 
First change has already been worked out in the relation between $H_{{\bf{+a}\,{\widetilde{\bf{+b}}}}}$ and $H_{{\bf{a}}\,\widetilde{\bf{b}}}$ in (\ref{redneck}). We also note that there are  $AdS$ contributions in equations (\ref{kukura1}) also in (\ref{vonadlo1}) and (\ref{vonadlo2}). The nontrivial contributions also appeared in equations (\ref{latenight}) and (\ref{latenight1}).

We could visualise the relations (\ref{kukura1}) and (\ref{vonadlo1}) and (\ref{vonadlo2}) and (\ref{latenight}) and (\ref{latenight1}) by the similar diagram as used in flat case, see (\ref{scheme1}). The structure is very similar just with more arrows between nodes. Since the $AdS$ diagram is messier we will not provide it. The idea is however the same as in the flat case. In order to determine the vanishing $D^{v}{}_{\alpha^{\prime}}$ derivative and the pre-potential we need to combine $D_{\alpha^{\prime}}$ and $D_{\widetilde{\alpha}^{\prime}}$ for $D^{v}$ derivative and $H_{{\bf{+a}}\,\widetilde{\bf{+b}}}$ together with $H_{\alpha\,\widetilde{\beta}}$ for pre-potential.

The only missing derivative in the set of $AdS$ equations: (\ref{kukura1}) and (\ref{vonadlo1}) and (\ref{vonadlo2}) and (\ref{latenight}) and (\ref{latenight1}), is an action of $D_{\alpha^{\prime}}$ on $H_{\beta\,\widetilde{\sigma}}$. This action can be calculated from $T_{D\,D\,\widetilde{D}}\,\equiv\,T_{\alpha^{\prime}\,\beta\,\widetilde{\sigma}}\,=\,0$ torsion constraint. The $AdS$ contribution in that constraint comes from $f_{\alpha^{\prime}\,\widetilde{\sigma}\,\go N}$ structure constant. We have already analysed this structure constant, see equations (\ref{duha}) and (\ref{mix1}) till (\ref{mix4}). We can thus directly write the constraint with an extra $AdS$ term:
\begin{eqnarray}
\label{k3}
T_{D\,D\,\widetilde{D}}\,\equiv\,T_{\alpha^{\prime}\,\beta\,\widetilde{\sigma}}\,=\,0&=&D_{[\alpha^{\prime}}H_{\beta\,\widetilde{\sigma})}\,+\,H_{[\alpha^{\prime}\,|\,\go M}\,\eta^{\go M \go N}\,f_{\beta\,\widetilde{\sigma}\,)\,\go N}\\
&=&D_{\alpha^{\prime}}\,H_{\beta\,\widetilde{\sigma}}\,+\,2\,(\gamma^{\bf{a}})_{\alpha^{\prime}\,\beta}\,H_{\widetilde{\sigma}\,{\bf{a}}}\,+\frac{1}{r_{AdS}}\,(\gamma^{\bf{[c}})_{\sigma\,\rho}\,(\widetilde{\Gamma}_{5})^{\rho\,\nu}\,(\gamma^{\bf{d}]})_{\nu\,\alpha^{\prime}}\,H_{\beta\,{\bf{cd}}}\nonumber
\end{eqnarray}
where we again note that the $\Sigma$ indices in the last expression of the (\ref{k3}) second line  are from the $SO(\,5\,)\,\otimes\,SO(\,5\,)$ diagonal subgroup. The $H_{\beta\,{\bf{cd}}}$ vielbein has nonzero both $H_{\beta\,{\bf{+b}}}$ and also $H_{\beta\,\widetilde{\bf{+b}}}$. The second vielbein is the term already in the matrix ${\mathbb{H}}$ from the flat section (ultimate goal is to rewrite the $AdS$ case in the terms of  matrix ${\mathbb{H}}$ and use the super-trace trick to get the pre-potential). The field $H_{\beta\,{\bf{+b}}}$ is related to the $H_{\widetilde{\rho}\,{\bf{+c}}}$ as we saw in table (\ref{hrncek3}).

There is one last piece in the equation (\ref{k3}) that we did not relate to the fields in the ${\mathbb{H}}$ matrix. The field $H_{\widetilde{\sigma}\,{\bf{a}}}$. As we saw in the flat case, that field should be related to $H_{\widetilde{\sigma}\,{\bf{+a}}}$ via $P_{-}$. We have seen however (for example in (\ref{redneck})) that such relations are a bit changed in the $AdS$ case.
Consider the following torsion constraint (and use mixed light-cone and $H_{-\,{\bf{+a}}}\,\rightsquigarrow\,0$):
\begin{eqnarray}
\label{yahoo}
T_{P\,\widetilde{D}\,S}\,\equiv\,T_{-\,\widetilde{\beta}\,{\bf{+a}}}\,=\,0\,&=&\,D_{[-}\,H_{\widetilde{\beta}\,{\bf{+a}})}\,+\,H_{[-\,|\,\go M}\,\eta^{\go M \go N}\,f_{\widetilde{\beta}\,{\bf{+a}})\go N}\\
&=&\,P_{-}\,H_{\widetilde{\beta}\,{\bf{+a}}}\,+\,H_{\widetilde{\beta}\,\go M}\,\eta^{\go M \go N}\,f_{-\,{\bf{+a}}\,\go N}\,+\,H_{{\bf{+a}}\,\go M}\,\eta^{\go M \go N}\,f_{-\,\widetilde{\beta}\,\go N}\nonumber\\
\label{twiter}
&=&\,P_{-}\,H_{\widetilde{\beta}\,{\bf{+a}}}\,+\,\eta_{-\,+}\,H_{\widetilde{\beta}\,{\bf{a}}}\,+\,\frac{1}{r_{AdS}}\,(\gamma_{-})_{\beta\,\nu}\,(\widetilde{\Gamma}_{5})^{\nu\,\sigma}\,H_{{\bf{+a}}\,\nu}
\end{eqnarray}
In the table (\ref{hrncek3}) we derived the relation between $H_{{\bf{+a}}\,\nu}$ and $H_{{\bf{+a}}\,\widetilde{\nu}}$. That result together with (\ref{twiter}) we get:
\be
\label{honduras}
{\Big{(}}P_{-}\,-\,\frac{1}{P_{\widetilde{-}}}\,\frac{1}{(r_{AdS})^{2}}{\Big{)}}\,H_{{\bf{+a}}\,\widetilde{\alpha}}\,=\,H_{{\bf{a}}\,\widetilde{\alpha}}
\ee

With the equation (\ref{honduras}) we succeeded to calculate the last missing derivative $D_{\alpha^{\prime}}\,H_{\beta\,\widetilde{\sigma}}$ in terms of ${\mathbb{H}}$ vielbeins:
\begin{eqnarray}
\label{adsYES2}
D_{\alpha^{\prime}}\,H_{\beta\,\widetilde{\sigma}}\,-\,2\,(\,P_{-}\,-\frac{1}{P_{\widetilde{-}}}\,\frac{1}{(r_{AdS})^{2}}\,)\,(\gamma^{\bf{a}})_{\alpha^{\prime}\,\beta}\,H_{{\bf{+a}}\,\widetilde{\sigma}}\,&&\\
\,-\,\frac{1}{2\,(r_{AdS})^{2}\,P_{\widetilde{-}}}\,(\widetilde{\Gamma}_{5})_{\sigma}{}^{\nu}\,(\gamma^{\bf{d}})_{\nu\,\alpha^{\prime}}\,(\widetilde{\Gamma}_{5})_{\beta}{}^{\rho}\,H_{\widetilde{\rho}\,{\bf{+d}}}\,-\,\frac{1}{2\,(r_{AdS})}\,(\widetilde{\Gamma}_{5})_{\sigma}{}^{\nu}\,(\gamma^{\bf{d}})_{\nu\,\alpha^{\prime}}\,H_{\beta\,\widetilde{\bf{+d}}}&=&0\nonumber\\
\label{adsYES3}
D_{\widetilde{\alpha}^{\prime}}\,H_{\widetilde{\beta}\,{\sigma}}\,-\,2\,(\,P_{\widetilde{-}}\,-\,\frac{1}{P_{{-}}}\,\frac{1}{(r_{AdS})^{2}}\,)\,(\gamma^{\bf{a}})_{\alpha^{\prime}\,\beta}\,H_{{\widetilde{\bf{+a}}}\,{\sigma}}\,&&\\
\,-\,\frac{1}{2\,(r_{AdS})^{2}\,P_{-}}\,(\widetilde{\Gamma}_{5})_{\sigma}{}^{\nu}\,(\gamma^{\bf{d}})_{\nu\,\alpha^{\prime}}\,(\widetilde{\Gamma}_{5})_{\beta}{}^{\rho}\,H_{{\rho}\,\widetilde{\bf{+d}}}\,-\,\frac{1}{2\,(r_{AdS})}\,(\widetilde{\Gamma}_{5})_{\sigma}{}^{\nu}\,(\gamma^{\bf{d}})_{\nu\,\alpha^{\prime}}\,H_{\widetilde{\beta}\,{\bf{+d}}}&=&0\nonumber
\end{eqnarray}
Where the $(\widetilde{\Gamma}_{5})_{\sigma}{}^{\nu}\,\defeq\,(\gamma^{+})_{\sigma\,\lambda}\,(\widetilde{\Gamma}_{5})^{\lambda\,\nu}$. The $AdS$ equations (\ref{kukura1}) and (\ref{vonadlo1}) and (\ref{vonadlo2}) and (\ref{latenight}) and (\ref{latenight1}) and (\ref{adsYES2}) and (\ref{adsYES3}) could be summarised in the following diagram:

\begin{equation}
\label{schemeADS3}
\xymatrix{
{(}\,H_{{\bf{+a}}\,\widetilde{\bf{+b}}},\,H_{\alpha\,\widetilde{\beta}}{)}   \ar@{-->}@/^2.5pc/[rr]^{D_{\widetilde{\alpha}^{\prime}}} \ar@{<--}@/^0.4pc/[rr]^{D_{\widetilde{\alpha}^{\prime}}}&&{(}H_{\beta\,\widetilde{\bf{+b}}},\,H_{{\bf{+a}\,\widetilde{\beta}}}{)} \ar@{<-}@/^2.5pc/[ll]^{D_{{\alpha}^{\prime}}} \ar@{->}@/^0.2pc/[ll]^{D_{{\alpha}^{\prime}}}  
}
\end{equation}
From the above diagram is obvious that we again have to combine $H_{{\bf{+a}}\,\widetilde{\bf{+b}}}$ and $H_{\alpha\,\widetilde{\beta}}$ and derivatives $D_{\widetilde{\alpha}^{\prime}}$ and $D_{\alpha^{\prime}}$ to get a vanishing derivative on some scalar. 
\subsection{\texorpdfstring{The ${\mathbb{H}}$ matrix in $AdS_{5}\,\times\,S^{5}$}{The H matrix in AdS\textfiveinferior{} x S\textfivesuperior{}}}

We want to repeat the chapter on the flat solution via the ${\mathbb{H}}$ matrix. The ${\mathbb{H}}$ matrix was defined in (\ref{sovicka}). We want to write the action $D_{\alpha^{\prime}}$ and $D_{\widetilde{\alpha}^{\prime}}$ on the ${\mathbb{H}}$. This was given in components in equations:  (\ref{kukura1}) and (\ref{vonadlo1}) and (\ref{vonadlo2}) and (\ref{latenight}) and (\ref{latenight1}) and (\ref{adsYES2}) and (\ref{adsYES3}) and also graphically in (\ref{schemeADS3}). We expect that the resulting matrix equations have pieces given by the flat equations (\ref{vintage1}) and (\ref{vintage2}) plus purely $AdS$ pieces (dependent as powers of $\frac{1}{r_{AdS}}$). We could write those equations in such explicit matrix form, but resulting equations are complicated and unnecessary for our purpose. We instead summarise the right hand side of $D_{\alpha^{\prime}}\,{\mathbb{H}}$ and $D_{\widetilde{\alpha}^{\prime}}\,{\mathbb{H}}$ using two new matrices ${\mathbb{X}_{\alpha^{\prime}}}$ and ${\mathbb{Y}_{\widetilde{\alpha}^{\prime}}}$ respectively. We propose matrix from of the $AdS$ equations:
\begin{eqnarray}
\label{zilla1}
D_{\alpha^{\prime}}\,{\mathbb{H}}&=&{\mathbb{X}_{\alpha^{\prime}}}\\
\label{zilla2}
D_{\widetilde{\alpha}^{\prime}}\,{\mathbb{H}}&=&{\mathbb{Y}_{\widetilde{\alpha}^{\prime}}}
\end{eqnarray}
The matrices ${\mathbb{X}_{\alpha^{\prime}}}$ and ${\mathbb{Y}_{\widetilde{\alpha}^{\prime}}}$ are fully fixed by (\ref{kukura1}) and (\ref{vonadlo1}) and (\ref{vonadlo2}) and (\ref{latenight}) and (\ref{latenight1}) and (\ref{adsYES2}) and (\ref{adsYES3}). In the $r_{AdS}\,\rightarrow\,\infty$ the ${\mathbb{X}_{\alpha^{\prime}}}\,\rightarrow\,{\mathcal{\Gamma}_{\alpha^{\prime}}}\,{\mathbb{H}}$ and ${\mathbb{Y}_{\widetilde{\alpha}^{\prime}}}\,\rightarrow\,{\mathbb{H}}\,\Gamma_{\widetilde{\alpha}^{\prime}}$, where the matrices $\Gamma_{\alpha^{\prime}}$ and $\Gamma_{\widetilde{\alpha}^{\prime}}$ are given in (\ref{vintage1}) and (\ref{vintage2}). 

\subsection{\texorpdfstring{Chiral and projective solutions for $AdS_{5}\,\times\,S^{5}$}{Chiral and projective solutions for AdS\textfiveinferior{} x S\textfivesuperior{}}}
In the next step we repeat the argument we gave in the flat case section but for the $AdS$ equations (\ref{zilla1}) and (\ref{zilla2}). We define:
\be
\label{muji22}
{\mathcal{D}_{v}}\,\equiv\,{\mathcal{D}^{v}{}_{\alpha^{\prime}}}\,\defeq\,(\,D_{\alpha^{\prime}}\,+\,{\mathcal{M}}_{\alpha^{\prime}}{}^{\beta^{\prime}}\,D_{\widetilde{\beta}^{\prime}}\,)
\ee
now we act by (\ref{muji22}) on ${\mathbb{H}}$:
\begin{eqnarray}
\label{mikro1}
{\mathcal{D}}^{v}{}_{\alpha^{\prime}}\,{\mathbb{H}}&=&(\,{\mathbb{X}}_{\alpha^{\prime}}\,+\,{\mathcal{M}}_{\alpha^{\prime}}{}^{\beta^{\prime}}\,{\mathbb{Y}}_{\widetilde{\beta}^{\prime}}\,)
\label{mikro2}
\end{eqnarray}
we multiply by $\mathbb{B}$ and apply $\mbox{Str}$:
\begin{eqnarray}
\label{mikro3}
{\mathcal{D}}^{v}{}_{\alpha^{\prime}}\,\mbox{Str}\,(\,{\mathbb{H}}\,{\mathbb{B}}\,)&=&\mbox{Str}\,{\big{(}}(\,{\mathbb{X}}_{\alpha^{\prime}}\,+\,{\mathcal{M}}_{\alpha^{\prime}}{}^{\beta^{\prime}}\,{\mathbb{Y}}_{\widetilde{\beta}^{\prime}}\,)\,{\mathbb{B}}\,{\big{)}}
\end{eqnarray}
We will further analyse the structure of (\ref{mikro3}) in next discussion but before we note one change with respect to (\ref{dv}). In (\ref{dv}) we used ${\mathbb{B}}^{-1}$ here we are using (yet to be determined) matrix ${\mathbb{B}}$, the difference is purely conventional. As in the flat case, we want to put the right hand side of (\ref{mikro3}) to zero and by that obtain vanishing ${\mathcal{D}}^{v}{}_{\alpha^{\prime}}$ on some scalar field $\mbox{Str}\,(\,{\mathbb{H}}\,{\mathbb{B}}\,)$, that will be called pre-potential. In the flat space it was crucial that we had the identity (\ref{iden}). It was used in the relation (\ref{act}). Similarly in the $AdS$ case the identity (\ref{iden}) will also be crucial. 

In the solution of the vanishing (\ref{mikro3}) right hand side we still want to maintain the $SO\,(\,4\,)\,\otimes\,SO\,(\,4\,)$ invariance. Therefore the  $\mathbb{B}$ matrix has a block-diagonal form:
\begin{eqnarray}
\label{bee2}
{\mathbb{B}}\defeq
\begin{pmatrix}
b_{{\bf{+a}}\,{\bf{+b}}}&0\\
0&b_{\alpha\,\beta}\\
\end{pmatrix}
\end{eqnarray}
Let us also simplify the notation for the constants appearing in the equations (\ref{kukura1}) and (\ref{adsYES2}) and similarly for their left-right conjugates. In (\ref{kukura1}) we redefine:
\begin{eqnarray}
\label{constantsX}
X_{1}\,\defeq\,-\,\frac{1}{g}\,\frac{1}{(r_{AdS})\,P_{\widetilde{-}}}\,\,\,&||&\,\,\,X_{2}\,\defeq\,\frac{1}{2\,g}\,(\,1\,-\,\frac{1}{f}\,\frac{1}{(r_{AdS})^{2}\,P_{-}\,P_{\widetilde{-}}}\,)\\
X_{3}\,\defeq\,+\,\frac{1}{f\,g}\,\frac{1}{2\,(r_{AdS})^{3}\,P_{-}\,(P_{\widetilde{-}})^{2}}&||&\nonumber
\end{eqnarray}
where the $f$ and $g$ were defined in (\ref{kukura2}). In (\ref{adsYES2}) we define:
\begin{eqnarray}
\label{constantsY}
Y_{1}\,\defeq\,-\,2\,{\Big{(}}\,P_{\widetilde{-}}\,-\,\frac{1}{P_{-}}\,\frac{1}{(r_{AdS})^{2}}\,{\Big{)}}\,\,\,||\,\,\,Y_{2}\,\defeq\,-\,\frac{1}{2\,(r_{AdS})^{2}\,P_{-}}\,\,\,||\,\,\,Y_{3}\,\defeq\,-\,\frac{1}{2\,(r_{AdS})}
\end{eqnarray}
With the definitions (\ref{bee2}), (\ref{constantsX}) and (\ref{constantsY}) let us rewrite the right hand side of (\ref{mikro3}) explicitly:
\begin{eqnarray}
\label{eu1}
0&=&(\,X_{1}\,-\,X_{3}\,)\,(\widetilde{\Gamma}_{5})^{\beta\,\sigma}\,b_{{\bf{+a}}\,{\bf{+b}}}\,(\gamma^{{\bf{b}}})_{\sigma\,\alpha^{\prime}}\,-\,{\mathcal{M}}_{\alpha^{\prime}}{}^{\sigma^{\prime}}\,X_{2}\,b_{{\bf{+a}}\,{\bf{+b}}}\,(\gamma^{{\bf{b}}})^{\beta}{}_{\sigma^{\prime}}\\
&&+\,Y_{1}\,b^{\beta\,\sigma}\,(\gamma_{{\bf{a}}})_{\sigma\,\alpha^{\prime}}\,+\,Y_{2}\,b^{\nu\,\sigma}\,(\widetilde{\Gamma}_{5})_{\sigma}{}^{\lambda}\,(\gamma_{{\bf{a}}})_{\lambda\,\alpha^{\prime}}\,(\widetilde{\Gamma}_{5})_{\nu}{}^{\beta}\,+\,{\mathcal{M}}_{\alpha^{\prime}}{}^{\sigma^{\prime}}\,Y_{3}\,b^{\beta\,\sigma}\,(\widetilde{\Gamma}_{5})_{\sigma}{}^{\lambda}\,(\gamma_{{\bf{a}}})_{\lambda\,\sigma^{\prime}}\nonumber\\
\label{eu2}
0&=&{\mathcal{M}}_{\alpha^{\prime}}{}^{\nu^{\prime}}\,(\,\tilde{X}_{1}\,-\,\tilde{X}_{3}\,)\,(\widetilde{\Gamma}_{5})^{\beta\,\sigma}\,b_{{\bf{+a}}\,{\bf{+b}}}\,(\gamma^{{\bf{b}}})_{\sigma\,\nu^{\prime}}\,-\,X_{2}\,b_{{\bf{+a}}\,{\bf{+b}}}\,(\gamma^{{\bf{b}}})^{\beta}{}_{\alpha^{\prime}}\\
&&+\,{\mathcal{M}}_{\alpha^{\prime}}{}^{\nu^{\prime}}\,\tilde{Y}_{1}\,b^{\beta\,\sigma}\,(\gamma_{{\bf{a}}})_{\sigma\,\nu^{\prime}}\,+\,{\mathcal{M}}_{\alpha^{\prime}}{}^{\nu^{\prime}}\,\tilde{Y}_{2}\,b^{\nu\,\sigma}\,(\widetilde{\Gamma}_{5})_{\sigma}{}^{\lambda}\,(\gamma_{{\bf{a}}})_{\lambda\,\nu^{\prime}}\,(\widetilde{\Gamma}_{5})_{\nu}{}^{\beta}\,+\,{Y}_{3}\,b^{\beta\,\sigma}\,(\widetilde{\Gamma}_{5})_{\sigma}{}^{\lambda}\,(\gamma_{{\bf{a}}})_{\lambda\,\alpha^{\prime}}\nonumber
\end{eqnarray}
where $\tilde{X}_{1}$, $\tilde{X_3}$ and $\tilde{Y}_{1}$, $\tilde{Y}_{2}$ are left-right conjugates of the constants defined in (\ref{constantsX}) and (\ref{constantsY}) and $X_{2}$ and $Y_3$ are the same after left-right swap.

The equations (\ref{eu1}) and (\ref{eu2}) are the $AdS$ analogies of the flat space equations (\ref{house}). To solve them we first multiply the equation (\ref{eu2}) by matrix ${\mathcal{M}}_{\alpha^{\prime}}{}^{\beta^{\prime}}$. Thus we get the equation (\ref{eu2}) into the form: 
\begin{eqnarray}
\label{eu3}
0&=&{\mathcal{M}}^{2}{}_{\alpha^{\prime}}{}^{\nu^{\prime}}\,(\,\tilde{X}_{1}\,-\,\tilde{X}_{3}\,)\,(\widetilde{\Gamma}_{5})^{\beta\,\sigma}\,b_{{\bf{+a}}\,{\bf{+b}}}\,(\gamma^{{\bf{b}}})_{\sigma\,\nu^{\prime}}\,-\,{\mathcal{M}}_{\alpha^{\prime}}{}^{\nu^{\prime}}\,X_{2}\,b_{{\bf{+a}}\,{\bf{+b}}}\,(\gamma^{{\bf{b}}})^{\beta}{}_{\nu^{\prime}}\\
&&+\,{\mathcal{M}}^{2}{}_{\alpha^{\prime}}{}^{\nu^{\prime}}\,\tilde{Y}_{1}\,b^{\beta\,\sigma}\,(\gamma_{{\bf{a}}})_{\sigma\,\nu^{\prime}}\,+\,{\mathcal{M}}^{2}{}_{\alpha^{\prime}}{}^{\nu^{\prime}}\,\tilde{Y}_{2}\,b^{\nu\,\sigma}\,(\widetilde{\Gamma}_{5})_{\sigma}{}^{\lambda}\,(\gamma_{{\bf{a}}})_{\lambda\,\nu^{\prime}}\,(\widetilde{\Gamma}_{5})_{\nu}{}^{\beta}\nonumber\\
&&+\,{\mathcal{M}}_{\alpha^{\prime}}{}^{\nu^{\prime}}\,{Y}_{3}\,b^{\beta\,\sigma}\,(\widetilde{\Gamma}_{5})_{\sigma}{}^{\lambda}\,(\gamma_{{\bf{a}}})_{\lambda\,\nu^{\prime}}\nonumber
\end{eqnarray}
The equation (\ref{eu3}) is almost identical to the (\ref{eu1}) except of the left-right swapped constants and ${\mathcal{M}}^{2}$ matrix. By suitable choice of the ${\mathcal{M}}$ matrix we can turn (\ref{eu3}) into (\ref{eu1}) and thus reduce number of equations by half. By that we get the condition on the matrix ${\mathcal{M}}$:
\begin{eqnarray}
\label{Mmatrix}
{\mathcal{M}}^{2}{}_{\alpha^{\prime}}{}^{\beta^{\prime}}\,=\,q^{2}\,\delta_{\alpha^{\prime}}{}^{\beta^{\prime}}
\end{eqnarray}
where the constant $q^{2}\,=\,\frac{P_{-}}{P_{\widetilde{-}}}$. By that choice of the matrix ${\mathcal{M}}^{2}$ and constant $q^{2}$ we turn equation (\ref{eu3}) into (\ref{eu1}). Furthermore we should solve relation (\ref{Mmatrix}) for the matrix ${\mathcal{M}}$. As in the whole $AdS$ section we ask for the $SO\,(\,4\,)\,\otimes\,SO\,(\,4\,)$ invariance. With that requirement we get two branches for the ${\mathcal{M}}$ matrix (actually we get four, as we will see, but the $\pm$ is not very important to us):
\begin{eqnarray}
\label{goose}
({\mathcal{M}}^2)_{\alpha^{\prime}}{}^{\beta^{\prime}}\,=\,\frac{P_{{-}}}{P_{\widetilde{-}}}\,\delta_{\alpha^{\prime}}{}^{\beta^{\prime}}\,\,\,\,\Rightarrow\,\,\,\,{\mathcal{M}}_{\alpha^{\prime}}{}^{\beta^{\prime}}\,=\,\pm\,\sqrt{\frac{P_{{-}}}{P_{\widetilde{-}}}}\,\begin{cases} 
       \delta_{\alpha^{\prime}}{}^{\beta^{\prime}}&   \\
        (\widetilde{\Gamma}_{5})_{\alpha^{\prime}}{}^{\beta^{\prime}} &
  \end{cases}
\end{eqnarray}
We first notice few nice properties of (\ref{goose}). The solution is actually the same as in the flat case, see (\ref{solsol}). We are in the $AdS$ space but the matrix ${\mathcal{M}}$ that combines $D_{\alpha^{\prime}}$ and $D_{\widetilde{\alpha}^{\prime}}$ does not depend on the $r_{AdS}$. Unfortunately we got the same not very nice square root factor in (\ref{goose}). We would need to find some way to deal with it. 

Having solved one half of equations (\ref{eu1}) and (\ref{eu2}).  We solve the second half, that is just relation (\ref{eu1}):
\begin{eqnarray}
\label{ginnipig}
0&=&(\,X_{1}\,-\,X_{3}\,)\,(\widetilde{\Gamma}_{5})^{\beta\,\sigma}\,b_{{\bf{+a}}\,{\bf{+b}}}\,(\gamma^{{\bf{b}}})_{\sigma\,\alpha^{\prime}}\,-\,{\mathcal{M}}_{\alpha^{\prime}}{}^{\sigma^{\prime}}\,X_{2}\,b_{{\bf{+a}}\,{\bf{+b}}}\,(\gamma^{{\bf{b}}})^{\beta}{}_{\sigma^{\prime}}\\
&&+\,Y_{1}\,b^{\beta\,\sigma}\,(\gamma_{{\bf{a}}})_{\sigma\,\alpha^{\prime}}\,+\,Y_{2}\,b^{\nu\,\sigma}\,(\widetilde{\Gamma}_{5})_{\sigma}{}^{\lambda}\,(\gamma_{{\bf{a}}})_{\lambda\,\alpha^{\prime}}\,(\widetilde{\Gamma}_{5})_{\nu}{}^{\beta}\,+\,{\mathcal{M}}_{\alpha^{\prime}}{}^{\sigma^{\prime}}\,Y_{3}\,b^{\beta\,\sigma}\,(\widetilde{\Gamma}_{5})_{\sigma}{}^{\lambda}\,(\gamma_{{\bf{a}}})_{\lambda\,\sigma^{\prime}}\nonumber
\end{eqnarray}
The claim is that given solution ${\mathcal{M}}$ the block matrices $b_{\bf{+a}\,{\bf{+b}}}$ and $b_{\alpha\,\beta}$ are fixed (up to the overall constant). We will again use the same identity (\ref{iden}) as in the flat case. We expect the solutions (we have two branches) will be certain $r_{AdS}$ dependent deformation of the original flat space solutions. We also require to maintain the $SO(\,4\,)\,\otimes\,SO\,(\,4\,)$ invariance of the solution so the most general ansatz for the equation (\ref{ginnipig}) is:
\begin{eqnarray}
\label{whale}
b_{{\bf{+a}}\,{\bf{+b}}}\,\defeq\,{A}\,\delta_{{\bf{a}}\,{\bf{b}}}\,+\,B\,(\widetilde{\Gamma}_{5})_{{\bf{a}}\,{\bf{b}}}\,\,\,\,||\,\,\,\,b_{\alpha\,\beta}\,\defeq\,C\,\delta_{\alpha\,\beta}\,+\,D\,(\widetilde{\Gamma}_{5})_{\alpha\,\beta}
\end{eqnarray} 

Because of later importance we will first solve the $(\widetilde{\Gamma}_{5})$ branch of the ${\mathcal{M}}$ solution (\ref{goose}). Later we will also provide solution for the $\delta$ branch of the (\ref{goose}). We plug ${\mathcal{M}}$ and (\ref{whale}) into (\ref{ginnipig}) and solve for $A,\,\,B,\,\,C\,\,\mbox{and}\,\,D$ using the identity (\ref{iden}), we remind that $q\,\defeq\,\pm\,\sqrt{\frac{P_{{-}}}{P_{\widetilde{-}}}}$. We get the following solutions: 
\begin{eqnarray}
\label{pidgeon}
\mbox{the $(\widetilde{\Gamma}_{5})_{\alpha^{\prime}}{}^{\beta^{\prime}}$ branch:}&&\\
det\,\defeq\,{\big{(}}\,(\,X_{1}\,-\,X_{3}\,)^{2}\,-\,(\,q\,X_{2}\,)^2\,{\big{)}}&||&\nonumber\\
A\,=\,\frac{D}{det}\,{\big{(}}\,(\,X_{1}\,-\,X_{3}\,)\,(\,Y_{1}\,+Y_{2}\,)\,+\,q^{2}\,X_{2}\,Y_{3}\,{\big{)}}&||&B\,=\,q\,\frac{D}{det}\,{\big{(}}\,X_{2}\,(\,Y_{1}\,+\,Y_{2}\,)\,+\,(X_{1}\,-\,X_{3})\,Y_{3}\,{\big{)}}\nonumber\\
C\,=\,0&||&\,\nonumber\\
\label{chicken}
\mbox{the $\delta_{\alpha^{\prime}}{}^{\beta^{\prime}}$ branch:}&&\\
det\,\defeq\,{\big{(}}\,(\,Y_{1}\,+\,Y_{2}\,)^{2}\,-\,(\,q\,Y_{3}\,)^{2}\,{\big{)}}&||&\nonumber\\
&||&B\,=\,0\nonumber\\
C\,=\,q\,\frac{A}{det}\,{\big{(}}\,(\,Y_{1}\,+\,Y_{2}\,)\,X_{2}\,+\,(\,X_{1}\,-\,X_{3}\,)\,Y_{3}\,{\big{)}}&||&D\,\defeq\,-\,\frac{A}{det}\,{\big{(}}\,(Y_{1}\,+\,Y_{2}\,)\,(\,X_{1}\,-\,X_{3}\,)\,+\,q^{2}\,X_{2}\,Y_{3}\,{\big{)}}\nonumber
\end{eqnarray}
We can again see that as we do $r_{AdS}\,\rightarrow\,\infty$ limit in (\ref{pidgeon}) we will get the flat solution (\ref{solsol}), keeping the $D$ ( or $A$ in $\delta$ branch ) $r_{AdS}$ independent in that limit. 
\subsection{\texorpdfstring{Near horizon limit}{Near horizon limit}}
In the previous section we found the structure of the linearised pre-potential (\ref{pidgeon}) and (\ref{chicken}) and also the construction of ${\mathcal{D}^{v}{}_{\alpha^{\prime}}}$ that vanishes on the pre-potential (\ref{muji22}) and (\ref{goose}). We will now introduce the complementary derivative ${\mathcal{D}}^{w}{}_{\alpha^{\prime}}$ that is constructed after picking the ${\mathcal{D}^{v}{}_{\alpha^{\prime}}}$ derivative (i.e. picking the matrix ${\mathcal{M}}$ in (\ref{goose})) and changing the sign in front of the ${\mathcal{M}}$ (the second linearly independent combination). Thus we have:
\begin{eqnarray}
\label{snake}
{\mathcal{D}}^{w}{}_{\alpha^{\prime}}\,\defeq\,D_{\alpha^{\prime}}\,-\,{\mathcal{M}}_{\alpha^{\prime}}{}^{\beta^{\prime}}\,D_{\widetilde{\beta}^{\prime}}
\end{eqnarray}
The notation for the upper indices $v$ and $w$ in ((\ref{muji22}) and (\ref{snake})) comes from equivalent notation for $d_{v}$ and $d_{w}$ derivatives used in \cite{sconf}, (and also for $d_{u}$ and $d_{\bar{u}}$, whose analogies are to be defined later). In analogy with the paper \cite{sconf} we want to define the ${\mathcal{P}}_{+}$ operator that has ${\mathcal{D}}^{v}{}_{\alpha^{\prime}}$ and ${\mathcal{D}}^{w}{}_{\alpha^{\prime}}$ as eigenvectors with nonzero eigenvalues. We can solve for ${\mathcal{P}}_{+}$ in full generality, i.e. keeping the non-local square root factors in derivatives ${\mathcal{D}}^{v}{}_{\alpha^{\prime}}$ and ${\mathcal{D}}^{w}{}_{\alpha^{\prime}}$. This would introduce the non-local square root factors also into the definition of ${\mathcal{P}}_{+}$ and would cause further problems. What we will do instead is to restrict the coordinate dependence of the pre-potential $V$ to be just the $PSU\,(2,\,2\,|\,4)$. This is the same algebra we wanted to use at the beginning of this project, but we were forced to extend it to the full $SO\,(10)\,\otimes\,SO\,(10)$ T-dually extended super-algebra. Now, we want to restrict just the coordinate dependence of the pre-potential. Doing so the $P_{-}\,=\,P_{\widetilde{-}}$ {\bf{on}} pre-potential, not everywhere. That is enough to get rid of the non-local factors in ${\mathcal{D}}^{v}{}_{\alpha^{\prime}}$ and ${\mathcal{D}}^{w}{}_{\alpha^{\prime}}$ as they act on pre-potential. Then we can redefine (\ref{muji22}) and (\ref{snake}) by saying that the new square root free ${\mathcal{D}}^{v}$ and ${\mathcal{D}}^{w}$ to be our new definitions. With this it is easy to see that the good definition of ${\mathcal{P}}_{+}$ is:
\begin{eqnarray}
\label{elephant}
{\mathcal{P}}_{+}\,\defeq\,\frac{1}{2}\,(\,P_{+}\,+\,P_{\widetilde{+}}\,)\,=\,P_{+}
\end{eqnarray}
where the last equality holds on pre-potential.

Following the definitions in \cite{sconf} of the $AdS$ boundary limit we propose that any operator ${\mathcal{K}}$ which is an eigenvector of ${\mathcal{P}}_{+}$ operator, i.e. $[\,{\mathcal{P}}_{+},\,{\mathcal{K}}\,]\,=\,{\bf{c}}\,{\mathcal{K}}$, scales as $R^{\bf{c}}$ as we approach the boundary, i.e. $R\,\rightarrow\,0$ limit, where $R$ is a radial coordinate on the Poincar\'e patch. Another way how to state the limit is that by putting the $R\,\rightarrow\,0$ we contract the isometry groups $SO\,(\,4,\,1\,)$ and $SO\,(\,4,\,1\,)$ to $ISO\,(\,3,1\,)$ and $ISO\,(\,3,1\,)$ (we Wick rotated the $S^{5}$ isometry group for the purpose of this limit). For more details on this limit (that can be stated also through the explicit coordinates on $AdS_{5}$ and $S^{5}$) see notes \cite{sconf}.

Using the previous definitions of the $AdS$ boundary limit we can analyse the different branches of the ${\mathcal{D}}^{v}$ solutions (\ref{goose}). Let's first pick the $\delta_{\alpha^{\prime}}{}^{\beta^{\prime}}$ branch (let's work with both $\pm$ sub-branches at once). Note that even on pre-potential the $D_{\alpha^{\prime}}\,\neq\,D_{\widetilde{\alpha}^{\prime}}$ as can be seen from the explicit construction of those derivatives in \cite{warren1} in the section ${\bf{5}}$. Then the commutator is: 
\begin{eqnarray}
\label{human}
[\,{\mathcal{P}}_{+},\,D_{\alpha^{\prime}}\,+\,D_{\widetilde{\alpha}^{\prime}}]\,=\,\pm\,\frac{1}{r_{AdS}}\,(\gamma_{+})_{\alpha^{\prime}\,\beta^{\prime}}\,(\widetilde{\Gamma}_{5})^{\beta^{\prime}\,\sigma^{\prime}}\,(D_{\sigma^{\prime}}\,\pm\,D_{\widetilde{\sigma}^{\prime}})\,+\,\dots
\end{eqnarray}
The $\dots$ part correspond to the current that vanishes in the super-gravity limit (i.e. we do not see string parameter $\sigma$) and on pre-potential. We also used the commutators from (\ref{algebra}) and the mixed $AdS$ commutators from (\ref{kurniksopa}). We also used the explicit solution for the $PSU\,(\,2,\,2\,|\,4\,)$ (we are on pre-potential) derivatives in terms of $\tau$ and $\sigma$ currents, see section ${\bf{5}}$ in \cite{warren1}. More specifically we used that $D_{\Omega}\,\equiv\,D^{\alpha^{\prime}}\,=\,\omega^{\alpha^{\prime}}\,+\,\frac{1}{2}\,\frac{1}{r_{AdS}}\,(\widetilde{\Gamma}_{5})^{\alpha^{\prime}\,\beta^{\prime}}\,D_{\widetilde{\beta}^{\prime}}$, where the $\omega^{\alpha^{\prime}}$ is the current proportional to $\sigma$ derivative and it has to vanish in the super-gravity limit. The equation (\ref{human}) is very interesting. It tells us how the ${\mathcal{D}}^{v}$ scales for the $\pm$ $\delta_{\alpha^{\prime}}{}^{\beta^{\prime}}$ branch of (\ref{goose}). We also notice that the scaling constant is $r_{AdS}$ dependent and vanishes for $r_{AdS}\,\rightarrow\,\infty$. More importantly because of the $(\widetilde{\Gamma}_{5})$ for fixed $r_{AdS}$ and for fixed sub-branch of $\pm\,\delta_{\alpha^{\prime}}{}^{\beta^{\prime}}$ the scaling constant ${\bf{c}}$ is either $+\,(\frac{1}{r_{AdS}})$ for one half of $SO\,(8)$ chiral index $\alpha^{\prime}$ or $-\,(\frac{1}{r_{AdS}})$ for second half. And this is not good because by \cite{sconf} the ${\mathcal{D}}^{v}$ derivative should scale like $\frac{1}{R}$ and ${\mathcal{D}}^{w}$ should scale like $R$ (put $r_{AdS}\,=\,1$ for simplicity). In (\ref{human}) we can see that just $\frac{1}{2}$ of derivatives scale properly. This boundary limit then distinguishes between two branches of (\ref{goose}). In the following we will see that the $(\widetilde{\Gamma}_{5})$ branch has exactly right scaling properties so it corresponds to the right solution. Without this boundary limit we did not have a way how to pick a branch in (\ref{goose}). In the case of $(\widetilde{\Gamma}_{5})$ branch we have one more $(\widetilde{\Gamma}_{5})$ matrix in (\ref{human}) thus we get:
\begin{eqnarray}
\label{eel}
[\,{\mathcal{P}}_{+},\,D_{\alpha^{\prime}}\,\pm\,(\widetilde{\Gamma}_{5})_{\alpha^{\prime}}{}^{\rho^{\prime}}\,D_{\widetilde{\rho}^{\prime}}]\,=\,\pm\,\frac{1}{r_{AdS}}\,(D_{\alpha^{\prime}}\,\pm\,(\widetilde{\Gamma}_{5})_{\alpha^{\prime}}{}^{\rho^{\prime}}\,D_{\widetilde{\rho}^{\prime}})\,+\,\dots
\end{eqnarray}
The equation (\ref{eel}) will give us the correct solution. From (\ref{eel}) we can see that for fixed $r_{AdS}$ and for fixed $(\widetilde{\Gamma}_{5})$ sub-branch we will have proper scaling for full $SO\,(\,8\,)$ chiral index $\alpha^{\prime}$. Because we require ${\mathcal{D}}^{v}$ to scale like $\frac{1}{R}$ and  ${\mathcal{D}}^{w}$ scale like $R$ we have ${\mathcal{D}}^{v}{}_{\alpha^{\prime}}\,=\,(D_{\alpha^{\prime}}\,-\,(\widetilde{\Gamma}_{5})_{\alpha^{\prime}}{}^{\rho^{\prime}}\,D_{\widetilde{\rho}^{\prime}})$ and ${\mathcal{D}}^{w}{}_{\alpha^{\prime}}\,=\,(D_{\alpha^{\prime}}\,+\,(\widetilde{\Gamma}_{5})_{\alpha^{\prime}}{}^{\rho^{\prime}}\,D_{\widetilde{\rho}^{\prime}})$. The positive news is that the blowing-up derivative ${\mathcal{D}}^{v}$ is zero on the pre-potential by our construction, so there is no possible singularity arising as we approach the boundary.

Its easy to see how the derivatives $D_{\alpha}$ and $D_{\widetilde{\alpha}}$ scale. Because the $(\gamma_{+})_{\alpha\,\beta}\,=\,0$ the $[\,{\mathcal{P}}_{+},\,D_{\alpha}\,]\,=\,[\,{\mathcal{P}}_{+},\,D_{\widetilde{\alpha}}\,]\,=\,0$. So they scale like $1$. Those derivatives are building up the ${\mathcal{D}}^{u}$ and ${\mathcal{D}}^{\bar{u}}$, analogous derivatives to paper \cite{sconf} derivatives $d_{u}$ and $d_{\bar{u}}$. The explicit forms of ${\mathcal{D}}^{u}$ and ${\mathcal{D}}^{\bar{u}}$  won't be needed in this paper so we do not provide them.

\subsection{\texorpdfstring{Near horizon limit and field equations}{Near horizon limit and field equations}}
Comparing result with \cite{sconf} we want to see that the field equations for the pre-potential in the near horizon limit (i.e. in the $R\,\rightarrow\,0$) is just of the form ${\mathcal{P}}_{+}\,V\,=\,0\,+\,{\mathcal{O}}(\,R\,)$. This will be our final confirmation that we discovered the right pre-potential. We first notice that the Lorentz generator scales like ${\mathcal{O}}(\,1\,)$, this can be seen from commutator $[\,S_{\bf{+a}},\,{\mathcal{P}}_{+}\,]\,=\,[\,S_{\widetilde{\bf{+a}}},\,{\mathcal{P}}_{+}\,]\,=\,0$. To see what is ${\mathcal{P}}_{+}$ on pre-potential we could directly use some appropriate torsions (remember pre-potential is a linear combination of fields). We found it easier however to use a different approach. Let's look at the torsion constraint (\ref{fnukadlo}) but for the $\alpha$ index instead of $\alpha^{\prime}$ (the $\alpha$ index is one of the $SO\,(8)$ chiral indices):
\begin{eqnarray}
\label{deer}
T_{D\,S\,\widetilde{S}}\,\equiv\,T_{\alpha\,{\bf{+a}}\,\widetilde{\bf{+b}}}\,=\,0&=&D_{[\alpha}\,H_{{\bf{+a}}\,\widetilde{\bf{+b}})}\,+\,H_{[\alpha\,|\,\go M}\,\eta^{\go M \go N}\,f_{\,{\bf{+a}}\,\widetilde{\bf{+b}})\,\go N}\\
\label{gianteye}
&=&D_{\alpha}\,H_{{\bf{+a}}\,\widetilde{\bf{+b}}}\,+\,S_{\widetilde{+b}}\,H_{\alpha\,{\bf{+a}}}\,+\,S_{\bf{+a}}H_{\widetilde{\bf{+b}}\,\alpha} \label{pecul}
\end{eqnarray}
First notice that the structure of (\ref{gianteye}) is very different than the structure of (\ref{fnukadlo}). There is no $f$ term in (\ref{gianteye}) and there is the full derivative term present. Even in the $AdS$ case the $f$ term is missing. This can be seen as follows. The $f_{\bf{+a}\,\widetilde{\bf{+b}}\,\go N}\,=\,0$ in $AdS$ and also in flat case and also $f_{\alpha\,\widetilde{\bf{+b}}\,\go N}\,=\,0$. The only possibly nonzero $f$ term is coming from $f_{\alpha\,{\bf{+a}}\,\go N}$. The $H_{\widetilde{\bf{+b}}\,\go M}\,\eta^{\go M \go N}\,f_{\alpha\,{\bf{+a}}\,\go N}\,\propto\,(\gamma_{\bf{+a}})_{\alpha}{}^{\nu^{\prime}}\,H_{\nu^{\prime}\,\widetilde{\bf{+b}}}$. The vielbein $H_{\nu^{\prime}\,\widetilde{\bf{+b}}}$ is zero (also in the $AdS$) as was shown in the analysis under (\ref{fnukadlo}). 
Next, we can recognise the term $H_{\alpha\,\widetilde{\bf{+a}}}$ as a part of ${\mathbb{H}}$ matrix (\ref{hmat}). The vielbein $H_{\alpha\,{\bf{+a}}}$ has also been analysed in table (\ref{hrncek3}). It is related to $H_{{\bf{+a}}\,\widetilde{\alpha}}$ , see table (\ref{hrncek3}). We need to be more careful with that relation because in (\ref{pecul}) we again discover the $S$ derivative peculiarity, we saw earlier. 

In general all fields in ${\mathbb{H}}$ (now better viewed as their irreducible pieces) could be obtained from the pre-potential $V$ by an action of appropriate (irreducible) combination of ${\mathcal{D}}^{w}$ on the pre-potential. One could analyse in full detail what is the exact structure of those pieces and reproduce famous field content of $AdS_{5}\,\times\,S^{5}$ supergravity first discovered in \cite{pvn} and later used in \cite{rast}. This would lead us away from this paper real aim, so we postpone this analysis to next paper. The aim of this section is to show that on pre-potential $V$ the operator ${\mathcal{P}}_{+}$ vanishes in the near horizon limit. 

For this reason we notice following expansions: 
\begin{eqnarray}
\label{genushark}
H_{{\bf{+a}}\,\widetilde{\bf{+b}}}&=&{\bf{c}}_{0}\,V\,+\,{\bf{c}}_{2}\,({\mathcal{D}}^{w})^{2}\,V\,+\,{\bf{c}}_{4}\,({\mathcal{D}}^{w})^{4}\,V\,+\,{\bf{c}}_{6}\,({\mathcal{D}}^{w})^{6}\,V\,+\,{\bf{c}}_{8}\,({\mathcal{D}}^{w})^{8}\,V \label{uno} \\
H_{{\bf{+a}}\,\widetilde{\alpha}} &=&{\bf{d}}_{1}\,{\mathcal{D}}^{w}\,V\,+\,{\bf{d}}_{3}\,({\mathcal{D}}^{w})^{3}\,V\,+\,{\bf{d}}_{5}\,({\mathcal{D}}^{w})^{5}\,V\,+\,{\bf{d}}_{7}\,({\mathcal{D}}^{w})^{7}\,V \label{duo}\label{genushark1}\\
H_{\widetilde{\bf{+a}}\,\alpha} &=&{\bf{e}}_{1}\,{\mathcal{D}}^{w}\,V\,+\,{\bf{e}}_{3}\,({\mathcal{D}}^{w})^{3}\,V\,+\,{\bf{e}}_{5}\,({\mathcal{D}}^{w})^{5}\,V\,+\,{\bf{e}}_{7}\,({\mathcal{D}}^{w})^{7}\,V \label{duo}\label{genushark2}
\end{eqnarray}
where factors ${\bf{c}_{0}},\,{\bf{c}_{2}}\,\dots,\,{\bf{d}_{1}},\,{\bf{d}_{3}}\,\dots$ and ${\bf{e}_{1}},\,{\bf{e}_{3}}\,\dots$  are constant factors with appropriate index structure. Note that the ${\bf{c}_{0}}$ is non-zero. The important observation is that for each term in (\ref{genushark}), (\ref{genushark1}) and (\ref{genushark2}) we know how it scales in the $R\,\rightarrow\,0$ limit, because we know that ${\mathcal{D}}^{w}$ scales like $R$.

Next, we want to combine (\ref{gianteye}) with known scalings of all (\ref{gianteye}) objects to get an information how $D_{\alpha}\,V$ scales. On one hand it should scale like ${\mathcal{O}}\,(1)$ on the other hand the relation (\ref{gianteye}) relates it to different fields. What we obtain is a nontrivial relation that $D_{\alpha}\,V\,=\,{\mathcal{O}}\,(\,R\,)$ as we go to the boundary. It just means that $D_{\alpha}\,V\,=\,0$ (and so also $D_{\widetilde{\alpha}}\,V\,=\,0$) in the near horizon limit. Because of the anti-commutator $\{\,D_{\alpha},\,D_{\beta}\}\,=\,2\,(\gamma_{-})_{\alpha\,\beta}\,P_{+}$. This is enough to see that $P_{+}\,V\,\equiv\,{\mathcal{P}}_{+}\,V\,=\,0$ in the near horizon limit. There are two crucial steps. One is to relate the (\ref{pecul}) term $S_{\widetilde{{\bf{+b}}}}\,H_{\alpha\,{\bf{+a}}}$ to $H_{{\bf{+a}}\,\widetilde{\alpha}}$. This is relatively straightforward using table (\ref{hrncek3}) and explicit $S_{\widetilde{{\bf{+b}}}}$ derivative. Second step is to plug expansions (\ref{genushark}), (\ref{genushark1}) and (\ref{genushark2}) and the scalings of particular pieces into (\ref{gianteye}). Doing that we get the following: 
\begin{eqnarray}
\label{fieldwreck}
0&=&D_{\alpha}\,H_{{\bf{+a}}\,\widetilde{\bf{+b}}}\,+\,S_{\widetilde{\bf{+b}}}\,H_{\alpha\,{\bf{+a}}}\,+\,S_{\bf{+a}}\,H_{\widetilde{\bf{+b}}\,\alpha}\\
&=&D_{\alpha}\,H_{\bf{+a}\,\widetilde{\bf{+b}}}\,+\,S_{\widetilde{\bf{+b}}}\,{\Big{(}}\,-\,f_{\widetilde{-}\,\alpha\,\go M}\,\eta^{\go M \go N}\,\frac{1}{P_{\widetilde{-}}}\,H_{{\bf{+a}}\,\go N}\,{\Big{)}}\,+\,S_{\bf{+a}}\,H_{\widetilde{{\bf{+b}}}\,\alpha}\nonumber\\
\label{eagle}
&=&D_{\alpha}\,H_{\bf{+a}\,\widetilde{\bf{+b}}}\,-\,\frac{{\bf{c}}}{r_{AdS}}\,(\gamma_{-})_{\alpha\,\nu}\,(\,\widetilde{\Gamma}_{5}\,)^{\nu\,\sigma}\,S_{\widetilde{{\bf{+b}}}}\,\frac{1}{P_{\widetilde{-}}}\,H_{{\bf{+a}}\,\widetilde{\sigma}}\,+\,S_{\bf{+a}}\,H_{\widetilde{{\bf{+b}}}\,\alpha}\\
&=&D_{\alpha}\,{\Big{(}}\,{\bf{c}}_{0}\,V\,+\,{\bf{c}}_{2}\,({\mathcal{D}}^{w})^{2}\,V\,+\,\dots\,{\Big{)}}\\
&-&\,\frac{1}{r_{AdS}}\,(\gamma_{-})_{\alpha\,\nu}\,(\,\widetilde{\Gamma}_{5}\,)^{\nu\,\sigma}\,S_{\widetilde{{\bf{+b}}}}\,\frac{1}{P_{\widetilde{-}}}\,{\Big{(}}\,{\bf{d}}_{1}\,{\mathcal{D}}^{w}\,V\,+\,{\bf{d}}_{3}\,({\mathcal{D}}^{w})^{3}\,V\,+\,\dots\,{\Big{)}}\nonumber\\
&+&S_{\bf{+a}}\,{\Big{(}}\,{\bf{e}}_{1}\,{\mathcal{D}}^{w}\,V\,+\,{\bf{e}}_{3}\,({\mathcal{D}}^{w})^{3}\,V\,+\,\dots\,{\Big{)}}\,\nonumber
\end{eqnarray}

The equation (\ref{eagle}) contains all the right expressions to establish the near horizon limit. By the discussion below (\ref{eel}) the ${\mathcal{D}}^{w}{}_{\alpha^{\prime}}$ derivative scales like ${\mathcal{O}}\,(\,R\,)$ (for the projective branch), we also have the scaling of $S_{\bf{+a}}$ and $S_{\widetilde{\bf{+b}}}$ that goes like a constant. Applying that knowledge we get the equation (\ref{eagle}) in the near horizon limit:
\begin{eqnarray}
\label{capricorn}
0&=&{\bf{c}}_{0}\,D_{\alpha}\,V\,+\,{\mathcal{O}}\,(R)\,
\end{eqnarray}
The ${\bf{c}_{0}}$ is nonzero constant (tensor) so it follows that $D_{\alpha}\,V\,=\,0$ at the $AdS$ boundary. From $\{\,D_{\alpha},\,D_{\beta}\,\}\,=\,(\gamma_{-})_{\alpha\,\beta}\,P_{+}$ we get the field equation for the pre-potential in the near horizon limit: 
\begin{eqnarray}
\label{elk}
0&=&P_{+}V\,+\,{\mathcal{O}}\,(R)\\
&\equiv&{\mathcal{P}}_{+}V\,+\,{\mathcal{O}}\,(R)
\end{eqnarray} 

\section{\texorpdfstring{Conclusion}{Conclusion}}
We outline results we have obtained: starting from the $10$ dimensional IIB string theory. We embedded the $AdS_{5}\,\times\,S^{5}$ background and expanded the theory around this background (we also considered a flat background, i.e. $AdS_{5}\,\times\,S^{5}$ with $r_{AdS}\,\rightarrow\,\infty$ ). Our aim was to obtain (linearised) pre-potential with desired properties in the case of $AdS_{5}\,\times\,S^{5}$ (also in the flat case). We succeeded and obtained pre-potential construction for flat and $AdS_{5}\,\times\,S^{5}$ background. We derived only the linearised form, but the vielbein construction makes non-linearisation straightforward perturbation. The pre-potential (in flat and also in $AdS_{5}\,\times\,S^{5}$, the projective and chiral) sits in the combination (without further derivatives) of vielbeins $H_{S\,\widetilde{S}}$ and $H_{D\,\widetilde{D}}$. By construction the $D^{v}$ derivative vanishes in bulk on the pre-potential and the (projective) pre-potential satisfies the near horizon limit field equation ${\mathcal{P}}_{+}\,V\,=\,0\,+\,{\mathcal{O}}\,(R)$ together with vanishing of ${\mathcal{D}}^{u}$ and ${\mathcal{D}}^{\bar{u}}$ on pre-potential in the near horizon limit. This near horizon limit picks out the projective pre-potential instead of chiral pre-potential (both were obtained as valid bulk solutions). 

The vanishing of ${\mathcal{P}}_{+}$ at the boundary fixes the difference between the conformal weights ($\equiv\,\Delta$) and $U(1)$ charges ($\equiv\,\Delta_{Y}$) of all boundary BPS operators, since ${\mathcal{P}}_{+}\,\propto\,\Delta\,-\,\Delta_{Y}$. The ${\mathcal{P}}_{-}\,\propto\,\Delta\,+\,\Delta_{Y}$ and known expansion of ${\mathbb{H}}$ in powers of ${\mathcal{D}}^{w}$ from $V$, fixes the conformal weights and the $U(1)$ charges for the boundary BPS operators, the relations important in the $AdS/CFT$ correspondence, see \cite{malda}, \cite{wittenE}.

\section*{\texorpdfstring{Acknowledgment}{Acknowledgment}}
This work was supported in part by National Science Foundation Grant No. PHY-1316617.

\newpage
\appendix
\numberwithin{equation}{section}

\section{\texorpdfstring{$AdS_{5}\,\times\,S^{5}$ structure of some vielbeins and their derivatives}{AdS\textfiveinferior{} x S\textfivesuperior{} structure of some vielbeins and their derivatives}}
\subsection{\texorpdfstring{The $H_{S\,\widetilde{S}}$ }{The H S S }}
\label{vareska1}
Using equations (\ref{torzor}), (\ref{torzor1}), (\ref{posero}) and the mixed light-cone gauge together with keeping the mixed structure constants and evaluating the explicit actions of the $S$ and $\widetilde{S}$ derivatives we derived the first important result for the structure of the $H_{\go S {\widetilde{\go S}}}$ vielbein (in the $AdS$ case). Note that by the symbol $\rightsquigarrow$ in the in the whole text we denoted the evaluation of the mixed structure constants in the sense described in chapter (\ref{torconI}):
\begin{table}[H]
{\centering
\begin{tabular}{|c|}
\noalign{\vspace{.1in}}
\hline
$H_{\bf{-a}\,\widetilde{\bf{-b}}}\,=\,-\,f_{-\,\widetilde{-}\,{\go M}}\,\eta^{\go M \go N}\,\frac{1}{P_{\widetilde{-}}}\,S_{-a}\,(\,\frac{1}{P_{{-}}}\,H_{\widetilde{-b}\,\go N}\,)\,\rightsquigarrow\,0$\\
\hline
$H_{\bf{-a}\,\widetilde{\bf{+b}}}\,=\,-\,f_{-\,\widetilde{-}\,{\go M}}\,\eta^{\go M \go N}\,\frac{1}{P_{{-}}}\,S_{\widetilde{\bf{+b}}}\,(\,\frac{1}{P_{\widetilde{-}}}\,H_{{\bf{-a}}\,\go N}\,)\,-\,f_{-\,\widetilde{\bf{b}}\,\go M}\,\eta^{\go M \go N}\,\frac{1}{P_{{-}}\,P_{\widetilde{-}}}\,H_{{\bf{-a}}\,\go N}\,$\\
$\rightsquigarrow\,-\,\frac{1}{2\,(r_{AdS})^{2}}\,\frac{1}{P_{{-}}\,P_{\widetilde{-}}}\,H_{{\bf{-a}}\widetilde{\bf{+b}}}\,\Rightarrow\,H_{\bf{-a}\,\widetilde{\bf{+b}}}\,=\,0$\\
\hline
$H_{\bf{+a}\,\widetilde{\bf{+b}}}\,=\,-\,f_{-\,\widetilde{-}\,{\go M}}\,\eta^{\go M \go N}\,\frac{1}{P_{{-}}}\,S_{\widetilde{\bf{+b}}}\,(\,\frac{1}{P_{\widetilde{-}}}\,H_{{\bf{+a}}\,\go N}\,)\,-\,f_{-\,\widetilde{\bf{b}}\,\go M}\,\eta^{\go M \go N}\,\frac{1}{P_{{-}}\,P_{\widetilde{-}}}\,H_{{\bf{+a}}\,\go N}\,+\,\frac{1}{P_{-}}\,H_{\widetilde{\bf{+b}}\,{\bf{a}}}\,$\\
$\rightsquigarrow\,-\,\frac{1}{2\,(r_{AdS})^{2}}\,\frac{1}{P_{{-}}\,P_{\widetilde{-}}}\,H_{{\bf{+a}}\widetilde{\bf{+b}}}\,+\,\frac{1}{P_{-}}\,H_{\widetilde{\bf{+b}}\,{\bf{a}}}\,\Rightarrow\,(\,1\,+\,\frac{1}{2\,(r_{AdS})^{2}}\,\frac{1}{P_{{-}}\,P_{\widetilde{-}}}\,)\,H_{{\bf{+a}}\widetilde{\bf{+b}}}\,=\,\frac{1}{P_{-}}\,H_{{\bf{a}}\,\widetilde{\bf{+b}}}$\\
\hline
$H_{{\bf{-a}}\,\widetilde{\bf{bc}}}\,=\,-\,f_{{\bf{-}}\,\widetilde{\bf{-}}\,\go M}\,\eta^{\go M \go N}\,\frac{1}{P_{-}}\,S_{\widetilde{\bf{bc}}}\,(\,\frac{1}{P_{\widetilde{-}}}\,H_{{\bf{-a}}\,\go N}\,)\,\rightsquigarrow\,0$\\
\hline
$H_{{\bf{+a}}\,\widetilde{\bf{bc}}}\,=\,f_{\bf{a}\,\widetilde{-}\,{\go M}}\,\eta^{\go M \go N}\,\frac{1}{P_{\widetilde{-}}\,P_{-}}\,H_{\widetilde{\bf{bc}}\,\go N}\,-\,f_{-\,\widetilde{-}\,\go M}\,\eta^{\go M \go N}\,\frac{1}{P_{\widetilde{-}}}\,S_{\bf{+a}}\,(\,\frac{1}{P_{-}}\,H_{\widetilde{\bf{bd}}\,\go N}\,)\,$\\
$\rightsquigarrow\,-\,\frac{1}{2\,(r_{AdS})^{2}}\,\frac{1}{P_{\widetilde{-}}\,P_{-}}\,H_{\widetilde{\bf{bc}}\,{\bf{+a}}}\Rightarrow\,H_{{\bf{+a}}\,\widetilde{\bf{bc}}}\,=\,0$\\
\hline
$H_{{\bf{ab}}\,\widetilde{\bf{cd}}}\,=\,-\,f_{{\bf{-}}\,\widetilde{\bf{-}}\,\go M}\,\eta^{\go M \go N}\,\frac{1}{P_{-}}\,S_{\widetilde{\bf{cd}}}\,(\,\frac{1}{P_{\widetilde{-}}}\,H_{{\bf{ab}}\,\go N}\,)\,\rightsquigarrow\,0$\\
\hline
$H_{{\bf{+-}}\,\widetilde{\bf{+-}}}\,=\,f_{{\bf{-}}\,\widetilde{\bf{-}}\,\go M}\,\eta^{\go M \go N}\,\frac{1}{P_{\widetilde{-}}\,P_{-}}\,H_{\widetilde{\bf{+-}}\,\go N}\,-\,f_{{\bf{-}}\,\widetilde{\bf{-}}\,\go M}\,\eta^{\go M \go N}\,\frac{1}{P_{\widetilde{-}}}\,S_{\bf{+-}}\,(\,\frac{1}{P_{{-}}}\,H_{\widetilde{+-}\,\go N}\,)\,\rightsquigarrow\,0$\\
\hline
$H_{{\bf{+-}}\,\widetilde{\bf{-a}}}\,=\,f_{{\bf{-}}\,\widetilde{\bf{-}}\,\go M}\,\eta^{\go M \go N}\,\frac{1}{P_{\widetilde{-}}\,P_{-}}\,H_{\widetilde{\bf{-a}}\,\go N}\,-\,f_{{\bf{-}}\,\widetilde{\bf{-}}\,\go M}\,\eta^{\go M \go N}\,\frac{1}{P_{\widetilde{-}}}\,S_{\bf{+-}}\,(\,\frac{1}{P_{{-}}}\,H_{\widetilde{\bf{-a}}\,\go N}\,)\,\rightsquigarrow\,0$\\
\hline
$H_{{\bf{+-}}\,\widetilde{\bf{+a}}}\,=\,f_{-\,\widetilde{\bf{a}}\,\go M}\,\eta^{\go M \go N}\,\frac{1}{P_{-}\,P_{\widetilde{-}}}\,H_{{\bf{+-}}\,\go N}\,-\,f_{-\,{\widetilde{-}}\,\go M}\,\eta^{\go M \go N}\,\frac{1}{P_{-}}\,S_{\widetilde{\bf{+a}}}\,(\,\frac{1}{P_{\widetilde{-}}}\,H_{{\bf{+-}}\,\go N}\,)$\\
$\rightsquigarrow\,\frac{1}{2\,(r_{AdS})^{2}}\,\frac{1}{P_{-}\,P_{\widetilde{-}}}\,H_{\bf{+-}\,\widetilde{\bf{+a}}}\,\Rightarrow\,H_{{\bf{+-}}\,\widetilde{\bf{+a}}}\,=\,0$\\
\hline
$H_{{\bf{+-}}\,\widetilde{\bf{ab}}}\,=\,f_{-\,\widetilde{-}\,\go M}\,\eta^{\go M \go N}\,\frac{1}{P_{\widetilde{\bf{-}}}\,P_{-}}\,H_{\widetilde{\bf{ab}}\,\go N}\,-\,f_{-\,\widetilde{-}\,\go M}\,\eta^{\go M \go N}\,\frac{1}{P_{\widetilde{-}}}\,S_{\bf{+-}}\,(\,\frac{1}{P_{-}}\,H_{\widetilde{\bf{ab}}\,\go N}\,)\,\rightsquigarrow\,0$\\
\hline
\end{tabular}
\caption{$H_{S\,\widetilde{S}}$ vielbein}
\protect\label{hrncek1}
}
\end{table}
In the table (\ref{hrncek1}) (and after the evaluation of mixed structure constants) we have heavily used the structure of the mixed structure constant $f_{{\bf{a}}\,{\widetilde{\bf{b}}}\,\go M}$ that is analysed in the main text, see analysis before equation (\ref{sedlak}). Moreover we used one more torsion constrain to fix $H_{P\,S}$ and $H_{\widetilde{P}\,\widetilde{S}}$ in the table (\ref{hrncek1}). Let's take an example $H_{P\,S}\,=H_{{\bf{a}}\,{\bf{bc}}}$. To fix that vielbein we consider $T_{\widetilde{-}\,{\bf{a}}\,{\bf{bc}}}\,=\,0$:
\begin{eqnarray}
\label{homedepo}
T_{\widetilde{P}\,P\,S}\,\equiv\,T_{\widetilde{-}\,{\bf{a}}\,{\bf{bc}}}&=&0\\
&=&P_{\widetilde{-}}\,H_{\bf{a}\,{\bf{bc}}}\,+\,S_{{\bf{bc}}}\,H_{\widetilde{-}\,\bf{a}}\,+\,P_{\bf{a}}\,H_{{\bf{bc}}\,\widetilde{\bf{-}}}\,+\,H_{{\bf{bc}}\,\go M}\,\eta^{\go M \go N}\,f_{\widetilde{\bf{-}}\,\bf{a}\,\go N}\nonumber\\
&=&P_{\widetilde{-}}\,H_{\bf{a}\,{\bf{bc}}}\,+\,f_{\widetilde{\bf{-}}\,\bf{a}\,\go M}\,\eta^{\go M \go N}\,H_{{\bf{bc}}\,\go N}\nonumber\\
&\Rightarrow&H_{\bf{a}\,{\bf{bc}}}\,=\,-\,f_{\widetilde{\bf{-}}\,\bf{a}\,\go M}\,\eta^{\go M \go N}\,\frac{1}{P_{\widetilde{-}}}\,H_{{\bf{bc}}\,\go N}\nonumber
\end{eqnarray}


\subsection{\texorpdfstring{The $H_{S\,\widetilde{D}}$  and $H_{\widetilde{S}\,D}$ }{The H S D}}

In the section (\ref{vareska2}) we analysed vielbein $H_{\alpha^{\prime}\,{\bf{+b}}}$. By the similar set of equations as in the section (\ref{vareska2}) we can fix $H_{\alpha^{\prime}\,\widetilde{\bf{-b}}}$. We summarise the structure of the fixed vielbeins from the section (\ref{vareska2}) discussion in the following table:
\begin{table}[H]
{\centering
\begin{tabular}{|c|}
\noalign{\vspace{.1in}}
\hline
$H_{\alpha^{\prime}\,{\bf{-a}}}\,=\,f_{\widetilde{-}\,\alpha^{\prime}\,\go M}\,\eta^{\go M \go N}\,\frac{1}{P_{\widetilde{-}}}\,H_{{\bf{-a}}\,\go N}\rightsquigarrow\,0$\\
\hline
$H_{\alpha^{\prime}\,{\bf{+a}}}\,=\,f_{\widetilde{-}\,\alpha^{\prime}\,\go M}\,\eta^{\go M \go N}\,\frac{1}{P_{\widetilde{-}}}\,H_{{\bf{+a}}\,\go N}\rightsquigarrow\,0$\\
\hline
$H_{\alpha^{\prime}\,{\widetilde{\bf{-a}}}}\,=\,-\,f_{-\,{\widetilde{-}}\,\go M}\eta^{\go M \go N}\,\frac{1}{P_{-}}\,S_{\widetilde{\bf{-a}}}\,\frac{1}{P_{\widetilde{-}}}\,H_{\alpha^{\prime}\,\go N}\,\rightsquigarrow\,0$\\
\hline
$H_{\alpha^{\prime}\,{\widetilde{\bf{+a}}}}\,=\,-\,S_{\widetilde{\bf{+a}}}\,(\,f_{-\,\widetilde{-}\,\go M}\,\eta^{\go M \go N}\,\frac{1}{P_{-}\,P_{\widetilde{-}}}\,H_{\alpha^{\prime}\,\go N}\,)\,\rightsquigarrow\,0$\\
\hline
\end{tabular}
\caption{$H_{D\,\widetilde{S}}$ vielbein}
\protect\label{hrncek2}
}
\end{table}
Similarly we can calculate what is the table (\ref{hrncek2}) with $\alpha^{\prime}$ swapped with $\alpha$. We will use the analogous analysis as in section (\ref{vareska2}) except sometimes instead of the equation (\ref{hena}) we use $T_{\widetilde{P}\,D\,\widetilde{S}}$ and also we fix the $H_{\widetilde{P}\,\widetilde{S}}$ using $T_{P\,\widetilde{P}\,\widetilde{S}}$ (or some $\mbox{left}-\mbox{right}$ swap of those). Let's look at two such examples and calculate what is $H_{\alpha\,{\bf{-a}}}$ and $H_{\widetilde{\alpha}\,{\bf{-a}}}$ respectively (we also use the mixed light-cone gauge): 
\begin{eqnarray}
T_{\widetilde{P}\,D\,{S}}\,\equiv\,T_{\widetilde{-}\,\alpha\,{\bf{-a}}}\,=\,0\,&=&\,P_{[\widetilde{-}}\,H_{\alpha\,{\bf{-a}}\,)}\,+\,H_{[\,\widetilde{-}\,|\,\go M}\,\eta^{\go M \go N}\,f_{\alpha\,{\bf{-a}}\,) \go N}\\
\label{miklos}
&=&P_{\widetilde{-}}\,H_{\alpha\,{\bf{-a}}}\,+\,H_{{\bf{-a}}\,\go M}\,\eta^{\go M \go N}\,f_{\alpha\,\widetilde{-}\,\go N}\\
\Rightarrow&&H_{\alpha\,{\bf{-a}}}\,=\,-\,f_{\alpha\,\widetilde{-}\,\go M}\,\eta^{\go M \go N}\,\frac{1}{P_{\widetilde{-}}}\,H_{{\bf{-a}}\,\go N}\,\rightsquigarrow\,(\gamma_{-})_{\alpha\,\nu}\,(\widetilde{\Gamma}_{5})^{\nu\,\sigma}\,\frac{1}{(r_{AdS})\,P_{\widetilde{-}}}\,H_{{\bf{-a}}\,\widetilde{\sigma}}\nonumber
\end{eqnarray}
Next, examine:
\begin{eqnarray}
\label{sakra}
T_{P\,S\,\widetilde{D}}\,\equiv\,T_{-\,{\bf{-a}}\,\widetilde{\alpha}}\,=\,0&=&P_{[-}\,H_{{\bf{-a}}\,\widetilde{\alpha}\,)}\,+\,H_{[-\,|\,\go M}\,\eta^{\go M \go N}\,f_{{\bf{-a}}\,\widetilde{\alpha}\,)\,\go N}\\
\label{vlhka}
&=&P_{-}\,H_{{\bf{-a}}\,\widetilde{\alpha}}\,+\,D_{\widetilde{\alpha}}\,H_{-\,{\bf{-a}}}\,+\,H_{\bf{-a}\,\go M}\,\eta^{\go M \go N}\,f_{\widetilde{\alpha}\,-\,\go N}
\end{eqnarray}
the $H_{-\,{\bf{-a}}}$ is fixed by $T_{\widetilde{P}\,P\,S}\,\equiv\,T_{\widetilde{-}\,-\,{\bf{-a}}}\,=\,0$:
\begin{eqnarray}
T_{\widetilde{P}\,P\,S}\,\equiv\,T_{\widetilde{-}\,-\,{\bf{-a}}}\,=\,0\,&=&P_{[\widetilde{-}}\,H_{-\,{\bf{-a}})}\,+\,H_{[\widetilde{-}\,|\,\go M}\,\eta^{\go M \go N}\,f_{-\,{\bf{-a}})\,\go N}\\
&=&P_{\widetilde{-}}\,H_{-\,{\bf{-a}}}\,+\,H_{{\bf{-a}}\,\go M}\,\eta^{\go M \go N}\,f_{\widetilde{-}\,-\,\go N}\\
\label{ina}
\Rightarrow&&H_{-\,{\bf{-a}}}\,=\,f_{-\,\widetilde{-}\,\go M}\,\eta^{\go M \go N}\,\frac{1}{P_{\widetilde{-}}}\,H_{{\bf{-a}}\,\go N}\,\rightsquigarrow\,0
\end{eqnarray}
plugging (\ref{ina}) into the (\ref{vlhka}) we get:
\begin{eqnarray}
H_{{\bf{-a}}\,\widetilde{\alpha}}&=&-\,f_{-\,\widetilde{-}\,\go M}\,\eta^{\go M \go N}\,\frac{1}{P_{-}}\,D_{\widetilde{\alpha}}\,\frac{1}{P_{\widetilde{-}}}\,H_{\bf{-a}\,\go N}\,-\,f_{-\,\widetilde{\alpha}\,\go M}\,\eta^{\go M \go N}\,\frac{1}{P_{-}}\,H_{{\bf{-a}}\,\go N}\\
\label{dzurinda}
\Rightarrow&&H_{{\bf{-a}}\,\widetilde{\alpha}}\rightsquigarrow\,-\,(\gamma_{-})_{\alpha\,\nu}\,(\widetilde{\Gamma}_{5})^{\nu\,\sigma}\,\frac{1}{(r_{AdS})\,P_{-}}\,H_{{\bf{-a}}\,{\sigma}}
\end{eqnarray}
We notice that combining the result (\ref{dzurinda}) with (\ref{miklos}) we get after the evaluation of the mixed structure constants that $H_{{\bf{-a}}\,\alpha}\,\rightsquigarrow\,0$ and so also $H_{{\bf{-a}}\,\widetilde{\alpha}}\,\rightsquigarrow\,0$. Similar analysis can be made for the rest of the vielbeins (we mean those from table (\ref{hrncek2}), except $\alpha^{\prime}$ switched with $\alpha$). Thus we get the following table (\ref{hrncek3}):
\begin{table}[H]
{\centering
\begin{tabular}{|c|}
\noalign{\vspace{.1in}}
\hline
$H_{\alpha\,{\bf{-a}}}\,=\,-\,f_{\alpha\,\widetilde{-}\,\go M}\,\eta^{\go M \go N}\,\frac{1}{P_{\widetilde{-}}}\,H_{{\bf{-a}}\,\go N}$\\
$\Rightarrow\,H_{\alpha\,{\bf{-a}}}\,\rightsquigarrow\,0$\\
\hline
$H_{\alpha\,{\bf{+a}}}\,=\,f_{\widetilde{-}\,\alpha\,\go M}\,\eta^{\go M \go N}\,\frac{1}{P_{\widetilde{-}}}\,H_{{\bf{+a}}\,\go N}\,\rightsquigarrow\,(\gamma_{-})_{\alpha\,\nu}\,(\widetilde{\Gamma}_{5})^{\nu\,\sigma}\,\frac{1}{(r_{AdS})\,P_{\widetilde{-}}}\,H_{{\bf{+a}}\,\widetilde{\sigma}}$\\
\hline
$H_{\widetilde{\alpha}\,{\bf{-a}}}\,=\,f_{-\,\widetilde{-}\,\go M}\,\eta^{\go M \go N}\,\frac{1}{P_{-}}\,D_{\widetilde{\alpha}}\,\frac{1}{P_{\widetilde{-}}}\,H_{\bf{-a}\,\go N}\,+\,f_{-\,\widetilde{\alpha}\,\go M}\,\eta^{\go M \go N}\,\frac{1}{P_{-}}\,H_{{\bf{-a}}\,\go N}$\\
$\Rightarrow\,H_{\widetilde{\alpha}\,{\bf{-a}}}\,\rightsquigarrow\,0$\\
\hline
$H_{\alpha\,\widetilde{\bf{+a}}}\,=\,-\,f_{-\,\widetilde{-}\,\go M}\,\eta^{\go M \go N}\,\frac{1}{P_{\widetilde{-}}}\,D_{\alpha}\,\frac{1}{P_{-}}\,H_{\widetilde{\bf{+a}}\,\go N}\,+\,f_{\widetilde{-}\,\alpha\,\go M}\,\eta^{\go M \go N}\,\frac{1}{P_{\widetilde{-}}}\,H_{\widetilde{\bf{+a}}\,\go N}\,-\,\eta_{\widetilde{-}\,\widetilde{+}}\,\frac{1}{P_{\widetilde{-}}}\,H_{\alpha\,\widetilde{\bf{a}}}$\\
$\Rightarrow\,H_{\alpha\,\widetilde{\bf{+a}}}\,\rightsquigarrow\,(\gamma_{-})_{\alpha\,\nu}\,(\widetilde{\Gamma}_{5})^{\nu\,\sigma}\,\frac{1}{(r_{AdS})\,P_{\widetilde{-}}}\,H_{\widetilde{\bf{+a}}\,\widetilde{\sigma}}\,+\,\frac{1}{P_{\widetilde{-}}}\,H_{\alpha\,\widetilde{\bf{a}}}$\\
\hline
\end{tabular}
\caption{$H_{D\,\widetilde{S}}$ vielbein}
\protect\label{hrncek3}
}
\end{table}
Let us repeat our goal. We wanted to determine the actions of $S_{\widetilde{\bf{+b}}}$ and $S_{\bf{+a}}$ on $H_{\alpha^{\prime}\,{\bf{+a}}}$ and $H_{\widetilde{\bf{+b}}\,\alpha^{\prime}}$ respectively. We wanted to do that because then the (\ref{fulltime}) gives the action of $D_{\alpha^{\prime}}$ on $H_{{\bf{+a}}\,\widetilde{\bf{+b}}}$ (where at least the part of the pre-potential sits). The action of $S_{\widetilde{\bf{+b}}}$ on $H_{\alpha^{\prime}\,{\bf{+a}}}$ is easily computed using our table (\ref{hrncek2}). Taking the second top relation from the table (\ref{hrncek2}) and by explicitly applying the $S_{\widetilde{\bf{+b}}}$ derivative we get:
\begin{eqnarray}
\label{att}
S_{\widetilde{\bf{+b}}}\,H_{\alpha^{\prime}\,{\bf{+a}}}\,&=&S_{\widetilde{\bf{+b}}}\,(\,f_{\widetilde{-}\,\alpha^{\prime}\,\go M}\,\eta^{\go M \go N}\,\frac{1}{P_{\widetilde{-}}}\,H_{{\bf{+a}}\,\go N})\\
&=&\eta_{\widetilde{-}\,\widetilde{+}}\,f_{\widetilde{\bf{b}}\,\alpha^{\prime}\,\go M}\,\eta^{\go M \go N}\,\frac{1}{P_{\widetilde{-}}}\,H_{{\bf{+a}}\,\go N}\,+\,f_{\widetilde{-}\,\alpha^{\prime}\,\go M}\,\eta^{\go M \go N}\,S_{\widetilde{\bf{+b}}}\,(\,\frac{1}{P_{\widetilde{-}}}\,H_{{\bf{+a}}\,\go N}\,)\\
\label{verizon}
\Rightarrow&&S_{\widetilde{\bf{+b}}}\,H_{\alpha^{\prime}\,{\bf{+a}}}\,\rightsquigarrow\,-\,(\gamma_{\bf{b}})_{\alpha^{\prime}\,\nu}\,(\widetilde{\Gamma}_{5})^{\nu\,\sigma}\,\frac{1}{(r_{AdS})\,P_{\widetilde{-}}}\,H_{{\bf{+a}}\,\widetilde{\sigma}}
\end{eqnarray}
To evaluate $S_{\bf{+a}}\,H_{\widetilde{\bf{+b}}\,\alpha^{\prime}}$ we need to work a bit more. One can directly use the last relation in the table (\ref{hrncek2}). We found an easier way however. For that we need an alternative fixing of the vielbein $H_{\widetilde{\bf{+b}}\,\alpha^{\prime}}$. This alternative fixing seems to be more suited for an explicit evaluation of the $S_{\bf{+a}}$ action (and $S_{\bf{-a}}$ action). An alternative way how to fix $H_{\alpha^{\prime}\,\widetilde{\bf{+a}}}$ is to use $T_{\widetilde{P}\,D\,\widetilde{S}}\,\equiv\,T_{\widetilde{-}\,\alpha^{\prime}\,\widetilde{\bf{+a}}}$ instead of one that we used in (\ref{hena}) and (\ref{puzdro}). Similarly it will be useful to find an alternative fixing for $H_{\alpha^{\prime}\,\widetilde{\bf{-a}}}$. Again that could be done by considering torsion $T_{\widetilde{P}\,D\,\widetilde{S}}\,\equiv\,T_{\widetilde{-}\,\alpha^{\prime}\,\widetilde{\bf{-a}}}$. Let's look at this alternative fixing more closely:
\begin{eqnarray}
\label{krivan}
T_{\widetilde{P}\,D\,\widetilde{S}}\,\equiv\,T_{\widetilde{-}\,\alpha^{\prime}\,\widetilde{\bf{-a}}}\,=\,0\,&=&P_{[\widetilde{-}}\,H_{\alpha^{\prime}\,\widetilde{\bf{-a}}\,)}\,+\,H_{[\widetilde{-}\,|\,\go M}\,\eta^{\go M \go N}\,f_{\alpha^{\prime}\,\widetilde{\bf{-a}}\,)\,\go N}\\
\label{krivan1}
&=&P_{\widetilde{-}}\,H_{\alpha^{\prime}\,\widetilde{\bf{-a}}}\,+\,D_{\alpha^{\prime}}\,H_{\widetilde{\bf{-a}}\,\widetilde{-}}\,+\,H_{\widetilde{\bf{-a}}\,\go M}\,\eta^{\go M \go N}\,f_{\widetilde{-}\,\alpha^{\prime}\,\go N}
\end{eqnarray}
The $H_{\widetilde{\bf{-a}}\,\widetilde{-}}$ type of vielbein has been fixed in (\ref{ina}). Plugging the fixing into (\ref{krivan1}) we get an alternative $H_{\alpha^{\prime}\,\widetilde{\bf{-a}}}$ fixing:
\begin{eqnarray}
\label{zvlhnutaTrava}
H_{\alpha^{\prime}\,\widetilde{\bf{-a}}}&=&-\,f_{-\,\widetilde{-}\,\go M}\,\eta^{\go M \go N}\,\frac{1}{P_{\widetilde{-}}}\,\,D_{\alpha^{\prime}}\,\frac{1}{P_{-}}\,H_{\widetilde{\bf{-a}}\,\go N}\,-\,f_{\widetilde{-}\,\alpha^{\prime}\,\go M}\,\eta^{\go M \go N}\,\frac{1}{P_{\widetilde{-}}}\,H_{\widetilde{\bf{-a}}\,\go N}\\
\label{zvlhnutaTrava1}
H_{\alpha^{\prime}\,\widetilde{\bf{-a}}}&\rightsquigarrow&\,0
\end{eqnarray}
again we can see the behaviour of the $H_{\alpha^{\prime}\,\widetilde{\bf{-a}}}$ in (\ref{zvlhnutaTrava1}) as we evaluate the theory, as it should be comparing with its behaviour from the fixing in the table (\ref{hrncek2}). The alternative fixing for the vielbein $H_{\widetilde{\bf{+b}}\,\alpha^{\prime}}$ is calculated similarly: 
\begin{eqnarray}
\label{albertina}
T_{\widetilde{P}\,D\,\widetilde{S}}\,\equiv\,T_{\widetilde{-}\,\alpha^{\prime}\,\widetilde{\bf{+a}}}\,=\,0\,&=&
P_{[\widetilde{-}}\,H_{\alpha^{\prime}\,\widetilde{\bf{+a}}\,)}\,+\,H_{[\widetilde{-}\,|\,\go M}\,\eta^{\go M \go N}\,f_{\alpha^{\prime}\,\widetilde{\bf{+a}}\,)\,\go N}\\
\label{albertina1}
&=&P_{\widetilde{-}}\,H_{\alpha^{\prime}\,\widetilde{\bf{+a}}}\,+\,D_{\alpha^{\prime}}\,H_{\widetilde{\bf{+a}}\,\widetilde{-}}\,+\,H_{\widetilde{\bf{+a}}\,\go M}\,\eta^{\go M \go N}\,f_{\widetilde{-}\,\alpha^{\prime}\,\go N}\,+\,H_{\alpha^{\prime}\,\widetilde{\bf{a}}}
\end{eqnarray}
The $H_{\widetilde{\bf{+a}}\,\widetilde{-}}$ is fixed similarly to (\ref{ina}) resulting in:
\begin{eqnarray}
\label{fishtail}
H_{\widetilde{\bf{+a}}\,\widetilde{-}}\,=\,f_{-\,\widetilde{-}\,\go M}\,\eta^{\go M \go N}\,\frac{1}{P_{-}}\,H_{\widetilde{\bf{+a}}\,\go N}\,\rightsquigarrow\,0
\end{eqnarray}
The $H_{\alpha^{\prime}\,\widetilde{\bf{a}}}$ is fixed by the dim $\frac{1}{2}$ torsion constraint $T_{{P}\,D\,\widetilde{P}}\equiv\,T_{{-}\,\alpha^{\prime}\,\widetilde{\bf{a}}}\,=\,0$:
\begin{eqnarray}
T_{{P}\,D\,\widetilde{P}}\equiv\,T_{{-}\,\alpha^{\prime}\,\widetilde{\bf{a}}}\,=\,0&=&P_{[{-}}\,H_{\alpha^{\prime}\,\widetilde{\bf{a}}\,)}\,+\,H_{[{-}\,|\,\go M}\,\eta^{\go M \go N}\,f_{\alpha^{\prime}\,\widetilde{\bf{a}}\,)\,\go N}\\
\label{chamurapi}
&=&P_{{-}}\,H_{\alpha^{\prime}\,\widetilde{\bf{a}}}\,+\,P_{\widetilde{\bf{a}}}\,H_{-\,\alpha^{\prime}}\,+\,H_{\alpha^{\prime}\,\go M}\,\eta^{\go M \go N}\,f_{\widetilde{\bf{a}}\,-\,\go N}
\end{eqnarray}
The last vielbein we need to fix is the $H_{-\,\alpha^{\prime}}$, that is again fixed by the dim $\frac{1}{2}$ torsion constraint $T_{\widetilde{P}\,P\,D}\,\equiv\,T_{\widetilde{-}\,-\,\alpha^{\prime}}\,=\,0$:
\begin{eqnarray}
T_{\widetilde{P}\,P\,D}\,\equiv\,T_{\widetilde{-}\,-\,\alpha^{\prime}}\,=\,0&=&P_{[\widetilde{-}}\,H_{-\,\alpha^{\prime}\,)}\,+\,H_{[\widetilde{-}\,|\,\go M}\,\eta^{\go M \go N}\,f_{-\,\alpha^{\prime}\,)\,\go N}\\
&=&P_{\widetilde{-}}\,H_{-\,\alpha^{\prime}}\,+\,H_{\alpha^{\prime}\,\go M}\,\eta^{\go M \go N}\,f_{\widetilde{-}\,-\,\go N}\\
\label{zvlhnutie}
\Rightarrow&&H_{-\,\alpha^{\prime}}\,=\,f_{-\,\widetilde{-}\,\go M}\,\eta^{\go M \go N}\,\frac{1}{P_{\widetilde{-}}}\,H_{\alpha^{\prime}\,\go N}\,\rightsquigarrow\,0
\end{eqnarray}
Plugging (\ref{zvlhnutie}) into (\ref{chamurapi}) and that into (\ref{albertina1}) we finally get an alternative fixing for the $H_{\alpha^{\prime}\,\widetilde{+a}}$:
\begin{eqnarray}
\label{snehulienka}
H_{\alpha^{\prime}\,\widetilde{\bf{+a}}}&=&-\,f_{-\,\widetilde{-}\,\go M}\,\eta^{\go M \go N}\,{\Big{(}}\,\frac{1}{P_{\widetilde{-}}\,P_{-}}\,P_{\widetilde{a}}\,\frac{1}{P_{\widetilde{-}}}\,H_{\alpha^{\prime}\,\go N}\,+\,\frac{1}{P_{\widetilde{-}}}\,D_{\alpha^{\prime}}\,\frac{1}{P_{-}}\,H_{\widetilde{\bf{+a}}\,\go N}\,{\Big{)}}\,-\,f_{\alpha^{\prime}\,\widetilde{-}\,\go M}\,\eta^{\go M \go N}\,\frac{1}{P_{\widetilde{-}}}\,H_{\widetilde{\bf{+a}}\,\go N}\nonumber\\
&&-\,f_{-\,\widetilde{\bf{a}}\,\go M}\,\eta^{\go M \go N}\,\frac{1}{P_{\widetilde{-}}\,P_{-}}\,H_{\alpha^{\prime}\,\go N}
\end{eqnarray}
Now we are ready to calculate an action of $S_{\widetilde{\bf{-a}}}$ and $S_{\bf{-a}}$ and $S_{\widetilde{\bf{+a}}}$ and $S_{\bf{+a}}$ on table (\ref{hrncek2}) vielbeins (with the exception of $S_{\bf{+a}}\,H_{\widetilde{\bf{+b}}\,\alpha^{\prime}}$ that we want to calculate in the end of this paragraph). 
We summarise those $S$ (and $\widetilde{S}$) actions in the next tables:
\begin{table}[H]
{\centering
\begin{tabular}{|c|}
\noalign{\vspace{.1in}}
\hline
$S_{{\bf{-b}}}\,H_{\alpha^{\prime}\,{\bf{-a}}}\,=\,-\,(\gamma_{{\bf{-b}}})_{\alpha^{\prime}}{}^{\nu}\,f_{\widetilde{-}\,\nu\,\go M}\,\eta^{\go M \go N}\,\frac{1}{P_{\widetilde{-}}}\,H_{{\bf{-a}}\,\go N}\,-\,f_{\widetilde{-}\,\alpha^{\prime}\,\go M}\,\eta^{\go M \go N}\,S_{\bf{-b}}\,\frac{1}{P_{\widetilde{-}}}\,H_{{\bf{-a}}\,\go N}$\\
$\Rightarrow\,S_{{\bf{-b}}}\,H_{\alpha^{\prime}\,{\bf{-a}}}\,\rightsquigarrow\,0$\\
\hline
$S_{\widetilde{\bf{-b}}}\,H_{\alpha^{\prime}\,{\bf{-a}}}\,=\,-\,f_{\widetilde{-}\,\alpha^{\prime}\,\go M}\,\eta^{\go M \go N}\,S_{\widetilde{\bf{-b}}}\,\frac{1}{P_{\widetilde{-}}}\,H_{{\bf{-a}}\,\go N}$\\
$\Rightarrow\,S_{\widetilde{\bf{-b}}}\,H_{\alpha^{\prime}\,{\bf{-a}}}\,\rightsquigarrow\,0$\\
\hline
$S_{{\bf{-b}}}\,H_{\alpha^{\prime}\,{\bf{+a}}}\,=\,-\,(\gamma_{\bf{-b}})_{\alpha^{\prime}}{}^{\nu}\,f_{\widetilde{-}\,\nu\,\go M}\,\eta^{\go M \go N}\,\frac{1}{P_{\widetilde{-}}}\,H_{{\bf{+a}}\,\go N}\,-\,f_{\widetilde{-}\,\alpha^{\prime}\,\go M}\,\eta^{\go M \go N}\,S_{\bf{-b}}\,\frac{1}{P_{\widetilde{-}}}\,H_{{\bf{+a}}\,\go N}$\\
$\Rightarrow\,S_{{\bf{-b}}}\,H_{\alpha^{\prime}\,{\bf{+a}}}\,\rightsquigarrow\,0$\\
\hline
$S_{\widetilde{\bf{-b}}}\,H_{\alpha^{\prime}\,{\bf{+a}}}\,=\,-\,f_{\widetilde{-}\,\alpha^{\prime}\,\go M}\,\eta^{\go M \go N}\,S_{\widetilde{\bf{-b}}}\,\frac{1}{P_{\widetilde{-}}}\,H_{{\bf{+a}}\,\go N}$\\
$\Rightarrow\,S_{\widetilde{\bf{-b}}}\,H_{\alpha^{\prime}\,{\bf{+a}}}\,\rightsquigarrow\,0$\\
\hline
$S_{\bf{-b}}\,H_{\alpha^{\prime}\,\widetilde{\bf{-a}}}\,=\,-\,f_{-\,\widetilde{-}\,\go M}\,\eta^{\go M \go N}\,S_{\bf{-b}}\,\frac{1}{P_{\widetilde{-}}}\,D_{\alpha^{\prime}}\,\frac{1}{P_{-}}\,H_{\widetilde{\bf{-a}}\,\go N}\,-\,(\gamma_{\bf{-b}})_{\alpha^{\prime}}{}^{\nu}\,f_{\widetilde{-}\,\nu\,\go M}\,\eta^{\go M \go N}\,\frac{1}{P_{\widetilde{-}}}\,H_{\widetilde{\bf{-a}}\,\go N}$\\
$-\,f_{\widetilde{-}\,\alpha^{\prime}\,\go M}\,\eta^{\go M \go N}\,S_{\bf{-b}}\,\frac{1}{P_{\widetilde{-}}}\,H_{\widetilde{\bf{-a}}\,\go N}$\\
$\Rightarrow\,S_{\bf{-b}}\,H_{\alpha^{\prime}\,\widetilde{\bf{-a}}}\,\rightsquigarrow\,0$\\
\hline
$S_{\widetilde{\bf{-b}}}\,H_{\alpha^{\prime}\,\widetilde{\bf{-a}}}\,=\,-\,f_{-\,\widetilde{-}\,\go M}\,\eta^{\go M \go N}\,S_{\widetilde{\bf{-b}}}\,\frac{1}{P_{\widetilde{-}}}\,D_{\alpha^{\prime}}\,\frac{1}{P_{-}}\,H_{\widetilde{\bf{-a}}\,\go N}\,-\,f_{\widetilde{-}\,\alpha^{\prime}\,\go M}\,\eta^{\go M \go N}\,S_{\widetilde{\bf{-b}}}\,\frac{1}{P_{\widetilde{-}}}\,H_{\widetilde{\bf{-a}}\,\go N}$\\
$\Rightarrow\,S_{\widetilde{\bf{-b}}}\,H_{\alpha^{\prime}\,\widetilde{\bf{-a}}}\,\rightsquigarrow\,0$\\
\hline 
\end{tabular}
\caption{the $S$ action on $H_{D\,\widetilde{S}}$ vielbein}
\protect\label{hrncek4}
}
\end{table}
Now we calculate $S_{\bf{-b}}\,H_{\alpha^{\prime}\,\widetilde{\bf{+a}}}$. The reasoning will be similar later for the final calculation of the $S_{\bf{+b}}\,H_{\alpha^{\prime}\,\widetilde{\bf{+a}}}$ so we first do the former in order to see how it works. Calculation of the $S_{\bf{-b}}$ action on $H_{\alpha^{\prime}\,\widetilde{\bf{+a}}}$ is straightforward. It's done using the relation (\ref{snehulienka}) and applying $S_{\bf{-b}}$, thus we get:
\begin{eqnarray}
\label{zlavila}
S_{\bf{-b}}\,H_{\alpha^{\prime}\,\widetilde{\bf{+a}}}&=&-\,f_{-\,\widetilde{-}\,\go M}\,\eta^{\go M \go N}\,S_{\bf{-b}}\,{\Big{(}}\,\frac{1}{P_{\widetilde{-}}\,P_{-}}\,P_{\widetilde{a}}\,\frac{1}{P_{\widetilde{-}}}\,H_{\alpha^{\prime}\,\go N}\,+\,\frac{1}{P_{\widetilde{-}}}\,D_{\alpha^{\prime}}\,\frac{1}{P_{-}}\,H_{\widetilde{\bf{+a}}\,\go N}\,{\Big{)}}\,\\
&&-\,\frac{1}{2}\,(\gamma_{\bf{-b}})_{\alpha^{\prime}}{}^{\nu}\,f_{\nu\,\widetilde{-}\,\go M}\,\eta^{\go M \go N}\,\frac{1}{P_{\widetilde{-}}}\,H_{\widetilde{\bf{+a}}\,\go N}\,+\,f_{\alpha^{\prime}\,\widetilde{-}\,\go M}\,\eta^{\go M \go N}\,S_{\bf{-b}}\,\frac{1}{P_{\widetilde{-}}}\,H_{\widetilde{\bf{+a}}\,\go N}\nonumber\\
&&-\,f_{-\,\widetilde{\bf{a}}\,\go M}\,\eta^{\go M \go N}\,S_{\bf{-b}}\,\frac{1}{P_{\widetilde{-}}\,P_{-}}\,H_{\alpha^{\prime}\,\go N}\nonumber
\end{eqnarray}
Now, we want to evaluate equation (\ref{zlavila}). The terms proportional to $f_{-\,\widetilde{-}\,\go M}$ and $f_{\alpha^{\prime}\,\widetilde{-}\,\go M}$ are vanishing by the $AdS$ algebra. We write what's left over after evaluation: 
\begin{eqnarray}
\label{ohnado}
S_{\bf{-b}}\,H_{\alpha^{\prime}\,\widetilde{\bf{+a}}}&=&\,\frac{1}{2}\,(\gamma_{\bf{-b}})_{\alpha^{\prime}}{}^{\nu}\,(\gamma_{-})_{\nu\,\sigma}\,(\widetilde{\Gamma}_{5})^{\sigma\,\epsilon}\,\frac{1}{(r_{AdS})}\,\frac{1}{P_{\widetilde{-}}}\,H_{\widetilde{\bf{+a}}\,\widetilde{\epsilon}}\,\\
&&+\,\frac{1}{2\,(r_{AdS})^{2}}\,\frac{1}{P_{\widetilde{-}}\,P_{-}}\,S_{\bf{-b}}\,(\,H_{\alpha^{\prime}\,{\bf{+a}}}\,+\,H_{\alpha^{\prime}\,\widetilde{\bf{+a}}}\,\pm\,H_{\alpha^{\prime}\,{\bf{-a}}}\,\pm\,H_{\alpha^{\prime}\,\widetilde{\bf{-a}}}\,)\nonumber
\end{eqnarray}
Note, the $\pm$ in last line in (\ref{ohnado}) is explained in the section above (\ref{sedlak}). According to the table (\ref{hrncek4}) all actions of $S_{\bf{-b}}$ in the second line of (\ref{ohnado}) are evaluated to zero except of the $S_{\bf{-b}}\,H_{\alpha^{\prime}\,{\bf{+a}}}$ that we want to determine. Then the (\ref{ohnado}) could be rewritten in a way that determines $S_{\bf{-b}}\,H_{\alpha^{\prime}\,\widetilde{\bf{+a}}}$ (after evaluation):
\begin{eqnarray}
\label{lyft}
{\Big{(}}\,1\,-\,\frac{1}{2\,(r_{AdS})^{2}}\,\frac{1}{P_{\widetilde{-}}\,P_{-}}\,{\Big{)}}\,S_{\bf{-b}}\,H_{\alpha^{\prime}\,\widetilde{\bf{+a}}}\,=\,\frac{1}{2}\,(\gamma_{\bf{-b}})_{\alpha^{\prime}}{}^{\nu}\,(\gamma_{-})_{\nu\,\sigma}\,(\widetilde{\Gamma}_{5})^{\sigma\,\epsilon}\,\frac{1}{(r_{AdS})}\,\frac{1}{P_{\widetilde{-}}}\,H_{\widetilde{\bf{+a}}\,\widetilde{\epsilon}}
\end{eqnarray}
We remind that the $H_{\widetilde{\bf{+a}}\,\widetilde{\epsilon}}$ vielbein is related to the $H_{\widetilde{\bf{+a}}\,{\epsilon}}$ vielbein by the second top line in the table (\ref{hrncek3}).
Similarly to the (\ref{zlavila}) and its evaluated version (\ref{lyft}) we can calculate an action of $S_{\widetilde{\bf{-b}}}\,H_{\alpha^{\prime}\,\widetilde{\bf{+a}}}$. The result is: 
\begin{eqnarray}
\label{digital}
S_{\widetilde{\bf{-b}}}\,H_{\alpha^{\prime}\,\widetilde{\bf{+a}}}\,&=&-\,f_{-\,\widetilde{-}\,\go M}\,\eta^{\go M \go N}\,S_{\widetilde{\bf{-b}}}\,{\Big{(}}\,\frac{1}{P_{\widetilde{-}}\,P_{-}}\,P_{\widetilde{a}}\,\frac{1}{P_{\widetilde{-}}}\,H_{\alpha^{\prime}\,\go N}\,+\,\frac{1}{P_{\widetilde{-}}}\,D_{\alpha^{\prime}}\,\frac{1}{P_{-}}\,H_{\widetilde{\bf{+a}}\,\go N}\,{\Big{)}}\\
&&-\,f_{\alpha^{\prime}\,\widetilde{-}\,\go M}\,\eta^{\go M \go N}\,S_{\widetilde{\bf{-b}}}\,\frac{1}{P_{\widetilde{-}}}\,H_{\widetilde{\bf{+a}}\,\go N}\,-\,\eta_{\widetilde{\bf{a}}\,\widetilde{\bf{b}}}\,f_{-\widetilde{-}\,\go M}\,\eta^{\go M \go N}\,\frac{1}{P_{\widetilde{-}}\,P_{-}}\,H_{\alpha^{\prime}\,\go N}\nonumber\\
&&-\,f_{-\,\widetilde{\bf{a}}\,\go M}\,\eta^{\go M \go N}\,S_{\widetilde{\bf{-b}}}\,\frac{1}{P_{\widetilde{-}}\,P_{-}}\,H_{\alpha^{\prime}\,\go N}
\end{eqnarray}
and after evaluation, where we again use the results from table (\ref{hrncek4}):
\begin{eqnarray}
S_{\widetilde{\bf{-b}}}\,H_{\alpha^{\prime}\,\widetilde{\bf{+a}}}\,\rightsquigarrow\,0
\end{eqnarray}
Finally the action of $S_{{\bf{+b}}}\,H_{\alpha^{\prime}\,\widetilde{\bf{+a}}}$ is calculated as in (\ref{zlavila}). Now we know that by an analogy with the (\ref{zlavila}) and its evaluation we would need to know analogy of the table (\ref{hrncek4}) except now for the $S_{\bf{+b}}$. Since calculations are very analogous to those that led to the table (\ref{hrncek4}) we list just the resulting table(s): (\ref{hrncek5}) and (\ref{hrncek6})
\begin{table}[H]
{\centering
\begin{tabular}{|c|}
\noalign{\vspace{.1in}}
\hline
$S_{{\bf{+b}}}\,H_{\alpha^{\prime}\,{\bf{-a}}}\,=\,-\,\frac{1}{2}\,(\gamma_{{\bf{+b}}})_{\alpha^{\prime}}{}^{\nu}\,f_{\widetilde{-}\,\nu\,\go M}\,\eta^{\go M \go N}\,\frac{1}{P_{\widetilde{-}}}\,H_{{\bf{-a}}\,\go N}\,+\,f_{\widetilde{-}\,\alpha^{\prime}\,\go M}\,\eta^{\go M \go N}\,S_{{\bf{+b}}}\,\frac{1}{P_{\widetilde{-}}}\,H_{{\bf{-a}}\,\go N}$\\
$\Rightarrow\,S_{{\bf{+b}}}\,H_{\alpha^{\prime}\,{\bf{-a}}}\,\rightsquigarrow\,0$\\
\hline
$S_{\widetilde{\bf{+b}}}\,H_{\alpha^{\prime}\,{\bf{-a}}}\,=\,\eta_{\widetilde{-}\,\widetilde{+}}\,f_{\widetilde{{\bf{b}}}\,\alpha^{\prime}\,\go M}\,\eta^{\go M \go N}\,\frac{1}{P_{\widetilde{-}}}\,H_{{\bf{-a}}\,\go N}\,+\,f_{\widetilde{-}\,\alpha^{\prime}\,\go M}\,\eta^{\go M \go N}\,S_{\widetilde{{\bf{+b}}}}\,\frac{1}{P_{\widetilde{-}}}\,H_{{\bf{-a}}\,\go N}$\\ 
$\Rightarrow\,S_{\widetilde{\bf{+b}}}\,H_{\alpha^{\prime}\,{\bf{-a}}}\,\rightsquigarrow\,0$\\
\hline
$S_{\bf{+b}}\,H_{\alpha^{\prime}\,{\bf{+a}}}\,=\,-\,\frac{1}{2}\,(\gamma_{\bf{+b}})_{\alpha^{\prime}}{}^{\nu}\,f_{\widetilde{-}\,\nu\,\go M}\,\eta^{\go M \go N}\,\frac{1}{P_{\widetilde{-}}}\,H_{{\bf{+a}}\,\go N}\,+\,f_{\widetilde{-}\,\alpha^{\prime}\,\go M}\,\eta^{\go M \go N}\,S_{\bf{+b}}\,\frac{1}{P_{\widetilde{-}}}\,H_{{\bf{+a}}\,\go N}$\\
$\Rightarrow\,S_{\bf{+b}}\,H_{\alpha^{\prime}\,{\bf{+a}}}\,\rightsquigarrow\,-\,(\gamma_{\bf{+b}})_{\alpha^{\prime}}{}^{\nu}\,(\gamma_{-})_{\nu\,\sigma}\,(\widetilde{\Gamma}_{5})^{\sigma\,\lambda}\,\frac{1}{(r_{AdS})}\,\frac{1}{P_{\widetilde{-}}}\,H_{{\bf{+a}}\,\widetilde{\lambda}}$\\
\hline
$S_{\widetilde{\bf{+b}}}\,H_{\alpha^{\prime}\,{\bf{+a}}}\,=\,\eta_{\widetilde{-}\,\widetilde{+}}\,f_{\widetilde{\bf{b}}\,\alpha^{\prime}\,\go M}\,\eta^{\go M \go N}\,\frac{1}{P_{\widetilde{-}}}\,H_{{\bf{+a}}\,\go N}\,+\,f_{\widetilde{-}\,\alpha^{\prime}\,\go M}\,\eta^{\go M \go N}\,S_{\widetilde{\bf{+b}}}\,\frac{1}{P_{\widetilde{-}}}\,H_{{\bf{+a}}\,\go N}$\\
$\Rightarrow\,S_{\widetilde{\bf{+b}}}\,H_{\alpha^{\prime}\,{\bf{+a}}}\,\rightsquigarrow\,(\gamma_{\bf{b}})_{\alpha^{\prime}\,\nu}\,(\widetilde{\Gamma}_{5})^{\nu\,\sigma}\,\frac{1}{(r_{AdS})}\,\frac{1}{P_{\widetilde{-}}}\,H_{{\bf{+a}}\,\widetilde{\sigma}}$\\
\hline
\end{tabular}
\caption{the $S$ action on $H_{D\,\widetilde{S}}$ vielbein}
\protect\label{hrncek5}
}
\end{table}

\begin{table}[H]
{\centering
\begin{tabular}{|c|}
\noalign{\vspace{.1in}}
\hline
$S_{\bf{+b}}\,H_{\alpha^{\prime}\,\widetilde{\bf{-a}}}\,=\,-\,\frac{1}{2}\,(\gamma_{\bf{+b}})_{\alpha^{\prime}}{}^{\nu}\,f_{\widetilde{-}\,\nu\,\go M}\,\eta^{\go M \go N}\,\frac{1}{P_{\widetilde{-}}}\,H_{\widetilde{\bf{-a}}\,\go N}\,+\,f_{\widetilde{-}\,\alpha^{\prime}\,\go M}\,\eta^{\go M \go N}\,S_{\bf{+b}}\,\frac{1}{P_{\widetilde{-}}}\,H_{\widetilde{\bf{-a}}\,\go N}$\\
$+\,\eta_{-+}\,f_{{\bf{b}}\,\widetilde{-}\,\go M}\,\eta^{\go M \go N}\,\frac{1}{P_{\widetilde{-}}}\,D_{\alpha^{\prime}}\,\frac{1}{P_{-}}\,H_{\widetilde{{\bf{-a}}}\,\go N}\,+\,f_{-\,\widetilde{-}\,\go M}\,\eta^{\go M \go N}\,S_{{\bf{+b}}}\,\frac{1}{P_{\widetilde{-}}}\,D_{\alpha^{\prime}}\,\frac{1}{P_{-}}\,H_{\widetilde{{\bf{-a}}}\,\go N}$\\
$\Rightarrow\,S_{\bf{+b}}\,H_{\alpha^{\prime}\,\widetilde{\bf{-a}}}\,\rightsquigarrow\,0$\\
\hline
$S_{\widetilde{\bf{+b}}}\,H_{\alpha^{\prime}\,\widetilde{\bf{-a}}}\,=\,\eta_{\widetilde{-}\,\widetilde{+}}\,f_{-\,\widetilde{\bf{b}}\,\go M}\,\eta^{\go M \go N}\,\frac{1}{P_{\widetilde{-}}}\,D_{\alpha^{\prime}}\,\frac{1}{P_{-}}\,H_{\widetilde{\bf{-a}}\,\go N}\,+\,f_{-\,\widetilde{-}\,\go M}\,\eta^{\go M \go N}\,S_{\widetilde{\bf{+b}}}\,\frac{1}{P_{\widetilde{-}}}\,D_{\alpha^{\prime}}\,\frac{1}{P_{-}}\,H_{\widetilde{\bf{-a}}\,\go N}$\\
$+\,\eta_{\widetilde{-}\,\widetilde{+}}\,f_{\widetilde{\bf{b}}\,\alpha^{\prime}\,\go M}\,\eta^{\go M \go N}\,\frac{1}{P_{\widetilde{-}}}\,H_{\widetilde{\bf{-a}}\,\go N}\,+\,f_{\widetilde{-}\,\alpha^{\prime}\,\go M}\,\eta^{\go M \go N}\,S_{\widetilde{\bf{+b}}}\,\frac{1}{P_{\widetilde{-}}}\,H_{{\widetilde{\bf{-a}}}\,\go N}$\\
$\Rightarrow\,S_{\widetilde{\bf{+b}}}\,H_{\alpha^{\prime}\,\widetilde{\bf{-a}}}\,\rightsquigarrow\,0$\\
\hline
\end{tabular}
\caption{the $S$ action on $H_{D\,\widetilde{S}}$ vielbein}
\protect\label{hrncek6}
}
\end{table}

Now, we calculate the the missing piece in the equation (\ref{fulltime}), i.e. $S_{\bf{+b}}\,H_{\alpha^{\prime}\,\widetilde{\bf{+a}}}$. In analogy with (\ref{zlavila}) we get:
\begin{eqnarray}
\label{dato}
S_{\bf{+b}}\,H_{\alpha^{\prime}\,\widetilde{\bf{+a}}}&=&-\,f_{-\,\widetilde{-}\,\go M}\,\eta^{\go M \go N}\,S_{\bf{+b}}\,{\Big{(}}\,\frac{1}{P_{\widetilde{-}}\,P_{-}}\,P_{\widetilde{a}}\,\frac{1}{P_{\widetilde{-}}}\,H_{\alpha^{\prime}\,\go N}\,+\,\frac{1}{P_{\widetilde{-}}}\,D_{\alpha^{\prime}}\,\frac{1}{P_{-}}\,H_{\widetilde{\bf{+a}}\,\go N}\,{\Big{)}}\,\\
&&-\,\eta_{-+}\,f_{{\bf{b}}\,\widetilde{-}\,\go M}\,\eta^{\go M \go N}\,{\Big{(}}\,\frac{1}{P_{\widetilde{-}}\,P_{-}}\,P_{\widetilde{a}}\,\frac{1}{P_{\widetilde{-}}}\,H_{\alpha^{\prime}\,\go N}\,+\,\frac{1}{P_{\widetilde{-}}}\,D_{\alpha^{\prime}}\,\frac{1}{P_{-}}\,H_{\widetilde{\bf{+a}}\,\go N}\,{\Big{)}}\nonumber\\
&&+\,\frac{1}{2}\,(\gamma_{\bf{+b}})_{\alpha^{\prime}}{}^{\nu}\,f_{\nu\,\widetilde{-}\,\go M}\,\eta^{\go M \go N}\,\frac{1}{P_{\widetilde{-}}}\,H_{\widetilde{\bf{+a}}\,\go N}\,-\,f_{\alpha^{\prime}\,\widetilde{-}\,\go M}\,\eta^{\go M \go N}\,S_{\bf{+b}}\,\frac{1}{P_{\widetilde{-}}}\,H_{\widetilde{\bf{+a}}\,\go N}\nonumber\\
&&-\,f_{-\,\widetilde{\bf{a}}\,\go M}\,\eta^{\go M \go N}\,S_{\bf{+b}}\,\frac{1}{P_{\widetilde{-}}\,P_{-}}\,H_{\alpha^{\prime}\,\go N}\,+\,\eta_{-+}\,f_{\widetilde{{\bf{a}}}\,{\bf{b}}\,\go M}\,\eta^{\go M \go N}\,\frac{1}{P_{\widetilde{-}}\,P_{-}}\,H_{\alpha^{\prime}\,\go M}\nonumber
\end{eqnarray}
and the (partially) evaluate version of (\ref{dato}):
\begin{eqnarray}
\label{hilda}
S_{\bf{+b}}\,H_{\alpha^{\prime}\,\widetilde{\bf{+a}}}&\rightsquigarrow&\,-\frac{1}{2\,(r_{AdS})^{2}}\,\frac{1}{P_{\widetilde{-}}}\,D_{\alpha^{\prime}}\,\frac{1}{P_{-}}\,H_{\widetilde{\bf{+a}}\,{\bf{+b}}}\,-\,(\gamma_{\bf{+b}})_{\alpha^{\prime}}{}^{\nu}\,(\gamma_{-})_{\nu\,\sigma}\,(\widetilde{\Gamma}_{5})^{\sigma\,\lambda}\,\frac{1}{(r_{AdS})}\,\frac{1}{P_{\widetilde{-}}}\,H_{\widetilde{\bf{+a}}\,\widetilde{\lambda}}\nonumber\\
&&+\,\frac{1}{2\,(r_{AdS})^{2}}\,S_{{\bf{+b}}}\,\frac{1}{P_{\widetilde{-}}\,P_{-}}\,{\Big{(}}\,H_{\alpha^{\prime}\,{\bf{+a}}}\,+\,H_{\alpha^{\prime}\,\widetilde{\bf{+a}}}\,\pm\,H_{\alpha^{\prime}\,{\bf{-a}}}\,\pm\,H_{\alpha^{\prime}\,\widetilde{\bf{-a}}}{\Big{)}}
\end{eqnarray}
We can see why we just partially evaluated the equation (\ref{dato}). The reason is that last term leads to an action of $S_{\bf{+b}}$. Fortunately for us we already computed all those actions in tables (\ref{hrncek5}) and (\ref{hrncek6}) except $S_{\bf{+b}}\,H_{\alpha^{\prime}\,\widetilde{\bf{+a}}}$ that we want to calculate. Therefore the (\ref{hilda}) leads to the evaluated version of the $S_{\bf{+b}}\,H_{\alpha^{\prime}\,\widetilde{\bf{+a}}}$:
\begin{eqnarray}
\label{vilda}
{\Big{(}}\,1\,-\,\frac{1}{2\,(r_{AdS})^{2}}\,\frac{1}{P_{\widetilde{-}}\,P_{-}}\,{\Big{)}}\,S_{\bf{+b}}\,H_{\alpha^{\prime}\,\widetilde{\bf{+a}}}&\rightsquigarrow&\,-\,\frac{1}{2\,(r_{AdS})^{2}}\,\frac{1}{P_{\widetilde{-}}}\,D_{\alpha^{\prime}}\,\frac{1}{P_{-}}\,H_{\widetilde{\bf{+a}}\,{\bf{+b}}}\,\\
&&-\,\frac{1}{2}\,(\gamma_{\bf{+b}})_{\alpha^{\prime}}{}^{\nu}\,(\gamma_{-})_{\nu\,\sigma}\,(\widetilde{\Gamma}_{5})^{\sigma\,\lambda}\,\frac{1}{(r_{AdS})}\,\frac{1}{P_{\widetilde{-}}}\,H_{\widetilde{\bf{+a}}\,\widetilde{\lambda}}\nonumber\\
&&\,-\,(\gamma_{\bf{+b}})_{\alpha^{\prime}}{}^{\nu}\,(\gamma_{-})_{\nu\,\sigma}\,(\widetilde{\Gamma}_{5})^{\sigma\,\lambda}\,\frac{1}{2\,(r_{AdS})^{3}}\,\frac{1}{P_{\widetilde{-}}\,P_{-}\,P_{\widetilde{-}}}\,H_{{\bf{+a}}\,\widetilde{\lambda}}\nonumber
\end{eqnarray}
where we used results of tables (\ref{hrncek5}) and (\ref{hrncek6}). For completeness we provide (just the evaluated version) the last remaining part of tables (\ref{hrncek5}) and (\ref{hrncek6}), i.e. $S_{\widetilde{\bf{+b}}}\,H_{\alpha^{\prime}\,\widetilde{\bf{+a}}}$: 
\begin{eqnarray}
\label{zrzava}
{\Big{(}}\,1\,-\,\frac{1}{2\,(r_{AdS})^{2}}\,\frac{1}{P_{\widetilde{-}}\,P_{-}}\,{\Big{)}}\,S_{\widetilde{\bf{+b}}}\,H_{\alpha^{\prime}\,\widetilde{\bf{+a}}}&\rightsquigarrow&\,-\,\frac{1}{2\,(r_{AdS})^{2}}\,\frac{1}{P_{\widetilde{-}}}\,D_{\alpha^{\prime}}\,\frac{1}{P_{-}}\,H_{\widetilde{\bf{+a}}\,{\bf{+b}}}\,\\
&&-\,\frac{1}{2}\,(\gamma_{\bf{b}})_{\alpha^{\prime}\,{\nu}}\,(\widetilde{\Gamma}_{5})^{\nu\,\lambda}\,\frac{1}{(r_{AdS})}\,\frac{1}{P_{\widetilde{-}}}\,H_{\widetilde{\bf{+a}}\,\widetilde{\lambda}}\nonumber\\
&&\,-\,(\gamma_{\bf{b}})_{\alpha^{\prime}\,\nu}\,(\widetilde{\Gamma}_{5})^{\nu\,\lambda}\,\frac{1}{2\,(r_{AdS})^{3}}\,\frac{1}{P_{\widetilde{-}}\,P_{-}\,P_{\widetilde{-}}}\,H_{{\bf{+a}}\,\widetilde{\lambda}}\nonumber
\end{eqnarray}

We repeat the first important relation we derived by the above analysis from (\ref{fnukadlo}) where we add results from (\ref{verizon}) and (\ref{vilda}):
\begin{eqnarray}
\label{kukura11}
D_{\alpha^{\prime}}\,H_{{\bf{+a}}\,\widetilde{\bf{+b}}}&=&\,-\,\frac{1}{g}\,\frac{1}{(r_{AdS})\,P_{\widetilde{-}}}\,(\gamma_{{\bf{b}}})_{\alpha^{\prime}\,\sigma}\,(\widetilde{\Gamma}_{5})^{\sigma\,\beta}\,H_{\widetilde{\beta}\,{\bf{+a}}}\,+\,\frac{1}{2\,g}\,(\,1\,-\,\frac{1}{f}\,\frac{1}{(r_{AdS})^{2}\,P_{\widetilde{-}}\,P_{-}}\,)(\gamma_{\bf{+a}})_{\alpha^{\prime}}{}^{\beta}\,H_{\beta\,\widetilde{\bf{+b}}}\,\nonumber\\
&&+\,\frac{1}{f\,g}\,\frac{1}{2\,(r_{AdS})^{3}\,P_{-}\,(P_{\widetilde{-}})^{2}}\,\,(\gamma_{\bf{a}})_{\alpha^{\prime}\,\nu}\,(\widetilde{\Gamma}_{5})^{\nu\,\beta}\,H_{\widetilde{\beta}\,{\bf{+b}}}
\end{eqnarray}
where $f$ and $g$ are defined as follows:
\begin{eqnarray}
\label{kukura12}
f&\defeq&(\,1\,-\,\frac{1}{2\,(r_{AdS})^{2}\,P_{\widetilde{-}}\,P_{-}}\,)\\
g&\defeq&(\,1\,-\,\frac{1}{f}\,\frac{1}{2\,(r_{AdS})^{2}\,P_{\widetilde{-}}\,P_{-}\,}\,) \nonumber
\end{eqnarray}


\subsection{\texorpdfstring{The $H_{D\,\widetilde{D}}$ }{The H D D  }} 

To obtain the $AdS$ equation (\ref{gulasik}), we need to fix $H_{\alpha^{\prime}\,\beta}$. This term is fixed by the zero dimensional torsion $T_{\widetilde{P}\,D\,D}\,\equiv\,T_{\widetilde{-}\,\alpha^{\prime}\,\beta}\,=\,0$:
\begin{eqnarray}
\label{hrozienko}
T_{\widetilde{-}\,\alpha^{\prime}\,\beta}\,=\,0&=&P_{[\widetilde{-}}\,H_{\alpha^{\prime}\,\beta)}\,+\,H_{[\widetilde{-}\,|\,\go M}\,\eta^{\go M \go N}\,f_{\,\alpha^{\prime}\,\beta\,)\,\go N}\\
&=&P_{\widetilde{-}}\,H_{\alpha^{\prime}\,\beta}\,+\,H_{(\beta\,|\,\go M}\,\eta^{\go M \go N}\,f_{\alpha^{\prime})\,\widetilde{-}\,\go N}\nonumber\\
\label{bukvica}
\Rightarrow&&H_{\alpha^{\prime}\,\beta}\,=\,-\,f_{\widetilde{-}\,(\,\alpha^{\prime}\,|\,\go M}\,\eta^{\go M \go N}\,\frac{1}{P_{\widetilde{-}}}\,H_{\beta\,)\,\go N}\,\rightsquigarrow\,0
\end{eqnarray} 
In (\ref{hrozienko}) we again used the mixed light-cone gauge. In the flat case the mixed $f$ terms are zero so is $H_{\alpha^{\prime}\,\beta}$. In the $AdS$ case (after evaluation), term proportional to $f_{\alpha^{\prime}\,\widetilde{-}\,\go N}$ is zero because of $(\gamma_{-})_{\alpha^{\prime}\,\beta^{\prime}}\,=\,0$. But the term proportional to $f_{\beta\,\widetilde{-}\,\go N}$ is nonzero. Luckily for us the $f_{\beta\,\widetilde{-}\,\go N}\,\propto\,\frac{1}{r_{AdS}}\,(\gamma_{-})_{\beta\,\sigma}\,(\widetilde{\Gamma}_{5})^{\sigma\,\nu}\,\eta_{\widetilde{\nu}\,\go N}$. That structure constant just eats up the $\beta$ index and returns $\widetilde{\nu}$ index with some fixed constant dependence. The torsion constraint $T_{P\,\widetilde{D}\,D}\,\equiv\,T_{-\,\widetilde{\sigma}\,\alpha^{\prime}}\,=\,0$ relates $H_{\widetilde{\sigma}\,\alpha^{\prime}}$ back to $H_{\sigma\,\alpha^{\prime}}$ (after the evaluation). From that and assuming some wider invertibility ($P_{-}$ and $P_{\widetilde{-}}$ are bigger than some constant lower bound in $AdS$) we get also in the $AdS$ space $H_{\alpha^{\prime}\,\beta}\,\rightsquigarrow\,0$ (after the evaluation). 

We apply $S_{\widetilde{\bf{+a}}}$ on the result of non-evaluated (\ref{bukvica}), thus get:
\begin{eqnarray}
\label{srdiecko}
S_{\widetilde{\bf{+a}}}\,H_{\alpha^{\prime}\,\beta}&=&-\,\eta_{\widetilde{-}\,\widetilde{+}}\,f_{\widetilde{\bf{a}}\,(\,\alpha^{\prime}\,|\,\go M}\,\eta^{\go M \go N}\,\frac{1}{P_{\widetilde{-}}}\,H_{\beta\,)\,\go N}\,-\,f_{\widetilde{-}\,(\,\alpha^{\prime}\,|\,\go M}\,\eta^{\go M \go N}\,S_{\widetilde{\bf{+a}}}\,\frac{1}{P_{\widetilde{-}}}\,H_{\beta\,)\,\go N}\\
\label{dlhan}
S_{\widetilde{\bf{+a}}}\,H_{\alpha^{\prime}\,\beta}&\rightsquigarrow&-\,(\gamma_{-})_{\beta\,\nu}\,(\widetilde{\Gamma}_{5})^{\nu\,\sigma}\,\frac{1}{r_{AdS}}\,S_{\widetilde{\bf{+a}}}\,\frac{1}{P_{\widetilde{-}}}\,H_{\alpha^{\prime}\,\widetilde{\sigma}}\,+\,(\gamma_{\bf{a}})_{\alpha^{\prime}\,\nu}\,(\widetilde{\Gamma}_{5})^{\nu\,\sigma}\,\frac{1}{r_{AdS}}\,\frac{1}{P_{\widetilde{-}}}\,H_{\beta\,\widetilde{\sigma}}
\end{eqnarray}
The last term in (\ref{dlhan}) does not bother us too much (it will be a part of the pre-potential), the first term in (\ref{dlhan}) is actually something we need to evaluate. For that we need to fix $H_{\alpha^{\prime}\,\widetilde{\sigma}}$. That could be done by the torsion constraint $T_{P\,\widetilde{D}\,D}\,\equiv\,T_{-\,\alpha^{\prime}\,\widetilde{\sigma}}\,=\,0$:
\begin{eqnarray}
\label{filialka}
T_{-\,\alpha^{\prime}\,\widetilde{\sigma}}\,=\,0\,&=&P_{[\,{-}}\,H_{\alpha^{\prime}\,\widetilde{\sigma}\,)}\,+\,H_{[\,-\,|\,\go M}\,\eta^{\go M \go N}\,f_{\alpha^{\prime}\,\widetilde{\sigma}\,)\,\go N}\\
\label{filialka1}
&=&P_{-}\,H_{\alpha^{\prime}\,\widetilde{\sigma}}\,-\,D_{\widetilde{\sigma}}\,H_{-\,\alpha^{\prime}}\,+\,H_{-\,\go M}\,\eta^{\go M \go N}\,f_{\alpha^{\prime}\,\widetilde{\sigma}\,\go N}\,+\,H_{\alpha^{\prime}\,\go M}\,\eta^{\go M \go N}\,f_{\widetilde{\sigma}\,-\,\go N}
\end{eqnarray}
moreover the $H_{-\,\alpha^{\prime}}$ has been fixed in (\ref{zvlhnutie}), plugging that into (\ref{filialka1}) we get fixing of $H_{\alpha^{\prime}\,\widetilde{\sigma}}$:
\begin{eqnarray}
\label{zmena}
H_{\alpha^{\prime}\,\widetilde{\sigma}}\,=\,f_{-\,\widetilde{-}\,\go M}\,\eta^{\go M \go N}\,\frac{1}{P_{{-}}}\,D_{\widetilde{\sigma}}\,\frac{1}{P_{\widetilde{-}}}\,H_{\alpha^{\prime}\,\go N}\,-\,f_{\alpha^{\prime}\,\widetilde{\sigma}\,\go M}\,\eta^{\go M \go N}\,\frac{1}{P_{-}}\,H_{-\,\go N}\,+\,f_{\widetilde{\sigma}\,-\,\go M}\,\eta^{\go M \go N}\,\frac{1}{P_{-}}\,H_{\alpha^{\prime}\,\go N}
\end{eqnarray}
We are ready to calculate $S_{\widetilde{\bf{+a}}}\,H_{\alpha^{\prime}\,\widetilde{\sigma}}$ i.e. the term needed in (\ref{dlhan}):
\begin{eqnarray}
\label{duha}
S_{\widetilde{\bf{+a}}}\,H_{\alpha^{\prime}\,\widetilde{\sigma}}&=&\,\eta_{\widetilde{-}\,\widetilde{+}}\,f_{-\,\widetilde{{\bf{a}}}\,\go M}\,\eta^{\go M \go N}\,\frac{1}{P_{{-}}}\,D_{\widetilde{\sigma}}\,\frac{1}{{P_{\widetilde{-}}}}\,H_{\alpha^{\prime}\,\go N}\,+\,f_{-\,\widetilde{-}\,\go M}\,\eta^{\go M \go N}\,S_{\widetilde{\bf{+a}}}\,\frac{1}{P_{{-}}}\,D_{\widetilde{\sigma}}\,\frac{1}{P_{\widetilde{-}}}\,H_{\alpha^{\prime}\,\go N}\\
&&+\,\frac{1}{2}\,(\gamma_{\bf{+a}})_{\sigma}{}^{\nu^{\prime}}\,f_{\alpha^{\prime}\,\widetilde{\nu}^{\prime}\,\go M}\,\eta^{\go M \go N}\,\frac{1}{P_{-}}\,H_{-\,\go N}\,-\,f_{\alpha^{\prime}\,\widetilde{\sigma}\,\go M}\,\eta^{\go M \go N}\,S_{\widetilde{\bf{+a}}}\,\frac{1}{P_{-}}\,H_{-\,\go N}\nonumber\\
&&-\,\frac{1}{2}\,(\gamma_{\bf{+a}})_{\sigma}{}^{\nu^{\prime}}\,f_{\widetilde{\nu}^{\prime}\,-\,\go M}\,\eta^{\go M \go N}\,\frac{1}{P_{-}}\,H_{\alpha^{\prime}\,\go N}\,+\,f_{\widetilde{\sigma}\,-\,\go M}\,\eta^{\go M \go N}\,S_{\widetilde{\bf{+a}}}\,\frac{1}{P_{-}}\,H_{\alpha^{\prime}\,\go N}\nonumber
\end{eqnarray}
Now, we can evaluate (\ref{duha}), for clearness we include terms that we already know are evaluated to zero or are zero by the mixed light-cone gauge:
\begin{eqnarray}
\label{mix1}
S_{\widetilde{\bf{+a}}}\,H_{\alpha^{\prime}\,\widetilde{\sigma}}&\rightsquigarrow&\frac{1}{2\,(r_{AdS})^{2}}\,\frac{1}{P_{{-}}}\,D_{\widetilde{\sigma}}\,\frac{1}{P_{\widetilde{-}}}\,(\,H_{\widetilde{\alpha}^{\prime}\,{\bf{+a}}}\,+\,H_{\widetilde{\alpha}^{\prime}\,\widetilde{\bf{+a}}}\,\pm\,H_{\widetilde{\alpha}^{\prime}\,{\bf{-a}}}\,\pm\,H_{\widetilde{\alpha}^{\prime}\,\widetilde{\bf{-a}}}\,)\\
\label{mix2}
&&(\,\pm\,)\,\frac{1}{2}\,\frac{1}{r_{AdS}}\,(\gamma_{\bf{+a}})_{\sigma}{}^{\nu^{\prime}}\,(\gamma^{\bf{cd}})_{\nu^{\prime}}{}^{\rho^{\prime}}\,(\widetilde{\Gamma}_{5})_{\rho^{\prime}\,\alpha^{\prime}}\,\frac{1}{2\,P_{-}}\,(\,H_{{-}\,{\bf{cd}}}\,+\,H_{{-}\,\widetilde{\bf{cd}}}\,)\\
\label{mix3}
&&-\,(\,\pm\,)\,\frac{1}{r_{AdS}}\,(\gamma^{\bf{cd}})_{\sigma}{}^{\rho^{\prime}}\,(\widetilde{\Gamma}_{5})_{\rho^{\prime}\,\alpha^{\prime}}\,S_{\widetilde{\bf{+a}}}\,\frac{1}{2\,P_{-}}\,(\,H_{{-}\,{\bf{cd}}}\,+\,H_{{-}\,\widetilde{\bf{cd}}}\,)\\
\label{mix4}
&&\,-\,\frac{1}{r_{AdS}}\,(\gamma_{-})_{\sigma\,\nu}\,(\widetilde{\Gamma}_{5})^{\nu\,\rho}\,S_{\widetilde{\bf{+a}}}\,\frac{1}{P_{-}}\,H_{\alpha^{\prime}\,\rho}
\end{eqnarray}
Note that in the lines (\ref{mix2}) and (\ref{mix3}) we have the $\pm$ symbol. It comes from the mixed structure constant $f_{D\,\widetilde{D}\,\Sigma^{{\bf{d}}}}\,\equiv\,f_{\underline{\alpha}\,\widetilde{\underline{\beta}}}{}^{{\bf{cd}}}$, where underline indices are now (and just now) the $SO(\,10\,)$ chiral indices (for the left and right algebra), and $\Sigma^{{\bf{d}}}$ is the $\Sigma$ index for the $SO(\,5\,)\,\otimes\,SO(\,5\,)$ diagonal subgroup of the original $SO(\,10\,)\,\otimes\,SO(\,10\,)$ group.  The $(\,\pm\,)$ symbol determines to which $SO(\,5\,)$ of the diagonal subgroup given $\Sigma^{{\bf{d}}}$ belongs. This mixed structure constant could be written without the $\pm$ symbols as $f_{\underline{\alpha}\,\widetilde{\underline{\beta}}}{}^{{\bf{cd}}}\,=\,\frac{1}{r_{AdS}}\,(\gamma^{\bf{[c}})_{\sigma\,\rho}\,(\widetilde{\Gamma}_{5})^{\rho\,\nu}\,(\gamma^{\bf{d}]})_{\nu\,\alpha^{\prime}}$. The $\pm$ then comes from the fact that by the construction $\widetilde{\Gamma}_{5}$ commutes with ${\bf{a}}\,\in\,\{\,10,\,1\,\dots\,4\,\}$ and anti-commutes with ${\bf{a}}\,\in\,\{\,5,\dots\,9\,\}$. We used the prior definition in this section for some convenience. In the final expressions we will always use the definition without the $\pm$ symbol.

Let's evaluate expressions (\ref{mix1}), (\ref{mix2}), (\ref{mix3}) and (\ref{mix4}). The line (\ref{mix1}) is evaluated to $0$ by the table (\ref{hrncek2}).  We note very important property in the lines (\ref{mix2}) and (\ref{mix3}). The summation over the ${\bf{cd}}$ indices is really just a summation over the $SO\,(\,5\,)\,\otimes\,SO\,(\,5\,)$ diagonal subgroup of the full $SO\,(\,10\,)\,\otimes\,SO\,(\,10\,)$. The line (\ref{mix2}) is evaluated to $0$ by the mixed light-cone gauge (second term) and by the following fixing of the $H_{-\,{\bf{cd}}}$ (coming from torsion constraint $T_{\widetilde{P}\,P\,S}\,\equiv\,T_{\widetilde{-}\,-\,{\bf{ab}}}\,=\,0$):
\begin{eqnarray}
\label{centohubka}
H_{-\,{\bf{ab}}}&=&f_{-\,\widetilde{-}\,\go M}\,\eta^{\go M \go N}\,\frac{1}{P_{\widetilde{-}}}\,H_{{\bf{ab}}\,\go N}\,\,\,\Rightarrow\,\,\,H_{-\,{\bf{ab}}}\,\rightsquigarrow\,0
\end{eqnarray}
The line (\ref{mix4}) has an action $S_{\widetilde{\bf{+a}}}\,H_{\alpha^{\prime}\,\rho}$ that is exactly what we want to determine. The line (\ref{mix3}) is fixed as follows. The vielbein $H_{-\,\widetilde{\bf{cd}}}\,=\,0$ by the mixed light-cone gauge. The action $S_{\widetilde{\bf{+a}}}\,H_{-\,{\bf{cd}}}$ is however nontrivial. We should take fixing (\ref{centohubka}) and apply $S_{\widetilde{\bf{+a}}}$:
\begin{eqnarray}
\label{datel}
S_{\widetilde{\bf{+a}}}\,H_{-\,{\bf{cd}}}&=&-\,f_{-\,\widetilde{\bf{a}}\,\go M}\,\eta^{\go M \go N}\,\frac{1}{P_{\widetilde{-}}}\,H_{{\bf{cd}}\,\go N}\,+\,f_{-\,\widetilde{-}\,\go M}\,\eta^{\go M \go N}\,S_{\bf{+a}}\,\frac{1}{P_{\widetilde{-}}}\,H_{{\bf{cd}}\,\go N}\\
\label{oradlo}
S_{\widetilde{\bf{+a}}}\,H_{-\,{\bf{cd}}}&\rightsquigarrow&\,\frac{1}{2\,(r_{AdS})^{2}}\,\frac{1}{P_{\widetilde{-}}}\,(\,H_{{\bf{cd}}\,{\bf{+a}}}\,+\,H_{{\bf{cd}}\,\widetilde{\bf{+a}}}\,\pm\,H_{{\bf{cd}}\,{\bf{-a}}}\,\pm\,H_{{\bf{cd}}\,\widetilde{\bf{-a}}}\,)
\end{eqnarray}
By the table (\ref{hrncek1}) the only nonzero term in (\ref{oradlo}) is $H_{{\bf{+d}}\,\widetilde{\bf{+a}}}$ thus we get:
\begin{eqnarray}
\label{slack}
S_{\widetilde{\bf{+a}}}\,H_{-\,{\bf{+d}}}&\rightsquigarrow&\,\frac{1}{2\,(r_{AdS})^{2}}\,\frac{1}{P_{\widetilde{-}}}\,H_{{\bf{+d}}\,\widetilde{\bf{+a}}}\,
\end{eqnarray}
Then finally the equation (\ref{mix1} till \ref{mix4}) is evaluated to:
\begin{eqnarray}
\label{smersd}
S_{\widetilde{\bf{+a}}}\,H_{\alpha^{\prime}\,\widetilde{\sigma}}&\rightsquigarrow&\,\pm\,\frac{(\,-\,1\,)}{4\,(r_{AdS})^{3}\,P_{-}\,P_{\widetilde{-}}}\,(\gamma^{\bf{+d}})_{\sigma}{}^{\rho^{\prime}}\,(\widetilde{\Gamma}_{5})_{\rho^{\prime}\,\alpha^{\prime}}\,H_{{\bf{+d}}\,\widetilde{\bf{+a}}}\,\\
&&-\,\frac{1}{(r_{AdS})\,P_{-}}\,(\gamma_{-})_{\sigma\,\nu}\,(\widetilde{\Gamma}_{5})^{\nu\,\rho}\,S_{\widetilde{\bf{+a}}}\,H_{\alpha^{\prime}\,\rho}\nonumber
\end{eqnarray}
Combining (\ref{smersd}) and (\ref{dlhan}) we will get the following relation for the evaluated action of $S_{\widetilde{\bf{+a}}}\,H_{\alpha^{\prime}\,\beta}$:
\begin{eqnarray}
\label{leniak}
(\,1\,-\,\frac{1}{P_{-}\,P_{\widetilde{-}}\,(r_{AdS})^{2}}\,)\,S_{\widetilde{\bf{+a}}}\,H_{\alpha^{\prime}\,\beta}&=&\,-\,\frac{1}{4\,(r_{AdS})^{4}\,P_{-}\,P_{\widetilde{-}}{}^{2}}\,(\widetilde{\Gamma}_{5})_{\beta\,\nu}\,(\gamma^{\bf{d}})^{\nu\,\rho^{\prime}}\,(\widetilde{\Gamma}_{5})_{\rho^{\prime}\,\alpha^{\prime}}\,(\widetilde{\Gamma}_{5})_{{\bf{d}}}{}^{{\bf{g}}}\,H_{{\bf{+g}}\,\widetilde{\bf{+a}}}\,\nonumber\\
&&+\,\frac{1}{(r_{AdS})\,P_{\widetilde{-}}}\,(\gamma_{\bf{a}})_{\alpha^{\prime}\,\nu}\,(\widetilde{\Gamma}_{5})^{\nu\,\sigma}\,H_{\beta\,\widetilde{\sigma}}
\end{eqnarray}
where we simplified (\ref{smersd}) by using the explicit property of $\gamma_{-}$ and $\gamma_{+}$ being the unit or zero matrix (depending on specific indices), see (\ref{con}). We used this simplification in another equations as well (for example in equation (\ref{kukura1})). In the equation (\ref{leniak}) we also used new matrix $(\widetilde{\Gamma}_{5})_{{\bf{d}}}{}^{{\bf{g}}}$, that was be introduced in (\ref{proj}). We also used the identity (\ref{iden}) to simplify (\ref{leniak}). Plugging the evaluated expression (\ref{leniak}) into (\ref{gulasik}) we will get the action of $D_{\alpha^{\prime}}\,H_{\beta\,\widetilde{\bf{+a}}}$:
\begin{eqnarray}
\label{vonadlo1}
0&=&D_{\alpha^{\prime}}\,H_{\beta\,\widetilde{\bf{+a}}}\,-\,\frac{1}{h\,4\,(r_{AdS})^{4}\,P_{-}\,P_{\widetilde{-}}{}^{2}}\,(\gamma^{\bf{c}})_{\alpha^{\prime}\,\beta}\,H_{{\bf{+c}}\,\widetilde{\bf{+a}}}\,\\
&&\,+\,2\,(\gamma^{\bf{c}})_{\alpha^{\prime}\,\beta}\,H_{\widetilde{\bf{+a}}\,{\bf{c}}}\,+\,\frac{1}{h\,(r_{AdS})\,P_{\widetilde{-}}}\,(\gamma_{\bf{a}})_{\alpha^{\prime}\,\nu}\,(\widetilde{\Gamma}_{5})^{\nu\,\sigma}\,H_{\beta\,\widetilde{\sigma}}\,\nonumber\\
\label{vonadlo2}
&\mbox{left}\,\leftrightarrow\,\mbox{right}&
\end{eqnarray} 
where $h$ is defined as:
\begin{eqnarray}
\label{definitionofh}
h\,\defeq\,(\,1\,-\,\frac{1}{P_{-}\,P_{\widetilde{-}}\,(r_{AdS})^{2}}\,)
\end{eqnarray}
The equations (\ref{vonadlo1}) and (\ref{vonadlo2}) are very interesting since after applying the $D_{\alpha^{\prime}}$ (or $D_{\widetilde{\alpha}^{\prime}}$) derivatives we are getting terms like $H_{\widetilde{\bf{+a}}\,{\bf{c}}}$ that is basically our original $H_{\widetilde{\bf{+a}}\,{\bf{+c}}}$, see (\ref{vidlak}). Moreover we got also term $H_{\beta\,\widetilde{\sigma}}$ that is a new term and was important in chapters where we constructed the pre-potential.

Another important derivatives are $D_{\widetilde{\alpha}^{\prime}}$ on $H_{\beta\,\widetilde{\bf{+a}}}$ and $D_{\alpha^{\prime}}$ on $H_{\widetilde{\beta}\,\bf{+a}}$. We will look at those closer:     
\begin{eqnarray}
\label{kikinka}
T_{\widetilde{D}\,D\,\widetilde{S}}\,\equiv\,T_{\widetilde{\alpha}^{\prime}\,\beta\,\widetilde{\bf{+a}}}\,=\,0&=&D_{[\widetilde{\alpha}^{\prime}}\,H_{\beta\,\widetilde{\bf{+a}})}\,+\,H_{[\widetilde{\alpha}^{\prime}\,|\,\go M}\,\eta^{\go M \go N}\,f_{\beta\,\widetilde{\bf{+a}}\,)\,\go N}\\
\label{mikinka}
&=&D_{\widetilde{\alpha}^{\prime}}\,H_{\beta\,\widetilde{\bf{+a}}}\,+\,S_{\widetilde{\bf{+a}}}\,H_{\widetilde{\alpha}^{\prime}\,\beta}\,-\,D_{\beta}\,H_{\widetilde{\bf{+a}}\,\widetilde{\alpha}^{\prime}}\,+\,H_{\widetilde{\alpha}^{\prime}\,\go M}\,\eta^{\go M \go N}\,f_{\beta\,\widetilde{\bf{+a}}\,\go N}\,\nonumber\\
&&+\,H_{\widetilde{\bf{+a}}\,\go M}\,\eta^{\go M \go N}\,f_{\widetilde{\alpha}^{\prime}\,\beta\,\go N}\,-\,H_{\beta\,\go M}\,\eta^{\go M \go N}\,f_{\widetilde{\bf{+a}}\,\widetilde{\alpha}^{\prime}\,\go N}
\end{eqnarray}
The mixed $f$ terms are zero in the flat superspace. In the $AdS$ case the $f_{\widetilde{\alpha}^{\prime}\,\beta\,\go N}\,\neq\,0$ and there is also $AdS$ contribution coming from $S_{\widetilde{\bf{+a}}}\,H_{\widetilde{\alpha}^{\prime}\,\beta}$. This contribution can be calculated by analogy with the equations (\ref{mix1}), (\ref{mix2}), (\ref{mix3}), (\ref{mix4}) and (\ref{leniak}). Thus getting evaluated action $S_{\widetilde{\bf{+a}}}\,H_{\widetilde{\alpha}^{\prime}\,\beta}$:
\begin{eqnarray}
\label{smutnysom}
S_{\widetilde{\bf{+a}}}\,H_{\widetilde{\alpha}^{\prime}\,\sigma}&=&\pm\,\frac{(\,-\,1\,)}{h\,4\,(r_{AdS})^{3}\,P_{\widetilde{-}}\,P_{-}}\,(\gamma^{\bf{+d}})_{\alpha^{\prime}}{}^{\nu}\,(\widetilde{\Gamma}_{5})_{\nu\,\sigma}\,H_{\widetilde{\bf{+d}}\,{\bf{+a}}}\\
&&+\,\frac{1}{h\,(r_{AdS})^{2}\,P_{\widetilde{-}}\,P_{-}}\,(\gamma_{-})_{\sigma\,\nu}\,(\widetilde{\Gamma}_{5})^{\nu\,\rho}\,H_{\widetilde{\rho}\,\beta}\,(\widetilde{\Gamma}_{5})^{\beta\,\epsilon}\,(\gamma_{\bf{a}})_{\epsilon\,\alpha^{\prime}}\,\nonumber
\end{eqnarray}
where $h$ was defined in (\ref{definitionofh}). The (\ref{mikinka}) mixed structure constant $f_{\beta\,\widetilde{\bf{+a}}\,\go N}\,=\,0$ and the $f_{\widetilde{\alpha}^{\prime}\,\beta\,\go N}$ has been discussed before (see equations (\ref{duha}) and (\ref{mix1}) till (\ref{mix4})). Moreover, the vielbein $H_{\widetilde{\bf{+a}}\,\widetilde{\alpha}^{\prime}}$ is evaluated to $0$, see table (\ref{hrncek2}). Evaluating everything in (\ref{smutnysom}) we get:
\begin{eqnarray}
\label{latenight}
0\,&=&D_{\widetilde{\alpha}^{\prime}}\,H_{\beta\,\widetilde{\bf{+a}}}\,-\,\frac{1}{h\,4\,(r_{AdS})^{3}\,P_{\widetilde{-}}\,P_{-}}\,(\gamma^{\bf{d}})_{\alpha^{\prime}\,\epsilon}\,(\widetilde{\Gamma}_{5})^{\epsilon\,\sigma}\,\gamma^{{\bf{+}}}{}_{\sigma\,\beta}\,H_{\widetilde{\bf{+d}}\,{\bf{+a}}}\,\\
&&+\,\frac{1}{r_{AdS}}\,(\gamma^{\bf{d}})_{\alpha^{\prime}\,\epsilon}\,(\widetilde{\Gamma}_{5})^{\epsilon\,\sigma}\,(\gamma^{+})_{\sigma\,\beta}\,H_{{\bf{+d}}\,\widetilde{{\bf{+a}}}}\nonumber\\
&&+\,\frac{1}{h\,(r_{AdS})^{2}\,P_{\widetilde{-}}\,P_{-}}\,(\gamma_{-})_{\beta\,\nu}\,(\widetilde{\Gamma}_{5})^{\nu\,\rho}\,H_{\widetilde{\rho}\,\lambda}\,(\widetilde{\Gamma}_{5})^{\lambda\,\epsilon}\,(\gamma_{\bf{a}})_{\epsilon\,\alpha^{\prime}}\,-\,\frac{1}{2}\,(\gamma_{\bf{+a}})_{\alpha^{\prime}}{}^{\nu}\,H_{\beta\,\widetilde{\nu}}\nonumber\\
\label{latenight1}
\mbox{left}&\leftrightarrow&\mbox{right}
\end{eqnarray}

\end{document}